\documentclass[aps,onecolumn,nofootinbib,preprintnumbers]{revtex4}
\usepackage{amsmath,amssymb,epsfig}
\usepackage{times}
\usepackage{latexsym}
\usepackage{multirow}
\usepackage{url}
\usepackage{color}

\usepackage{graphicx}% Include figure files
\usepackage{epsfig}% Include figure files
\usepackage{amssymb}
\usepackage{color}
\usepackage{slashed}
\usepackage{soul}
\DeclareMathAlphabet   {\mathsc}{OT1}{cmr}{m}{sc}

\def\[{\left [}
\def\]{\right ]}
\def\({\left (}
\def\){\right )}
\newcommand{\lang}{\left\langle}
\newcommand{\rang}{\right\rangle}

\newcommand{\oline}[1]{\overline{#1}}

\newcommand{\GeV}      {~\mathrm{GeV}}
\newcommand{\TeV}      {~\mathrm{TeV}}

\newcommand{\STR}      {\mathsc{str}}
\newcommand{\SUSY}     {\mathsc{susy}}

\newcommand{\order}{\mathcal{O}}
\newcommand{\re}{{\rm Re}}

\newcommand{\gappeq}{\mathrel{\rlap {\raise.5ex\hbox{$>$}}
{\lower.5ex\hbox{$\sim$}}}}
\newcommand{\lappeq}{\mathrel{\rlap{\raise.5ex\hbox{$<$}}
{\lower.5ex\hbox{$\sim$}}}}

\newcommand{\comment}[1]{}
\newcommand{\MET}{\slashed{E}_T}

\begin{document}

\thispagestyle{empty}

\title{Mirage Models Confront the LHC:\\ III. Deflected Mirage Mediation}
\author{Lisa L. Everett$^1$, Todd Garon$^1$, Bryan L. Kaufman$^2$, and Brent D. Nelson$^2$}
\affiliation{$^1$Department of Physics, University of Wisconsin-Madison, Madison, WI 53706, USA\\ $^2$Department of Physics, Northeastern University, Boston, MA 02115, USA
}

\begin{abstract}
We complete the study of a class of string-motivated effective supergravity theories in which modulus-induced soft supersymmetry breaking is sufficiently suppressed in the observable sector so as to be competitive with anomaly-mediated supersymmetry breaking. Here we consider deflected mirage mediation (DMM), where contributions from gauge mediation are added to those arising from gravity mediation and anomaly mediation. We update previous work that surveyed the rich parameter space of such theories, in light of data from the CERN Large Hadron Collider (LHC) and recent dark matter detection experiments. Constraints arising from LHC superpartner searches at $\sqrt{s} = 8\,{\rm TeV}$ are considered, and discovery prospects at $\sqrt{s} = 14\,{\rm TeV}$ are evaluated. We find that deflected mirage mediation generally allows for SU(3)-charged superpartners of significantly lower mass (given current knowledge of the Higgs mass and neutralino relic density) than was found for the `pure' mirage mediation models of Kachru et al.~\cite{Kachru:2003aw}. Consequently, discovery prospects are enhanced for many combinations of matter multiplet modular weights. We examine the experimental challenges that will arise due to the prospect of highly compressed spectra in DMM, and the correlation between accessibility at the LHC and discovery prospects at large-scale liquid xenon dark matter detectors.
\end{abstract}

\maketitle

\section{Introduction}

With the resumption of data-taking at the CERN Large Hadron Collider (LHC), time is running short for the theoretical community to examine the impact that searches for superpartners has had on well-motivated models of supersymmetry breaking. 
Of the models with some theoretical support, one of the most well-studied is the so-called mirage model~\cite{Choi:2004sx,Choi:2005ge}. In this scenario, the dynamical supersymmetry breaking triggered by strong coupling in a hidden sector is connected to the observable sector in a manner that is suppressed, thus allowing loop-induced Weyl anomaly contributions to soft supersymmetry breaking to be of comparable size to tree-level contributions. Surprisingly, this is a common outcome of many well-motivated string constructions~\cite{Choi:2007ka,Gaillard:2007jr,Nath:2010zj}. The phenomenology of these models, in terms of LHC observables, has been recently described in~\cite{Kaufman:2013pya} for heterotic models, and in~\cite{Kaufman:2013oaa} for Type~IIB orientifold models. In this paper we generalize these results to the case of ``deflected mirage mediation''~\cite{Everett:2008qy,Everett:2008ey}. In this paradigm, a direct connection between a hidden sector, in which supersymmetry is broken, and the observable MSSM sector is contemplated, in which gauge-mediated contributions to soft supersymmetry breaking are of the same magnitude as those from gravity-induced terms. As such, deflected mirage mediation (DMM) is a natural generalization of the simple mirage models, and produces a theory space with the greatest possible richness for exploring current and future LHC supersymmetry searches.

The $\sqrt{s}=7$~TeV and $\sqrt{s}=8$~TeV runs of the LHC resulted in the triumphant discovery of the Higgs boson. However, as the LHC paused to upgrade to higher energies and luminosities, the various searches for TeV-scale supersymmetry have thus far been fruitless. 
Previous research in the area of mirage models has suggested the following broad observations. 
K\"ahler-stabilized heterotic models involve very few free parameters, and thus robust predictions are possible. If the hidden sector gaugino condensate involves $E_6$ or smaller-rank gauge groups, then the gluino is generally well below 3~TeV in mass.  As such, much of the parameter space that remains after the $\sqrt{s}=8$~TeV searches will be quickly probed in the first year or two after the LHC resumes operations~\cite{Kaufman:2013pya}.
In contrast, the Type~IIB orientifold models, of the type contemplated first by Kachru et al. (KKLT)~\cite{Kachru:2003aw} have a much more constrained parameter space. Achieving the observed CP-even Higgs mass of $m_h \simeq 126\,{\rm GeV}$ tends to require far more massive gluinos and squarks. As such, much of the nominally allowed parameter space resides in areas in which no superpartners are accessible at the early runs at the LHC -- and in many cases it is doubtful that superpartners would ever be accessible at the LHC~\cite{Kaufman:2013oaa}.

Given the above statements, it is of interest to ask whether the inclusion of some amount of gauge mediation can affect these conclusions. Gross properties of the DMM model were studied in~\cite{Choi:2009jn,Altunkaynak:2010tn}, with LHC implications and dark matter detection studied in~\cite{Altunkaynak:2010xe} and~\cite{Holmes:2009mx}, respectively. All of these studies, however, were performed prior to the supersymmetry searches at the LHC at $\sqrt{s}=8$~TeV and dark matter searches at the $\order(100\,{\rm kg})$ target level. At that time, the primary conclusion was that the LSP is likely to be heavy ($\order(1\,{\rm TeV})$) and that gluinos were likely to be much lighter than that predicted in the KKLT model without gauge mediation. It is of singular importance to revisit these early conclusions in light of the Higgs mass determination, and refine the predictions for the next run of the LHC, and larger dark matter detection experiments.

We begin our discussion in Section~\ref{sec:theory} with an overview of the theoretical structure that supports both mirage mediation and its deflected variant. We then exhibit the soft supersymmetry breaking terms, and identify the parameter space that defines the DMM model, in Section~\ref{sec:parameter}. This parameter space is quite a bit larger than that of the mirage/KKLT model, and we proceed to identify general features of this space, and correlations with physical observables, in Section~\ref{sec:scan}. This will allow us to identify representative benchmark examples to study in greater detail in Section~\ref{sec:collider}, where we focus on supersymmetry searches at the LHC. This is followed by a discussion of dark matter direct detection at current and future experiments in Section~\ref{sec:DM}. We will find that the DMM~paradigm spans cases that resemble the so-called `simplified models', as well as compressed-spectrum models often motivated from appeals to `naturalness'~\cite{Abe:2014kla}. We estimate the reach of the LHC at both $\sqrt{s} = 8\,{\rm TeV}$ and 14~TeV and suggest cases in which the current search strategies can be approved to address the specific challenges of the DMM model framework.

\section{Theoretical Framework}
\label{sec:theory}

\subsection{KKLT and K\"ahler Modulus Stabilization}

In what follows we review K\"ahler modulus stabilization in minimal $N=1$ supergravity, where we have in mind Type~IIB string theory compactified on a Calabi-Yau (CY) manifold in the presence of background fluxes. At the level of effective field theory, the precise origin of the various components of the effective Lagrangian is often irrelevant, so we will work in a simplified limit considered in~\cite{Kachru:2003aw}, in which a single K\"ahler modulus $T$ parameterizes the overall size of the compact space. It will be the non-vanishing vacuum expectation value $\lang F^T \rang$ that will set the scale of soft supersymmetry breaking in the absence of gauge mediation. 
The K\"ahler potential for the modulus $T$ is taken to be $K(T,\oline{T}) = -3\ln (T+\oline{T})$. For gauge theories with group $\mathcal{G}_a$, living on $D7$ branes which wrap four-cycles in the CY manifold, the gauge coupling is determined by the K\"ahler modulus $T$ via the (universal) gauge kinetic function $f_a = T$. Note that, with these assumptions, 
\begin{equation}
 <\re \; t>\; = 1/g_{\STR}^{2} , \label{f2}
\end{equation}
where $t=T|_{\theta=0}$ is the lowest component of the superfield $T$, and $g_{\STR}$ is the universal gauge coupling at the string scale.

In the effective supergravity theory just below the string compactification scale, the presence of the three-form fluxes is represented by a constant $w_0$ in the effective superpotential. It is presumed that these fluxes fix the value of the dilaton and the complex structure moduli, leaving only the K\"ahler moduli in the low-energy four-dimensional effective theory~\cite{Giddings:2001yu}.
Combined with the effect of gaugino condensation in the hidden sector the total effective superpotential is then
\begin{equation}
W_0 = w_0 +  A e^{-a T}\, , \label{WKKLT} \end{equation}
where there is a single gaugino condensate, for simplicity, and the constant $a$ is related to the beta-function coefficient of the hidden sector gauge group, with a normalization such that $a = 8\pi^2/N$ for the group $SU(N)$. 

In $N=1$ supergravity theories the scalar potential is determined by the auxiliary fields $F^N$, associated with the chiral supermultiplet $Z^N$, and the auxiliary field $M$ of the supergravity multiplet. The equations of motion for these auxiliary fields are given by
\begin{equation}
F^M = - e^{K/2} K^{M\oline{N}} \left(\oline{W}_{\oline{N}} + K_{\oline{N}} \oline{W} \right), \; \; \oline{M} = -3e^{K/2} \oline{W} \label{EQM}
\end{equation}
with $W_{\oline{N}} = \partial W / \partial \oline{Z}^{\oline{N}}$, $K_{\oline{N}} = \partial K / \partial \oline{Z}^{\oline{N}}$ and $K^{M \oline{N}}$ being the inverse of the K\"ahler metric $K_{M\oline{N}}= \partial^2 K / \partial Z^M \partial \oline{Z}^{\oline{N}}$.  
Note that the gravitino mass is determined via the vacuum relation
\begin{equation} \lang M \rang = -3\lang e^{K/2} W \rang = -3m_{3/2} \, . \label{Mvev} \end{equation}
Restoring the explicit Planck mass $M_P$, the scalar potential is then given by
\begin{equation}
V= K_{M\oline{N}} F^M  \oline{F}^{\oline{N}} - 3 m_{3/2}^2 M_{P}^2 \, , \label{pot}
\end{equation}
where repeated indices are summed. 

Minimizing the resulting scalar potential $V(t,\bar{t})$ %with $t= T\lowest$ 
generates a non-vanishing value for $\lang t + \bar{t} \rang$ at which the auxiliary field $F^T$ vanishes and the vacuum has an energy density given by $\lang V \rang = -3 m_{3/2}^2 M_{P}^2$. The size of the VEV for ${\rm Re}\,t$, as well as the size of the gravitino mass $m_{3/2}$, are determined by the size of the constant term $w_0$ in~(\ref{WKKLT}), which must be tuned to a value $w_0 \sim \order(10^{-13})$ in Planck units to obtain an acceptable phenomenology. In particular one has~\cite{Choi:2004sx}
\begin{eqnarray} \lang a\, {\rm Re}\,t \rang &\simeq& \ln(A/w_0) \nonumber \\
m_{3/2} & \simeq & M_{P} \frac{w_0}{(2\lang{\rm
Re}\,t\rang)^{3/2}}\, . \label{relations} \end{eqnarray}
%
%\textbf{CJO have $\lang a_+ {\rm Re}\,t \rang \simeq \ln(A/W_0)$.} 
Combining these relations in~(\ref{relations}) produces
\begin{equation} \lang a\, {\rm Re}\,t \rang \simeq \ln(M_{P}/m_{3/2}) \, .
\label{aReT}
 \end{equation}
Much of the phenomenology that has come to be known as ``mirage mediation'' is dependent only on the emergence of the parameteric relation in~(\ref{aReT}), and not on the particulars of any constants that may appear in the non-perturbative stabilizing superpotential, such as the one in~(\ref{WKKLT}).

To discuss supersymmetry-breaking, it is necessary to first address the vacuum energy problem, by adding some additional `uplift' sector which generates supersymmetry breaking in the observable sector while producing a Minkowski vacuum. Many such suggestions exist in the literature~\cite{Achucarro:2006zf,Choi:2006bh,Dudas:2006vc,Dudas:2006gr,Abe:2006xp}, and the precise choice will not affect our results provided that (a) the K\"ahler modulus dependence of the added terms in the Lagrangian is dictated solely by consistency of supergravity under K\"ahler $U(1)$ transformations, and (b) the vacuum expectation value $\lang {\rm Re}\,t \rang$ is not perturbed greatly by the addition of the uplift sector~\cite{Choi:2005uz}. 
If these conditions are satisfied, then the auxiliary field for the K\"ahler modulus no longer vanishes in the `lifted' vaccum, but instead satisfies the approximate solution
\begin{equation} M_0 \equiv \lang \frac{F^T}{t+\bar{t}} \rang \simeq \frac{2m_{3/2}}{a \lang t + \bar{t}\rang}\, . \label{MKKLT}
\end{equation}
This quantity $M_0$ then serves as an order parameter of supersymmetry breaking in the observable sector. 

 The derivation of these soft supersymmetry terms is made considerably more transparent if one employs the chiral compensator technique for generating anomaly-mediated contributions to supersymmetry breaking. If $C$ represents the conformal compensator of the supergravity multiplet, and $F^C$ is its corresponding auxiliary component, then $\lang F^C/C \rang \simeq m_{3/2}$ and there is
\begin{equation}
\lang \frac{F^T}{t +\bar{t}} \rang \simeq \lang  \frac{1}{a\, {\rm Re\, t}} \frac{F^C}{C} \rang.
\end{equation}
When working out soft terms, it is convenient to write the above expression as an equality by introducing the parameter $\alpha_m$~\cite{Choi:2005uz} via
\begin{equation} \alpha_m \equiv
\frac{m_{3/2}}{M_0\ln\(M_{P}/m_{3/2}\)} \, , \label{alpha}
\end{equation}
and thus
\begin{equation}
\lang \frac{F^C}{C} \rang = \alpha_m \ln\left( \frac{M_P}{m_{3/2}}\right) \lang \frac{F^T}{T+\bar{T}} \rang \, ,
\label{FCFT}
\end{equation}
where we have used the vacuum condition in~(\ref{aReT}).

\subsection{The Additional Singlets}
\label{sec:gauge}

In many string motivated models, additional pairs of fields $\Psi$, $\overline{\Psi}$ with SM gauge quantum number are not uncommon. Such vector-like pairs often have superpotential interactions with one or more SM singlets (here denoted by $X$) which can potentially serve as supersymmetric mass terms. Here the KKLT formalism is extended to include these fields acting as messengers, and show that they couple to a moduli field that gets a vev at the right scale to produce potentially large gauge mediation-like deflection.

The superpotential is assumed to be of the form
\begin{equation}
\label{superpot}
W=W_0+W_1(X)+\lambda X\Psi\overline{\Psi} +W_{\rm MSSM}(\Phi), 
\end{equation}
in which $W_0$ is given by~(\ref{WKKLT}), $W_1(X)$ denotes the singlet self-interaction superpotential terms, and $W_{\rm MSSM}(\Phi)$ is the standard MSSM superpotential, involving the observable sector fields $\Phi_i$. The fields $\Psi_i$ and $\overline{\Psi}_i$ are here after taken to be one or more pairs of SU(5) $\mathbf{5}$ and $\bar{\mathbf{5}}$ multiplets. Different forms of the singlet self-interaction $W_1(X)$, correspond to different ways of stabilizing the modulus $X$. The  K\"{a}hler potential will be taken to be
\begin{equation}
\label{fullkahler}
K = -3\ln (T+\oline{T}) + Z_{X} (T, \overline T ) X \overline X + Z_{i} (T,\overline{T}) \Phi_i \overline{\Phi}_i + 
{\mathcal O} \left((|\Phi|^4, |X|^4)\right), 
\end{equation}
in which  $Z_X$ and $Z_i$ are the K\"ahler metrics of $X$ and $\Phi^i$, respectively.  The K\"ahler metric for the messenger states $\Psi_i$ and $\overline{\Psi}_i$ will not be relevant for our discussion, but can be taken to be of the same form as the observable sector states.
The K\"{a}hler metrics $Z_X(T,\overline{T})$ and $Z_i(T,\overline{T})$ will be assumed to be of the standard form
\begin{equation}
\label{kahlermet}
Z_X=\frac{1}{(T+\overline{T})^{n_X}},\;\;\;Z_i=\frac{1}{(T+\overline{T})^{n_i}},
\end{equation}
in which $n_X$ and $n_i$ are the modular weights of $X$ and $\Phi_i$, respectively. 

Successful gauge mediation will require the dynamical generation of a vacuum expectation value (vev) for both the lowest component $\lang X \rang \neq 0$ and the highest component $\lang F^X \rang \neq 0$ of the singlet chiral superfield. The simplest case is to have $W_1(X) =0$ and assume that the coupling between $X$ and $\Psi$,$\overline{\Psi}$ generates $\lang X \rang \neq 0$ at low energies, as in the electroweak sector of the Standard Model. In this case an F-term vev of approximately the right size is automatically generated~\cite{Kane:2002qp}
\begin{equation}
F^X \simeq - e^{K_0/2} K^{X\bar{X}} D_{\bar{X}} W\simeq - e^{K_0/2} K^{X\bar{X}} K_{\bar{X}} W_0  \simeq  -m_{3/2} X,
\end{equation}
such that 
\begin{equation}\frac{F^X}{X} = - \frac{F^C}{C} \approx -m_{3/2}.
\label{radstab}
\end{equation}

Alternatively, one can consider a very simple form for $W_1$ in Eq.~(\ref{superpot}), such as 
\begin{equation}
W_1 =  \lambda_n \frac{ X^n }{\Lambda^{n-3}} \, , 
\end{equation}
in which $\Lambda$ is some cutoff scale. In principle, the exponent $n$ can have positive or negative values; a negative exponent would indicate that this term originates from nonperturbative dynamics. In the case where $n>3$ (stabilization by higher order terms), and the case where $n<0$ (stabilziation by nonperturbative dynamics), the resulting non-vanishing F-term vev is of the form~\cite{Everett:2008ey}
\begin{equation}
\frac{F^X}{X} = -\frac{2}{n-1}\frac{F^C}{C}\, .
\label{higherordstab}
\end{equation}
Since the modulus and anomaly contributions are already comparable, this result indicates that all three contributions should be roughly equal for a very general class of superpotentials.\footnote{This result is the same as that obtained in the case of deflected anomaly mediation~\cite{Pomarol:1999ie}. Further details of the calculation, and the case of renormalizable $W_1(X)$, can be found in~\cite{Everett:2008ey}.}

\section{Soft Supersymmetry Breaking}

\label{sec:parameter}

To derive the observable sector soft terms, it is convenient to use the spurion technique, in which the 
couplings of the effective supergravity Lagrangian are regarded as functions in superspace, with the $\theta$-dependent parts of these couplings generated by the F-term vevs of the theory (for a review, see \cite{Giudice:1998bp}).  The MSSM soft supersymmetry-breaking Lagrangian includes terms of the form
\begin{equation}
\mathcal{L}_{\rm soft}=-m^2_i |\Phi^i|^2-\left [ \frac{1}{2} M_a \lambda^a \lambda^a + A_{ijk} y_{ijk} \Phi^i \Phi^j \Phi^k +\mbox{h.c.} \right ],
\end{equation} 
in which $m^2_i$ are the soft scalar mass-squared parameters, $M_a$ are the gaugino masses, and $A_{ijk}$ are trilinear scalar interaction parameters.
These terms are defined in the field basis in which the kinetic terms are canonically normalized. 
The expressions for the soft supersymmetry-breaking terms take the standard supergravity form
\begin{eqnarray}
M_a &=& F^A \partial_A \log ({\rm Re\,}f_a),\label{Ma} \\
A_{ijk} &=& -F^A \partial_A \log \left( \frac{y^0_{ijk}}
{Y_i Y_j Y_k}\right), \nonumber \label{Aijk} \\
m_i^2 &=& - F^A \overline{F}^{\overline{B}} \partial_A \partial_{\overline{B}} \log
Y_i \, , \label{mi2}
\end{eqnarray}
where $f_a$ is the field-dependent, gauge kinetic function for the gauge group $\mathcal{G}_a$, $y^0_{ijk}$ is the bare Yukawa coupling appearing in the superpotential, and the function $Y_i$ is defined by
\begin{equation}
Y_i =  \frac{1}{(T+\bar{T})^{n_i-1}}\, .
\label{Yi}
\end{equation}
From here, one merely needs to specify the dependence of the relevant quantities on the fields $X$ and $T$, as well as the spurious conformal compensator, $C$. The latter follows standard computations familiar from the study of anomaly mediation~\cite{Giudice:1998xp,Randall:1998uk,Pomarol:1999ie,Gaillard:2000fk}. For the gauge kinetic function we take
\begin{equation} f_a(M_G) = T^{\ell_a} \, \label{fa}
\end{equation}
where $M_G$ is the boundary condition scale (taken as the energy scale at which $g_1^2(M_G)=g_2^2(M_G)$), and $l_a=0,1$ depending on the type of D branes from which the gauge groups originate. Since we wish to maintain gauge coupling unification at the GUT scale, we  assume that $l_a =1$ for each of the SM gauge group factors.  
For the unnormalized Yukawa couplings $y^0_{ijk}$, there is no $C$ dependence due to the supersymmetric nonrenormalization theorem. Since $y^0_{ijk}$ is also assumed to be independent of $T$ and $X$, the expression for trilinear terms~(\ref{Aijk}) can be  reduced to 
\begin{equation}
A_{ijk} = A_i + A_j + A_k\, , 
\end{equation}
in which
\begin{equation}
A_i = F^A \partial_A \log Y_i\, ,
\end{equation}
and $Y_i$ is given by~(\ref{Yi}).

Let $M_{\rm mess}$ be the mass of the messenger fields, with $M_{\rm mess} \equiv \lambda \lang X \rang$. Recalling that above the mass scale of the messengers the beta functions depend on not only the MSSM fields, but also on the messenger pairs,   the soft terms at the GUT scale $M_{\rm G}$ and the messenger threshold effects at $M_{\rm mess}$ are as follows:\\

\noindent {\bf Gaugino Masses.} The gaugino mass parameters are given by
\begin{eqnarray}
  M_a (\mu = M_{\rm G}) &=& \frac{F^T}{T+\overline{T}}
  + \frac{g_0^2}{16\pi^2}  b'_a\frac{F^C}{C} \\ \label{softgaugino}
  M_a (\mu = M_{\rm mess}^-)& =& M_a(\mu = M_{\rm mess}^+)
+ \Delta M_a,
\end{eqnarray}
in which the threshold corrections are
\begin{eqnarray}
  \Delta M_a = - N\frac{g_a^2(M_{\rm mess})}{16\pi^2} \left( \frac{F^C}{C}
  + \frac{F^X}{X} \right).
   \label{softgauginothresh}
\end{eqnarray}
Here $g_0$ is the unified gauge coupling at $M_{\rm G}$, $M^{\pm}_{\rm mess}$ represents an energy scale just above (just below) the messenger mass scale, and the beta functions $b^\prime_a$ are related to their MSSM counterparts by 
$b^\prime_a = b_a + N$,
with  
$(b_3, b_2, b_1) = (-3, 1, \frac{33}{5})$ 
(in our conventions, $b_a<0$ for asymptotically free theories), and $N$ the number of messenger fields $\Psi_i$, $\overline{\Psi}_i$, where $i=1,\dots,N$.\\

\noindent {\bf  Trilinear terms.}
The trilinear terms are $A_{ijk} = A_i+ A_j + A_k$, with
\begin{eqnarray}
A_i (\mu = M_{\rm G}) &=& \left (1-n_i \right) \frac{F^T}{T+\overline{T}}
- \frac{\gamma_i}{16\pi^2} \frac{F^C}{C},  \label{softA}
\end{eqnarray}
where $\gamma_i$ is the anomalous dimension of $\Phi_i$. \\

\noindent {\bf Soft scalar masses. }
The scalar mass-squared parameters are given by
\begin{eqnarray}
m_i^2 (\mu = M_{\rm G}) = \left (1-n_i \right) \left|\frac{F^T}{T+\overline{T}}\right|^2
%\nonumber \\
%&& 
-\frac{\theta'_i}{32 \pi^2} \left( \frac{F^T}{T+\overline{T}}\frac{F^{\overline{C}}}{\overline{C}} + \mbox{h.c.} \right) \nonumber \
\label{softscalar}
 - \frac{\dot{\gamma}'_i}{(16\pi^2)^2} \left|\frac{F^C}{C}\right|^2,
%\nonumber \\
\end{eqnarray}
\begin{eqnarray}
m_i^2 (\mu = M_{\rm mess}^-) &=& m_i^2 (\mu = M_{\rm mess}^+)
+ \Delta m_i^2,
\end{eqnarray}
where the threshold corrections are
\begin{eqnarray}
\Delta m_i^2 & = &\sum_a 2 c_a N
\frac{g_a^4 (M_{\rm mess}) }{(16\pi^2)^2}
 \left( \left|\frac{F^X}{X}\right|^2 + \left|\frac{F^C}{C}
\right|^2
+ \frac{F^X}{X}
 \frac{F^{\overline{C}}}{\overline{C}} +\mbox{h.c.} \right). \label{softscalarthresh}
\end{eqnarray}
In the above, $c_a$ is the quadratic Casimir, and $\gamma_i$, $\dot{\gamma}_i$, $\theta_i$ ($\gamma_i'$, $\dot{\gamma}_i'$, $\theta_i'$) are listed in Appendix A.
%\end{itemize}

We now replace the F terms with the parameterization given in~\cite{Everett:2008qy,Everett:2008ey}, as follows:
\begin{eqnarray}
\label{param1}
\frac{F^C}{C}&=&\alpha_m\ln \frac{M_P}{m_{3/2}} \frac{F^T}{T+\overline{T}}= \alpha_m\ln \frac{M_P}{m_{3/2}} M_0 \\
\label{param2}
\frac{F^X}{X}&=&\alpha_g \frac{F^C}{C}=\alpha_g\alpha_m \ln\frac{M_P}{m_{3/2}} M_0,
\end{eqnarray}
in which $M_0 \equiv F^T/(T+\overline{T}) $ sets the overall scale of the soft terms.  The dimensionless parameter $\alpha_m$ is the $\alpha$ parameter of mirage mediation: it denotes the relative importance of anomaly mediation with respect to gravity mediation.  In the specific scenario considered by~KKLT, $\alpha_m=1$.  The dimensionless parameter $\alpha_g$ denotes the relative importance of the gauge-mediated terms with respect to the anomaly-mediated terms.  The values of $\alpha_g$ depend on the details of the stabilization of $X$, as described in Section~\ref{sec:gauge}. 

With the parametrization given in Eqs.~(\ref{param1})--(\ref{param2}), the soft terms at $M_G$ take the form
\begin{eqnarray}
\label{gaugino1}
M_a(\mu = M_{\rm G})&=&M_0\left [1+\frac{g_0^2}{16\pi^2}b_a^\prime \alpha_m \ln \frac{M_P}{m_{3/2}}\right ] ,\\
\label{trilinear1}
A_i(\mu = M_{\rm G})&=&M_0 \left [(1-n_i)-\frac{\gamma_i}{16\pi^2} \alpha_m \ln \frac{M_P}{m_{3/2}}\right ] ,\\
\label{mssq1}
m_i^2(\mu = M_{\rm G})&=&M_0 ^2 \left [(1-n_i)-\frac{\theta'_i}{16 \pi^2} \alpha_m \ln \frac{M_P}{m_{3/2}} -\frac{\dot{\gamma}'_i}{(16\pi^2)^2}\left (\alpha_m \ln \frac{M_P}{m_{3/2}}\right )^2 \right ],
\end{eqnarray}
where the anomalous dimensions are given in Appendix \ref{Anomalous}, and the threshold terms are given by
\begin{eqnarray}
\label{threshgaug1}
\Delta M_a(\mu = M_{\rm mess})&=& -M_0 N\frac{g_a^2(M_{\rm mess})}{16\pi^2}   \alpha_m \left (1 +\alpha_g \right ) \ln \frac{M_P}{m_{3/2}} ,\\
\label{threshmssq1}
\Delta m_i^2(\mu = M_{\rm mess})&=&M_0^2\sum_a 2 c_a N
\frac{g_a^4 (M_{\rm mess}) }{(16\pi^2)^2} \left [\alpha_m  (1+\alpha_g)  \ln \frac{M_P}{m_{3/2}}\right ]^2.
\end{eqnarray}
The parameters of the model are the mass scales $M_0$ and $M_{\rm mess}$, as well as the dimensionless quantities $\alpha_m$, $\alpha_g$, the number of $SU(5)$ messenger pairs $N$, the modular weights $n_i$,  $\tan\beta$, and ${\rm sign}\ \mu$ (the model-dependent $\mu$ and $B\mu$ parameters are replaced as usual by the $Z$ boson mass, $\tan\beta$, and the sign of $\mu$).

In the mirage mediation scenario, one of the most distinctive features of the soft terms is the unification of the gaugino masses at the mirage unification scale $M_{\rm mirage}$:
% which is given by
\begin{equation}
M_{\rm mirage}=M_{\rm G}\left (\frac{m_{3/2}}{M_P} \right )^\frac{\alpha_m}{2}.
\label{mirageunif}
\end{equation}
In deflected mirage mediation, one finds a similar mirage unification phenomenon for the gaugino masses.  From the form of the soft terms of Eq.~(\ref{gaugino1}) and Eq.~(\ref{threshgaug1}), the new mirage unification scale for the gauginos (see also \cite{Everett:2008qy}) is
\begin{equation}
M_{\rm mirage}=M_{\rm G}\left (\frac{m_{3/2}}{M_P} \right )^\frac{\alpha_m\rho }{2},
\label{deflmirageunif}
\end{equation}
in which $\rho$ is given by
\begin{equation}
\rho= \frac{1+ \frac{2Ng_0^2}{16\pi^2}  \ln \frac{M_{\rm G}}{M_{\rm mess}}}{1- \frac{\alpha_{\rm m} \alpha_{\rm g} Ng_0^2}{16\pi^2} 
 \ln \frac{M_P}{m_{3/2}}}.
% \right )^{\alpha_{\rm m} \alpha_{\rm g}}}.
\label{rhodef}
\end{equation}
The mirage unification scale of the gauginos is thus deflected from the mirage mediation result.  The size of the deflection is dependent on $\alpha_g$, $N$, and $M_{\rm mess}$, which govern the size of the messenger thresholds.  

For the $A$ terms and the soft scalar mass-squares, the mirage unification behavior no longer happens in general in the presence of the messengers.  The exception is when the messenger scale is below the scale of mirage unification which would occur in the absence of the messenger thresholds, since the theory is then effectively the same as mirage mediation below $M_{\rm mess}$.   In Eqs.~(\ref{deflmirageunif})--(\ref{rhodef}), the mirage mediation result of Eq.~(\ref{mirageunif}) is obtained only if $N=0$.  This demonstrates that the mirage mediation limit is not reached when gauge mediation is switched off ($\alpha_g\rightarrow 0$); it only occurs when the messengers are removed from the theory at all scales ($N=0$).  The reason is that the messengers affect the MSSM beta functions above the messenger scale, which in turn affects the boundary conditions for the anomaly-mediated terms.

%%%%%%%%%%%%%%%%%%%%%%%%%%%%%%%%%%%%%%%%%%%%%%%%%%%%%%%%%%%
%%%%%%%%%%%%%%%%%%%%%%%%%%%%%%%%%%%%%%%%%%%%%%%%%%%%%%%%%%%
%%%%%%%%%%%%%%%%%%%%%%%%%%%%%%%%%%%%%%%%%%%%%%%%%%%%%%%%%%%
%%%%%%%%%%%%%%%%%%%%%%%%%%%%%%%%%%%%%%%%%%%%%%%%%%%%%%%%%%%
\section{Constraints on DMM Parameter Space}
\label{sec:scan}

In deflected mirage mediation there are two distinct contributions to soft supersymmetry breaking, a KKLT-like contribution at the GUT or string scale, followed by a deflection at some messenger scale $M_{\rm mess}$. For the first contribution, there are two independent mass scales, given by the (normalized) gravitino mass $m_{3/2}$  and the modulus contribution $M_0$. Alternatively, one can work with either of the mass scales and the derived parameter $\alpha_m$. We use the latter, and will use $M_0$ as the independent mass scale. The value of $m_{3/2}$ is computed by fitting to the expression in~(\ref{alpha}), and the calculated value will then be input into the high scale soft term expressions in~(\ref{gaugino1}), (\ref{trilinear1}) and~(\ref{mssq1}). 

In addition, one must specify the modular weights for the chiral supermultiplets that make up the MSSM field content. In this work we will allow only a limited amount of non-universality in assigning these weights. In particular, we will always assume that all matter multiplets arise from the same sector of the theory, so that they carry a universal modular weight $n_M$, while the two Higgs doublets may carry an independent modular weight which we will denote $n_H$. We let both take half integer values between zero and one. Under these assumptions there are then nine possible combinations of modular weights to consider, which we can represent by the pair of weights $(n_M, n_H)$. A theory defined solely by the choice of modular weights, $\tan\beta$, $\alpha_m$, and $M_0$, represents a general mirage mediation model. Nevertheless, we will often refer to it (somewhat inappropriately) as a ``KKLT model'', to distinguish it more clearly from a deflected mirage mediation model.

In DMM, there is an additional mass scale $M_{\rm mess}$, where $N$ SU(5) \textbf 5 $\bar{\textbf 5}$ messengers integrate out with a strength given by a derived parameter $\alpha_g$. The quantities $\alpha_g$, $N$, and the KKLT parameters are input to the deflected contributions given by \eqref{threshgaug1} and  \eqref{threshmssq1}, which may be sizeable. 
Thus, the parameter space we will consider consists of a discrete choice of modular weights and three continuous parameters: $M_0$, $\alpha_m$ and $\tan\beta$. These quantities define a KKLT-like `base point'. The deflection from this base point will be characterized by the three parameters $\alpha_g$, $M_{\rm mess}$, and $N$. These DMM parameters can be set by considering an explicit model, but here we choose to let them vary continuously.

To fully explore the entirety of this parameter space would require a scan over the five continuous parameters $M_0$, $M_{\rm mess}$, $\alpha_m$, $\alpha_g$ and $\tan\beta$ and the three discrete parameters $n_M$, $n_H$, and $N$. A comprehensive scan would quickly become unrealistically computationally expensive, as the parameter space is large. Because the DMM framework is an extension of the KKLT framework, we approach our scan in the same fashion. For each set of modular weights we will randomly select the three continuous parameters of the KKLT framework in the ranges $M_0\in[1,5]$ TeV, $\tan\beta\in[5,50]$ and $\alpha_m\in[0,2]$. We then build a three-dimensional scan in the DMM parameter space around each base point, scanning $\alpha_g \in [-1,1]$ in steps of 0.05, $\log_{10}\left[M_{\rm mess}/{\rm GeV}\right] \in[5,14]$ in unit steps, and $N \in [1,5]$ in unit steps. 

The range in $\alpha_g$ is chosen as in~\cite{Altunkaynak:2010xe} to reflect a range of possible moduli stabilization mechanisms. The UV cutoff on the range in the messenger scale is chosen to avoid possible GUT threshold contributions, while the lower bound is meant to avoid large contributions to flavor-changing neutral current processes. The number of messengers, $N$, is chosen between 0 (where the model is identical to KKLT), and $5$, which is the maximum number of messengers before couplings tend to run to nonperturbative values with an $\mathcal{O}(\text{TeV})$ messenger scale. We note that for the $N=3$ case, the strong coupling does not run between the GUT and messenger scales at one loop order. 

For each choice of KKLT input parameters $(n_M, n_H, M_0, \alpha_m, \tan\beta)$, the soft terms are computed from~(\ref{gaugino1}), (\ref{trilinear1}) and~(\ref{mssq1}). The renormalization group (RG) equations are solved from the boundary condition scale to the electroweak scale using a version of the package SOFTSUSY~3.3.9~\cite{Allanach:2001kg} that has been modified to account for the gauge mediation contributions~~\cite{Everett:2008qy,Everett:2008ey}.
The modification introduces an intermediate messenger scale where the threshold corrections~\eqref{threshgaug1} and~\eqref{threshmssq1}, determined by the DMM parameters $(N, \alpha_g, M_{\rm mess})$, are added to the running masses. The software uses a modified set of renormalization group equations to include the effect of messengers above the scale determined by $M_{\rm mess}$, at the two loop level.

At the electroweak scale, a combination of input parameters will be excluded from the data set if the soft supersymmetry breaking scalar mass-squared parameter is negative for one or more of the matter fields. At this stage, the radiatively-corrected Higgs potential is minimized and physical masses are calculated. We again eliminate a combination of input parameters if no solution to the conditions for electroweak symmetry breaking can be found, or if the solution fails to converge adequately. Finally, we then ask that the lightest supersymmetric particle (LSP) for each model point be a neutralino, though stau, gluino and stop LSPs are all possible in various regions of parameter space.  In total, 6.1~million  points were generated from 2700~KKLT base points, evenly distributed across the nine modular weight combinations. After application of all phenomenological constraints, slightly less than 390,000 DMM points survive, originating from just over 2500~KKLT base points.

Having passed the minimal phenomenological requirements, the electroweak scale spectrum is then passed to MicrOMEGAs~2.2~\cite{Belanger:2001fz} where the thermal relic abundance $\Omega_{\chi} h^2$ is computed for the stable neutralino. In addressing the issue of cold dark matter, we take a conservative approach and impose only an upper bound on the neutralino relic density. We use a cut on a three-sigma upper bound on the calculation from MicrOMEGAs of $\Omega_{\chi}h^2\leq0.128$ taken from~\cite{Ade:2013zuv}. We further require the mass of the lightest chargino to exceed the LEP bound ($m_{\chi^{\pm}_1} \geq 103.5 \,{\rm GeV}$) and that the value of ${\rm Br}(B_s \to \mu^+ \mu^-)$ to be within~$3\sigma$ of the LHCb measurement~\cite{Aaij:2012nna,Aaij:2013aka}. Finally, we take into account the recent measurements of the Higgs scalar mass~\cite{ATLAS:2013mma,Chatrchyan:2013lba,Aad:2015zhl} by requiring $123\,{\rm GeV}\leq m_h \leq 127\,{\rm GeV}$, which represents a rather generous mass range for this parameter, given ongoing efforts to improve the reliability of Higgs mass calculations~\cite{Buchmueller:2013psa,Hahn:2014qla,Borowka:2015ura}. 

We display the results of our scan in two subsections. In the first subsection, we comment on some generic features of the DMM parameter space, using all of the data generated by our scan. In the following subsection, we focus on specific KKLT base points with reasonable low-energy phenomenology, and discuss modifications to the spectra arising from the introduction of gauge-charged messenger fields.

%%%%%%%%%%%%%%%%%%%%%%%%%%%%%%%%%%%%%%%%%%%%%%%%%%%%%%%%%%%%
%%%%%%%%%%%%%%%%%%%%%%%%%%%%%%%%%%%%%%%%%%%%%%%%%%%%%%%%%%%%
%%%%%%%%%%%%%%%%%%%%%%%%%%%%%%%%%%%%%%%%%%%%%%%%%%%%%%%%%%%%
\subsection{Generic Properties}

%%%%%%%%%%%%%%%% FIGURE 1 %%%%%%%%%%%%%%%%%
\begin{figure}[t]
\begin{center}
%\begin{subfigure}[b]{0.45\textwidth}
\includegraphics[width=0.45\textwidth]{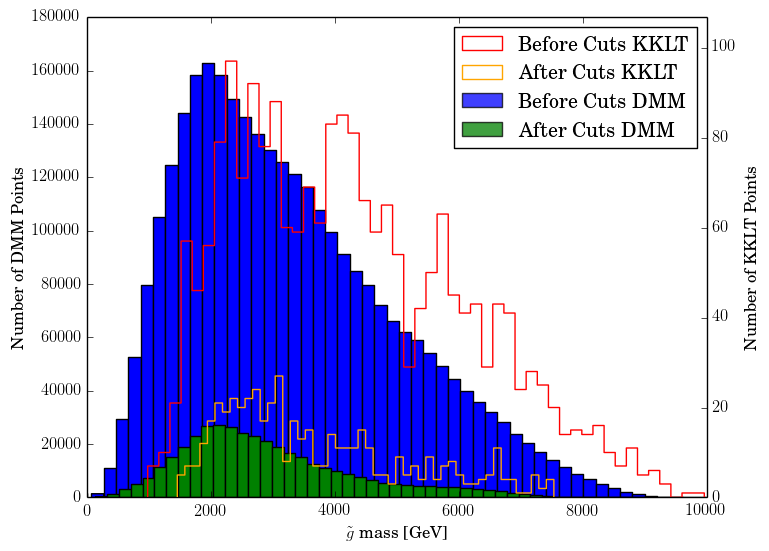}
%\end{subfigure}
%\begin{subfigure}[b]{0.45\textwidth}
\includegraphics[width=0.45\textwidth]{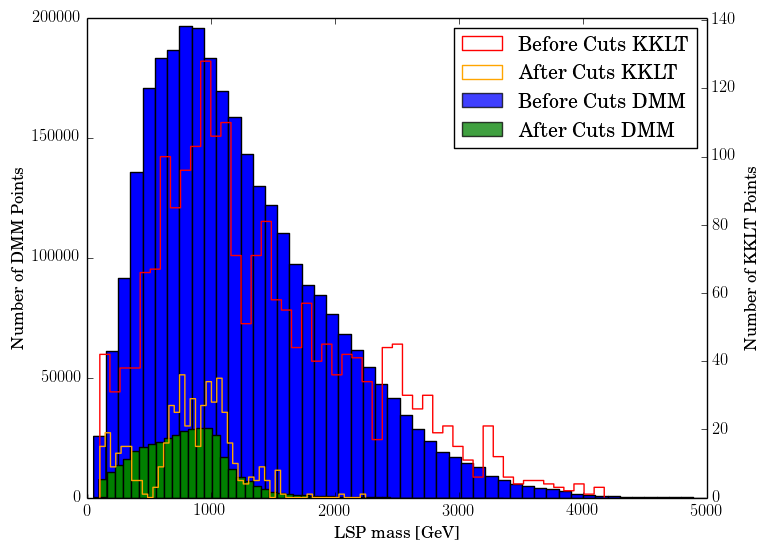}
%\end{subfigure}
\caption{Distribution of gluino masses (left) and LSP masses (right), in units of~GeV, for all modular weights in DMM and for the KKLT base points. The blue histogram represents all DMM points with an acceptable minimum and a neutralino LSP. Points in the green histogram also have an acceptable Higgs mass and neutralino relic density taken from the blue. The red and orange outlines represent the distributions for the KKLT base points where the red has an acceptable minimum and a neutralino LSP, and the orange has an acceptable Higgs mass and relic density, with the scale on the right side.}
\label{plot:gluinodist}
\end{center}
\end{figure}

In Figure~\ref{plot:gluinodist}, we see the effect of our constraints on the parameter space. The left panel represents the distributions in the gluino mass, while the right panel gives the distributions in the LSP mass. In both panels, the blue histogram represents all points with an acceptable minimum and a neutralino LSP. Points in the green histogram also have an acceptable Higgs mass and neutralino relic density. The red and orange outlines are the equivalent distributions for the KKLT basepoints. Although statistics are low for both KKLT distributions, we can see that the gluino masses of the base points are shifted to the right in relation to the solid DMM distributions. For the base points the minimum mass is roughly 1.5~TeV, with a peak roughly around 3~TeV, whereas for DMM, we have a peak closer to 2~TeV, with the possibility of very low-mass gluinos. Many, but not all, of these low mass points would have given a detectable signal at the past LHC run at $\sqrt{s} = 8\,{\rm TeV}$. The overall shift in these distributions means that more of the parameter space for DMM will be probed in the current LHC run, but there remains a long tail that extends beyond the expected reach of the LHC, even after 3000~fb$^{-1}$ of data-taking. For the mass of the lightest neutralino, the distribution of KKLT base points and DMM points are not significantly different. 
When the current limits on the neutralino relic density and Higgs mass are taken into account, we expect that the entirety of the neutralino and gluino mass ranges should be accessible at a future 100~TeV collider~\cite{Cohen:2013xda,Jung:2013zya}. 

%%%%%%%%%%%%%%%% FIGURE 2 %%%%%%%%%%%%%%%%%

\begin{figure}[th]
\centering
%\begin{subfigure}[b]{0.45\textwidth}
\includegraphics[width=0.45\textwidth]{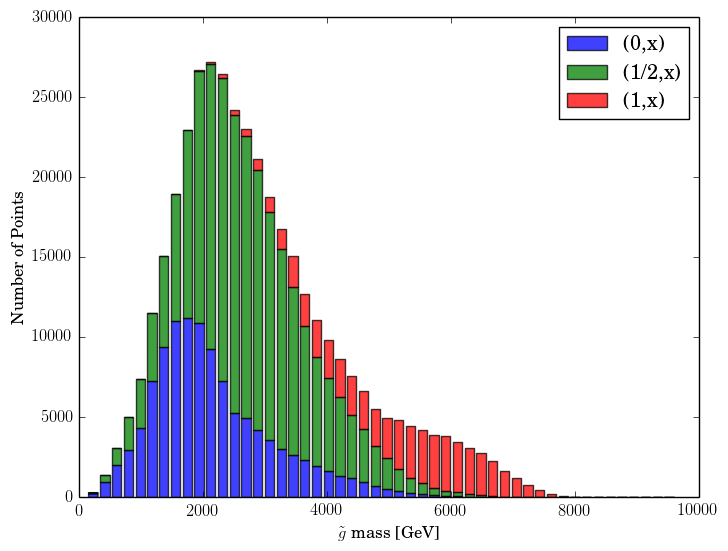}
%\label{fig:gbynM}
%\end{subfigure}
%\begin{subfigure}[b]{0.45\textwidth}
\includegraphics[width=0.45\textwidth]{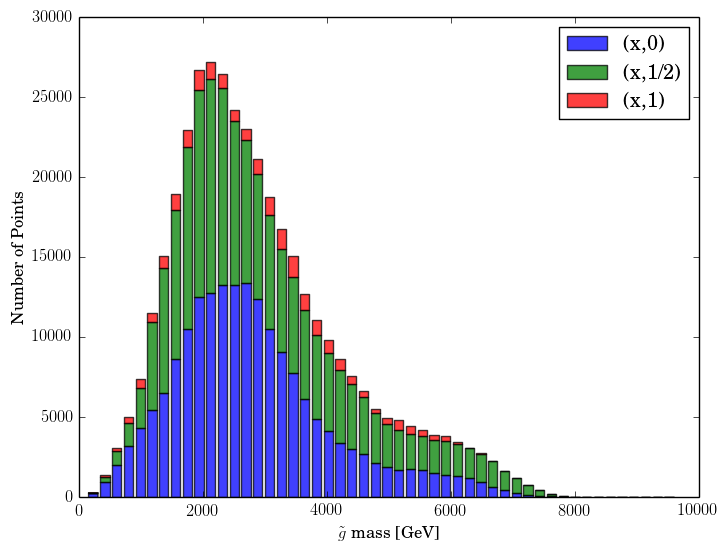}
%\label{fig:gbynH}
%\end{subfigure}
\caption{Distribution of gluino mass (in GeV), broken down by modular weight combinations. The left plot is broken down by $n_M$ and the right is by $n_H$. All points have an acceptable EW vacuum, Higgs mass, and neutralino relic density.}
\label{plot:gbymodularweight}
\end{figure}

Figure~\ref{plot:gbymodularweight} breaks down the allowed region (green histogram) in the left panel of Figure~\ref{plot:gluinodist} by modular weight. In the left (right) panel, data is aggregated over various $n_H$ ($n_M$) values, for particular $n_M$ ($n_H$) values held constant. The typical value of the gluino mass correlates strongly with the matter modular weight, with the distribution moving to larger values with increasing $n_M$. The origin of this behavior lies in the high scale boundary condition for the soft masses of the matter fields~(\ref{mssq1}), which decreases with increasing $n_M$. The relatively large Higgs mass $m_h \simeq 125\,{\rm GeV}$ requires a relatively heavy stop mass, and thus larger values of $n_M$ require a larger value of $M_0$ to compensate. 
The effect is enhanced by the fact that larger modular weights $n_M$ reduce the size of the trilinear $A$-terms~(\ref{trilinear1}), thereby reducing the left-right mixing in the stop sector. Conversely, as the values of $n_H$ only affect the boundary conditions of the Higgs scalar masses, we do not expect the overall mass scale to be dependent on this parameter, and indeed the three distributions in the right panel of Figure~\ref{plot:gbymodularweight} are qualitatively similar.
A desire for a lighter, and hopefully LHC accessible, spectrum motivates model-building efforts in which MSSM fields are localized on stacks of $D7$ branes for which $n_i=0$.

%%%%%%%%%%%%%%%% FIGURE 3 %%%%%%%%%%%%%%%%%

\begin{figure}[th]
\centering
%\begin{subfigure}[b]{0.45\textwidth}
\includegraphics[width=0.45\textwidth]{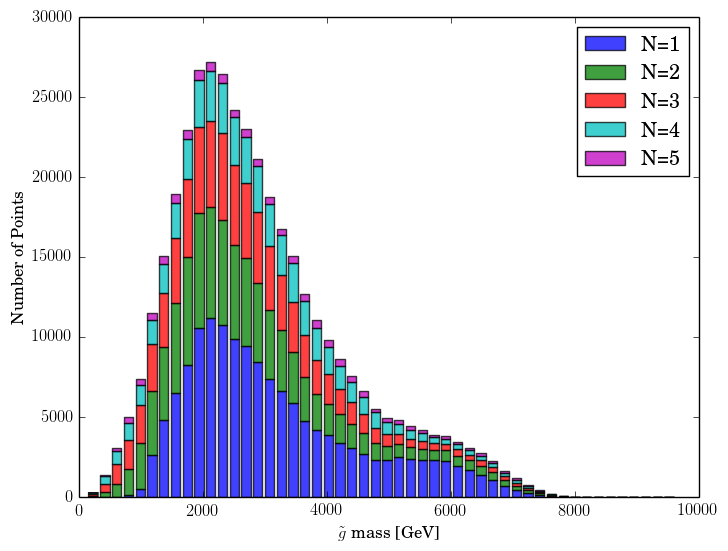}
%\label{fig:gbynM}
%\end{subfigure}
%\begin{subfigure}[b]{0.45\textwidth}
\includegraphics[width=0.45\textwidth]{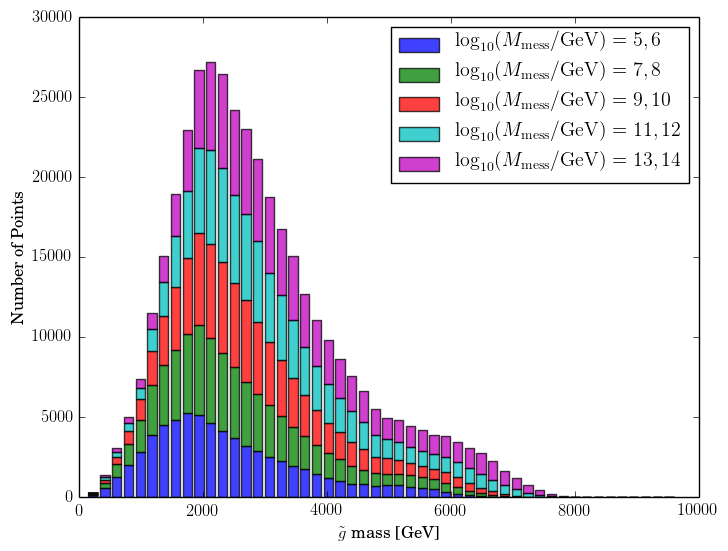}
%\label{fig:gbynH}
%\end{subfigure}
\caption{Distribution of gluino mass (in GeV), broken down by $N$ (left) and $\log_{10}(M_{\rm mess}/{\rm GeV})$ (right). All points have an acceptable EW vacuum, Higgs mass, and neutralino relic density.}
\label{plot:gbyN}
\end{figure}

Figure~\ref{plot:gbyN} further studies the influence of the messenger sector on the predicted gluino mass. The left panel of Figure~\ref{plot:gbyN}  breaks down the distribution by number of messengers, while the right panel addresses the messenger scale. Points with lighter gluino masses tend to have two or more messengers and low messenger scales. For these points, the deflection contribution is comparable in size to the running mass itself, allowing for a cancellation to occur, while the small messenger scale prevents large corrections from RG effects. In contrast, for large gluino masses, the messenger scale tends to be on the high end, with $\alpha_g\sim-1$, so that the gluino experiences the largest possible mass increase through RG evolution.

%%%%%%%%%%%%%%%% FIGURE 4 %%%%%%%%%%%%%%%%%

\begin{figure}[th]
\centering
%\begin{subfigure}[b]{0.45\textwidth}
\includegraphics[width=0.45\textwidth]{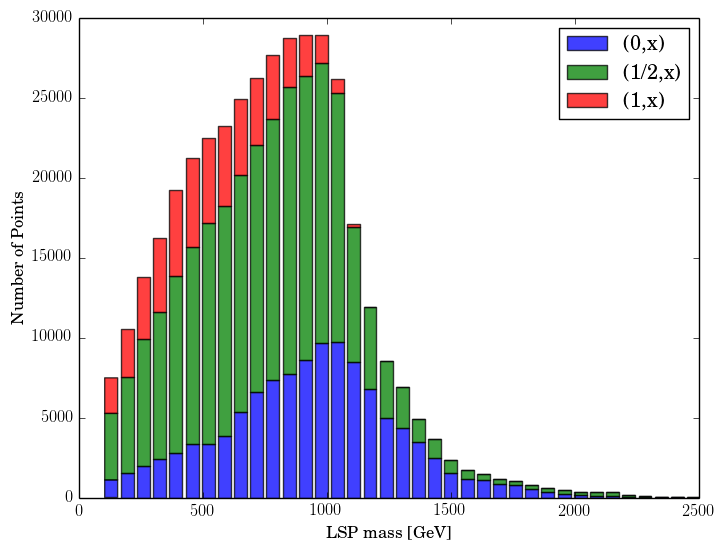}
%\label{fig:LSPbynM}
%\end{subfigure}
%\begin{subfigure}[b]{0.45\textwidth}
\includegraphics[width=0.45\textwidth]{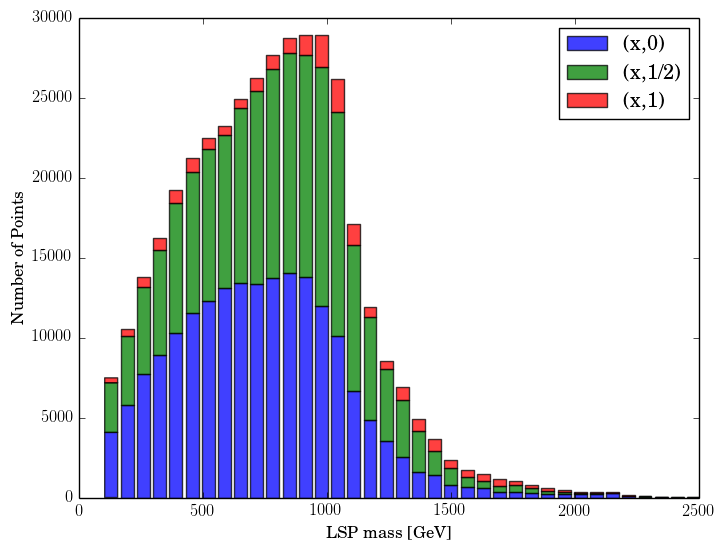}
%\label{fig:LSPbynH}
%\end{subfigure}
\caption{Distribution of the LSP mass (in GeV), broken down by modular weights. The left plot is broken down by $n_M$ and the right is by $n_H$. All points have an acceptable EW vacuum, Higgs mass, and neutralino relic density.}
\label{plot:LSPbymodularweight}
\end{figure}

In the right half of Figure~\ref{plot:gluinodist}, we see that the majority of possible LSP masses are $\mathcal{O}(1\,{\rm TeV})$ and likely accessible at the LHC. The sharp cutoff at 100~GeV is the result of the LEP limit on the chargino mass. The KKLT distributions for the LSP mass mirror those for DMM. Figure~\ref{plot:LSPbymodularweight} breaks this plot down by modular weight combination, similar to Figure~\ref{plot:gbymodularweight}. There is a weak dependence on $n_H$, with smaller $n_H$ preferring smaller values of the LSP mass, but a strong dependence on $n_M$. Larger values of $n_M$ push us towards smaller LSP masses, where the neutralino is almost exclusively Higgsino-like. For $n_M=1$, the up-type Higgs soft mass is generally large at the GUT scale, and it must run to a negative value to achieve proper EWSB; in this parameter range, $m^2_{H_u}$ achieves a small negative value, and the $\mu$ term needs to be small enough to satisfy the $Z$-mass constraint. For $n_M=0$ or $\frac 1 2$ there are points where the LSP has a bino or wino-like wave-function. The wino-like points tend to come from the negative gauge contribution~(\ref{threshgaug1}) pushing the value of $M_2$ below zero, where $|M_2| < |M_1|,\, |\mu|$. We will come back to these wino-like points in subsequent sections.

%%%%%%%%%%%%%%%% FIGURE 5 %%%%%%%%%%%%%%%%%

\begin{figure}[htbp]
\centering
%\begin{subfigure}[b]{0.45\textwidth}
\includegraphics[width=0.45\textwidth]{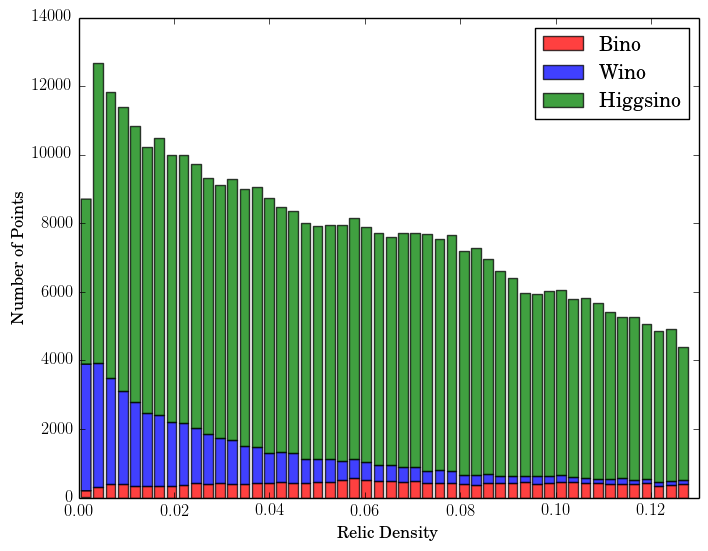}
%\label{fig:RelicbyLSP}
%\end{subfigure}
%\begin{subfigure}[b]{0.45\textwidth}
\includegraphics[width=0.45\textwidth]{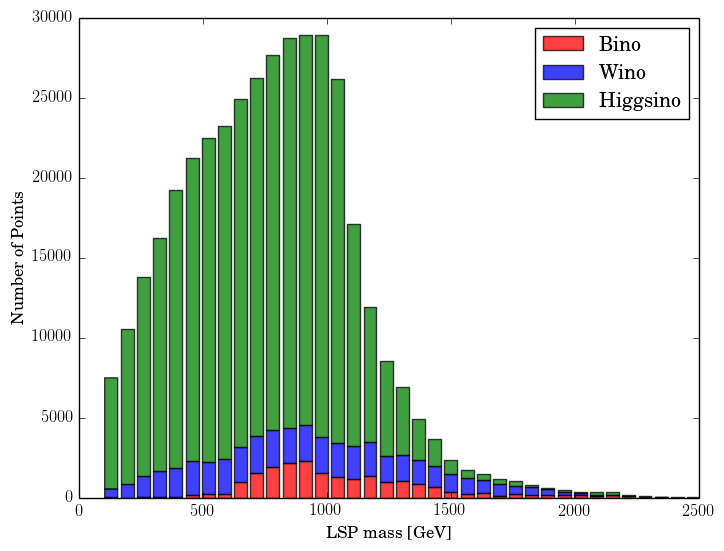}
%\label{fig:LSPbyLSP}
%\end{subfigure}
\caption{Distribution of the neutralino relic density $\Omega_{\chi}h^2$, left panel, and LSP mass (in GeV), right panel, aggregated for all modular weights. Both histograms are broken down by wave-function composition of the LSP. All points have an acceptable EW vacuum, Higgs mass, and neutralino relic density.}
\label{plot:relic}
\end{figure}

The distribution of relic density as a function of LSP mass, Figure~\ref{plot:relic}, shows that wino-like LSPs generally fail to saturate the Planck-preferred value of $\Omega_{\chi}h^2 \simeq 0.12$, with a proper relic density coming from Higgsino and bino-like points. The distribution of LSP masses, in the right panel of Figure~\ref{plot:relic}, shows that the majority of possible LSPs are Higgsino-like with an average mass of approximately 800~GeV. Higgsino annihilation in the early universe becomes too inefficient to achieve $\Omega_{\chi}h^2 \lappeq 0.12$ when the mass exceeds about 1~TeV, as is clear from the sharp cutoff in the distribution. Meanwhile, most bino-like points with an acceptable relic density arise from co-annihilation, primarily with stops, but also occasionally with gluinos. Low mass wino-like cases involve co-annihilation with other low-mass gauginos~\cite{BirkedalHansen:2001is, BirkedalHansen:2002am}, while higher-mass winos involve standard thermal freeze-out. 
Compared to KKLT, there is a relative paucity, about 5\% of the the sample, of bino-like LSPs in Figure~\ref{plot:relic}. As messengers are introduced, the values of the GUT scale coupling and beta function coefficients increase, leading to heavier Majorana masses at the GUT scale. At the messenger scale, the bino will experience the smallest deflection, so a point with a bino-like LSP needs some sort of conspiracy in the RG-flow to get the bino lighter than the wino and Higgsino.

The wino-like points, though relatively few in number, are worth exploring further as such an outcome does not occur in the KKLT scenario~\cite{Kaufman:2013oaa}. Wino-like points tend to have large values for the parameter $\alpha_g$, which controls the size of the correction in~(\ref{threshgaug1}). When $\alpha_g \simeq 1$, the wino mass $M_2$ is pushed to values which are below that for $M_1$ and $M_3$.
There is some correlation with the modular weights as well. These wino-like points are common at small $n_M$, admitting the entire range in $\alpha_g$, and vanish when $n_M=1$, as the one-loop Higgs mass corrections are not large enough. These points tend to have larger values of $n_H$, though this is a weaker effect, likely the result of needing a lighter Higgs soft mass to get electroweak symmetry breaking to occur properly. There are points that yield an acceptable low-energy spectrum for the entire range in $M_{\rm mess}$, and for $\alpha_m>0.5$. 
This relative lack of points for small $\alpha_m$ is seen globally in the DMM parameter space, because small or vanishing $\alpha_m$ corresponds to the limit where the model looks like minimal supergravity with a single mass parameter.

\begin{figure}[th]
\centering
%\begin{subfigure}[b]{0.45\textwidth}
\includegraphics[width=0.55\textwidth]{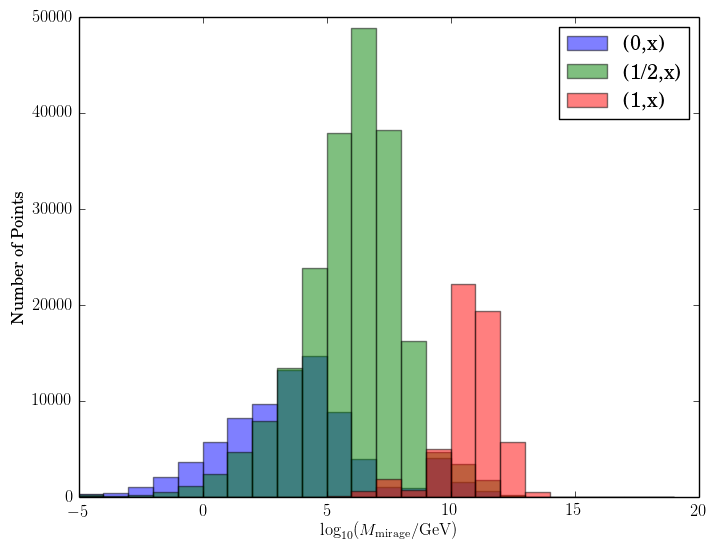}
%\label{fig:gbynM}
%\end{subfigure}
\caption{Distribution of the mirage scale, $M_{\rm mirage}$, defined by equation~(\ref{deflmirageunif}). All points have an acceptable EW vacuum, Higgs mass, and neutralino relic density. The red distribution represents all cases with $n_M=1$, the green those cases with $n_M = 1/2$, and the blue those cases with $n_M=0$. Not shown is a long tail of cases with $n_M=0,1/2$, extending to very small mirage scales ($M_{\rm mirage} \sim 10^{-30}\,{\rm GeV}$). Note that these histograms overlap, with the darker shaded green and red colors indicating the presence of cases with $n_M=0$ and $n_M=1/2$, respectively.}
\label{plot:MirageDist}
\end{figure}

Finally, Figure~\ref{plot:MirageDist} gives the distribution in the mirage scale, $M_{\rm mirage}$, from equation~(\ref{deflmirageunif}), broken down by the value of the matter field modular weight $n_M$. Though the mirage scale is not itself directly measurable, it can be inferred from a successful extraction of the gaugino mass hierarchy via measurements at the LHC~\cite{Altunkaynak:2009tg}, followed by RG evolution to discover the scale of unification~\cite{Baer:2006tb,Huitu:2015zqa}. It is intriguing to note that such an exercise provides some information on the matter modulus weight; for instance, points with large mirage scales are more often points with $n_M=1$. If instead one were to find that gaugino masses would unify at scales {\em below} a~TeV, then the matter modular weight is more likely $n_M=0$ or $n_M=1/2$.

%%%%%%%%%%%%%%%%%%%%%%%%%%%%%%%%%%%%%%%%%%%%%%%%%%%%%%%%%%%%
%%%%%%%%%%%%%%%%%%%%%%%%%%%%%%%%%%%%%%%%%%%%%%%%%%%%%%%%%%%%
%%%%%%%%%%%%%%%%%%%%%%%%%%%%%%%%%%%%%%%%%%%%%%%%%%%%%%%%%%%%

\subsection{DMM Perturbations on KKLT Base Points}

The previous subsection identified one particular qualitative difference between the pure mirage mediation/KKLT framework, and the allowed possibilities for deflected mirage mediation: the possibility of wino-like dark matter. In this section we will pursue other qualitative distinctions that arise from the addition of gauge-charged messenger fields.
Previous work~\cite{Kaufman:2013oaa} performed an exhaustive search in the KKLT framework, from which we will make our departures from KKLT into DMM. This work scanned $M_0$, $\alpha_m$, and $\tan\beta$ for each combination of modular weights $(n_M,n_H)$. From these results, we can choose a small number of benchmarks of the most phenomenologically relevant points in this framework. The benchmark points chosen are listed in Table~\ref{table:KKLTBenchmarks}, and are representative of the most phenomenologically interesting points within the KKLT framework across a span of modular weight combinations. 
We can then use these particular combinations of $M_0$, $\alpha_m$, $\tan\beta$, and the modular weights, then scan over $\alpha_g$, $M_{\rm mess}$, and $N$.

\begin{table}[t]
\begin{center}
\begin{tabular}{|c|ccccc|cccccccc|c|} \hline
&\multicolumn{5}{c|}{KKLT Parameters}&\multicolumn{8}{c|}{Key Masses [GeV]}&\\
\hline
\parbox{0.4in}{Point} & \parbox{0.3in}{$n_M$} & \parbox{0.3in}{$n_H$} & \parbox{0.4in}{$M_0$} & \parbox{0.4in}{$\alpha_M$} & \parbox{0.3in}{$\tan\beta$} & \parbox{0.4in}{$m_h$} & \parbox{0.4in}{$m_{\tilde{\chi}^0_1}$} & \parbox{0.4in}{$m_{\tilde{\chi}^0_2}$} & \parbox{0.4in}{$m_{\tilde{\chi}^{\pm}_1}$} & \parbox{0.4in}{$m_A$} & \parbox{0.4in}{$m_{\tilde \tau_1}$} & \parbox{0.4in}{$m_{\tilde g}$} & \parbox{0.4in}{$m_{\tilde t_1}$} &\parbox{0.5in}{$\Omega_{\chi} h^2$}\\
\hline 
1 & 0 & 0 & 1900 & 1.05 & 9 & 125.1 & 1406 & 1715 & 1715 & 2966 & 1910 & 2873 & 1434 & 0.062 \\
2 & 0 & 0 & 2900 & 1.80 & 9 & 123.8 & 1547 & 1553 & 1550 & 3224 & 2821 & 3084 & 1554 & 0.077 \\
3 & 0 & 0.5 & 1950 & 1.65 & 27 & 125.2 & 1415 & 1429 & 1420 & 1647 & 1749 & 2264 & 1500 & 0.124 \\
4 & 0 & 1 & 1350 & 0.63 & 29 & 123.5 & 837 & 1177 & 1177 & 1680 & 1217 & 2417 & 1685 & 0.114 \\
5 & 0.5 & 0 & 2000 & 1.25 & 28 & 125.5 & 676 & 683 & 679 & 1825 & 1219 & 2727 & 1461 & 0.055 \\
6 & 0.5 & 0.5 & 1800 & 0.70 & 9 & 123.3 & 1150 & 1554 & 1554 & 2327 & 1360 & 3055 & 1978 & 0.069 \\
7 & 0.5 & 0.5 & 3200 & 1.45 & 7 & 123.9 & 974 & 978 & 976 & 2628 & 2286 & 3924 & 2478 & 0.106 \\
8 & 0.5 & 1 & 4100 & 1.85 & 9 & 123.4 & 1090 & 1093 & 1092 & 855 & 2806 & 4072 & 2878 & 0.124 \\
9 & 1 & 0 & 4000 & 0.65 & 6 & 124.1 & 667 & 669 & 668 & 4596 & 1181 & 6517 & 3683 & 0.048 \\
10 & 1 & 0.5 & 3600 & 0.80 & 20 & 125.1 & 763 & 766 & 765 & 2987 & 891 & 5578 & 3473 & 0.063 \\ \hline
\end{tabular}
\caption{KKLT Benchmark Points. These cases with $N=0$ will serve as reference points for our exploration of the much richer DMM parameter space.}
\label{table:KKLTBenchmarks}
\end{center}
\end{table}%

\begin{table}[t]
\begin{center}
\begin{tabular}{|c|cc|cc|cc|cc|cc|cc|cc|}
\hline
&\multicolumn{12}{c|}{Ranges [GeV]}& \multicolumn{2}{c|}{  }\\
\hline
 &  \multicolumn{2}{c|}{$m_{\tilde{\chi}^0_1}$} & \multicolumn{2}{c|}{$m_{\tilde{\chi}^0_2}-m_{\tilde{\chi}^0_1}$} & \multicolumn{2}{c|}{$m_{\tilde{\chi}^{\pm}_1}-m_{\tilde{\chi}^0_1}$} & \multicolumn{2}{c|}{$m_{\tilde g}$} & \multicolumn{2}{c|}{$m_{\tilde \tau_1}$} & \multicolumn{2}{c|}{$m_{\tilde t_1}$} & \multicolumn{2}{c|}{$\Omega h^2$} \\
\parbox{0.4in}{Point} &  \parbox{0.4in}{Min} & \parbox{0.4in}{Max} & \parbox{0.4in}{Min} & \parbox{0.4in}{Max} & \parbox{0.4in}{Min} & \parbox{0.4in}{Max} & \parbox{0.4in}{Min} & \parbox{0.4in}{Max} & \parbox{0.4in}{Min} & \parbox{0.4in}{Max} & \parbox{0.4in}{Min} & \parbox{0.4in}{Max} & \parbox{0.4in}{Min} & \parbox{0.4in}{Max} \\
\hline
1 & 234 & 1485 & 4.8 & 442 & 0.2 & 358 & 526 & 3162 & 1400 & 1910 & 514 & 1844 & 0.001 & 0.122 \\
2 & 116 & 1638 & 4.6 & 1241 & 0.2 & 2.9 & 439 & 4094 & 1830 & 2824 & 982 & 4859 & 0.001 & 0.127 \\
3 & 968 & 1422 & 7.9 & 33 & 3.3 & 5.9 & 1074 & 2672 & 1217 & 1753 & 989 & 1661 & 0.014 & 0.123 \\
4 & 437 & 839 & 3.0 & 340 & 0.2 & 340 & 448 & 2421 & 1157 & 1217 & 1074 & 1687 & 0.006 & 0.128 \\
5 & 64 & 1101 & 5.2 & 100 & 1.6 & 4.8 & 440 & 3398 & 879 & 1241 & 668 & 2387 & 0.001 & 0.127 \\
6 & 654 & 1272 & 4.8 & 498 & 0.7 & 498 & 713 & 3560 & 1154 & 1385 & 1098 & 2334 & 0.019 & 0.128 \\
7 & 74 & 1453 & 3.0 & 7.9 & 1.6 & 2.7 & 1228 & 4956 &  1640 & 2313 &  1366 & 3126 &  0.001 & 0.128 \\
8 & 318 & 1111 & 2.5 & 1876 & 0.2 & 1.4 & 3767 & 5642 & 1779 & 2806 & 2700 & 7035 & 0.003 & 0.128 \\
9 & 86 & 1083 & 2.7 & 6.3 & 1.4 & 2.7 & 3029 & 6843 & 898 & 1214 & 2537 & 3929 & 0.001 & 0.127 \\
10 & 79 & 885 & 2.9 & 5.5 & 1.6 & 2.6 & 2169 & 5682 & 543 & 895 & 2707 & 3549 & 0.001 & 0.116 \\ \hline
\end{tabular}
\caption{Ranges for the superpartner masses in DMM for the KKLT Benchmark Points presented in Table~\ref{table:KKLTBenchmarks}. The minimum and maximum values give the observed range in each quantity over the three-dimensional scan in $(N, \alpha_g, M_{\rm mess})$. \label{table:DMMKKLT}}
\end{center}
\end{table}%

% LSP Mass
For a given KKLT point, deflection can lead to large changes in the spectrum.
Table~\ref{table:KKLTBenchmarks} gives the values for our benchmark KKLT points, and Table~\ref{table:DMMKKLT} shows the effect of scanning over the DMM extension of the parameter space for these points. That is, Table~\ref{table:DMMKKLT} gives the maximum and minimum value of each quantity, over the three-dimensional scan in $(N, \alpha_g, M_{\rm mess})$ described above.
Consider, for example, point~1 in Table~\ref{table:KKLTBenchmarks}. The KKLT parameter set predicts a $1400\,{\rm GeV}$ LSP neutralino. However, the range of lightest neutralino masses for point~1 in Table~\ref{table:DMMKKLT} indicates that the deflection can reduce the neutralino mass down to $\sim 250\, {\rm GeV}$, or (alternatively) push other superpartner masses to very low values. For all ten benchmarks the minimum LSP mass found is generally quite lower than that predicted by the KKLT base point.

% LSP Composition
The LSP for KKLT base point~1 is 99.8\% bino-like. The neutralino relic density is acceptable as the result of co-annihilation between this bino-like state and the nearly degenerate stop. From Table~2 we see that over the range of DMM variants of this point, there is a transition from bino-like to wino-like and Higgsino-like LSPs. This is evidenced by the minimum values obtained in the mass differences $m_{\tilde{\chi}^0_2}-m_{\tilde{\chi}^0_1}$ and $m_{\tilde{\chi}^{\pm}_1}-m_{\tilde{\chi}^0_1}$. As a consequence, co-annihilation with light gauginos can produce a greatly reduced neutralino relic density. The LSP for KKLT base point~2 is 99.7\% Higgsino-like, and again very close in mass to the lightest top squark. Much of the DMM parameter space based on this point also yields a Higgsino-like LSP, but the possibility of getting large $m_{\tilde{\chi}^0_2}-m_{\tilde{\chi}^0_1}$ for certain $N$ and $\alpha_g$ combinations suggests that wino-like LSPs are also possible. In such cases, the stop mass can be quite a bit larger than for the KKLT point. Similar behavior is seen with base point~5 in Tables~\ref{table:KKLTBenchmarks} and~\ref{table:DMMKKLT}.

%%%%%%%%%%%%%%%%% Table 2 Bar Charts: Points 1 and 2 %%%%%%%%%%%%%%%%%%
%
\begin{figure}[t]
\begin{center}
\includegraphics[width=0.45\textwidth]{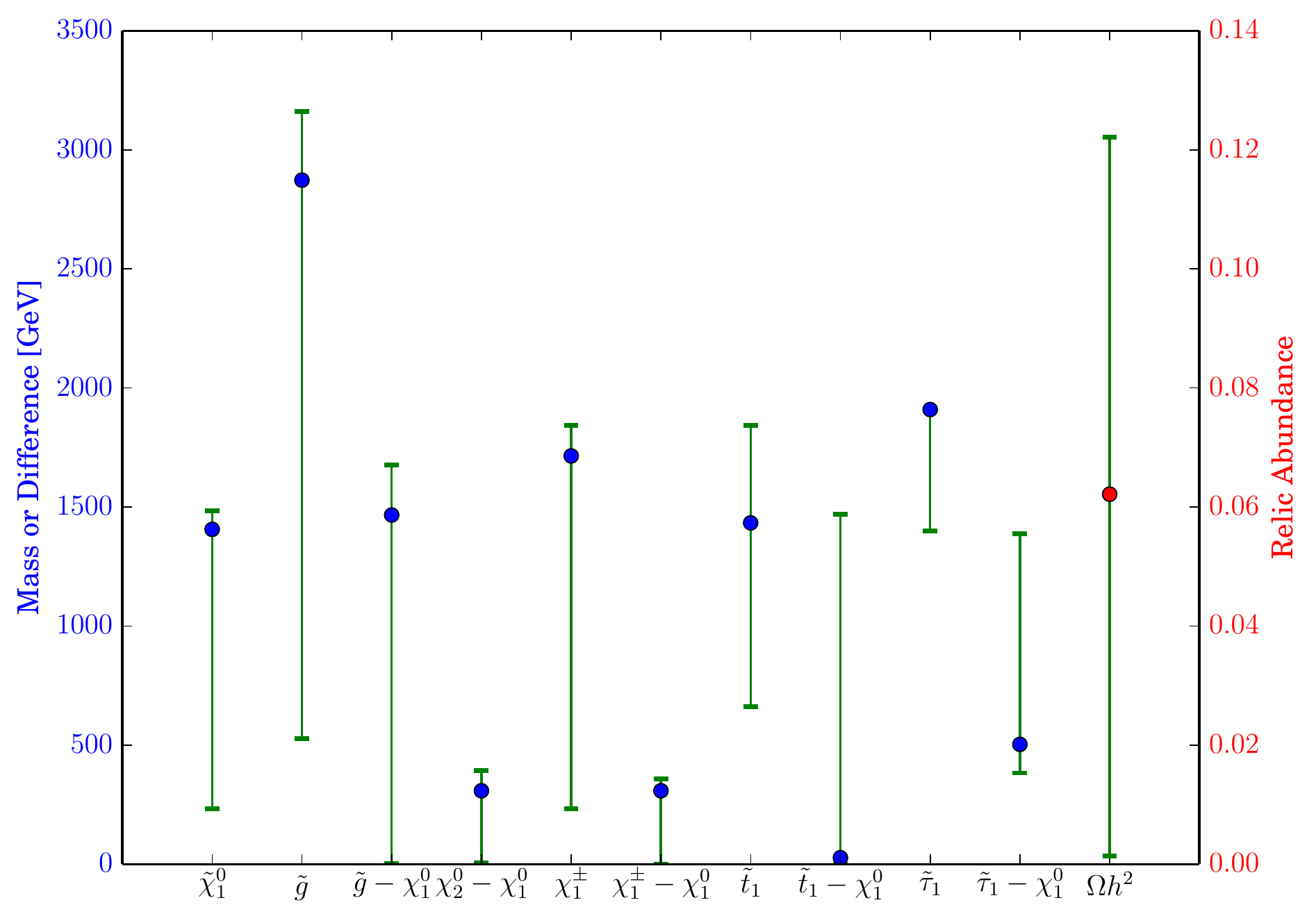}
\includegraphics[width=0.45\textwidth]{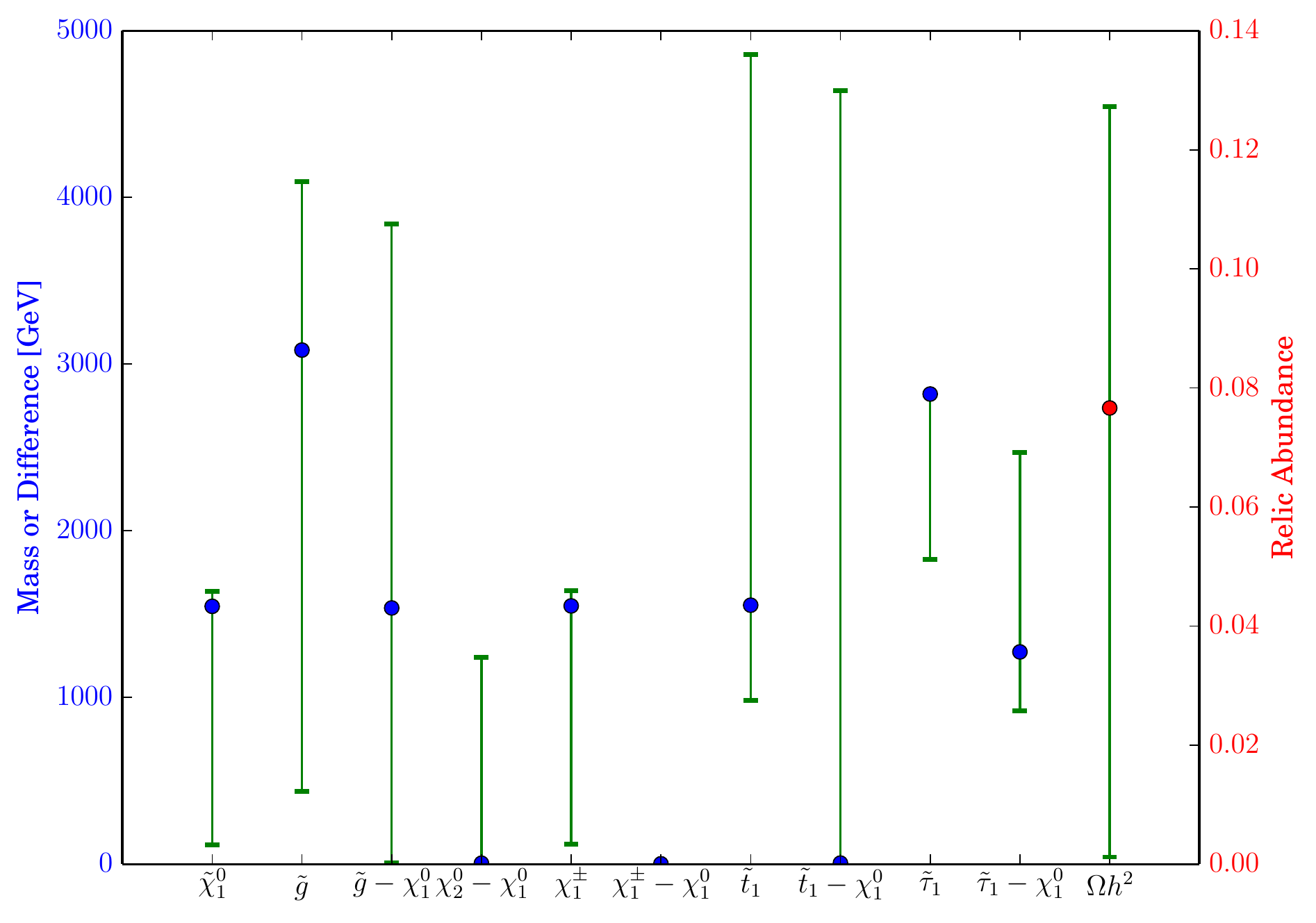}
\caption{Low energy mass ranges for point~1 (left) and point 2 (right) for the quantities in Table~\ref{table:DMMKKLT}. The dots represent the values for the corresponding KKLT base point with zero messengers, from Table~\ref{table:KKLTBenchmarks}.}
\label{plot:n00}
\end{center}
\end{figure}
%%%%%%%%%%%%%%%%%%%%%%%%%%%%%%%%%%%%%%%%%%%%%%%%%%%%%%

% Lighter gluinos and compressed gluino/LSP
We can visualize the content of these two tables by looking at Figure~\ref{plot:n00}, which depicts the minimum and maximum values of Table~\ref{table:DMMKKLT} for point~1 (left panel), and point~2 (right panel), both involving the modular weight set $(n_M,n_H)=(0,0)$. 
The heavy dot represents the KKLT base point from Table~\ref{table:KKLTBenchmarks}. The magnitude of the DMM corrections~(\ref{threshgaug1}) and~(\ref{threshmssq1}) increase as $\alpha_g$ moves from negative to positive values. 
In both cases we see the striking effects that the gauge messenger fields can have on the resulting low-energy spectrum. Notable is the great reduction in gluino mass that is possible, relative to the KKLT base point. This typically comes in conjunction with a great compression of the spectrum, with the mass difference $m_{\tilde{g}}-m_{\tilde{\chi}^0_1}$ often approaching zero in the extreme DMM~limit.

%%%%%%%%%%%%%%%%% Table 2 Bar Charts: Points 5 and 8 %%%%%%%%%%%%%%%%%%
%
\begin{figure}[t]
\begin{center}
\includegraphics[width=0.45\textwidth]{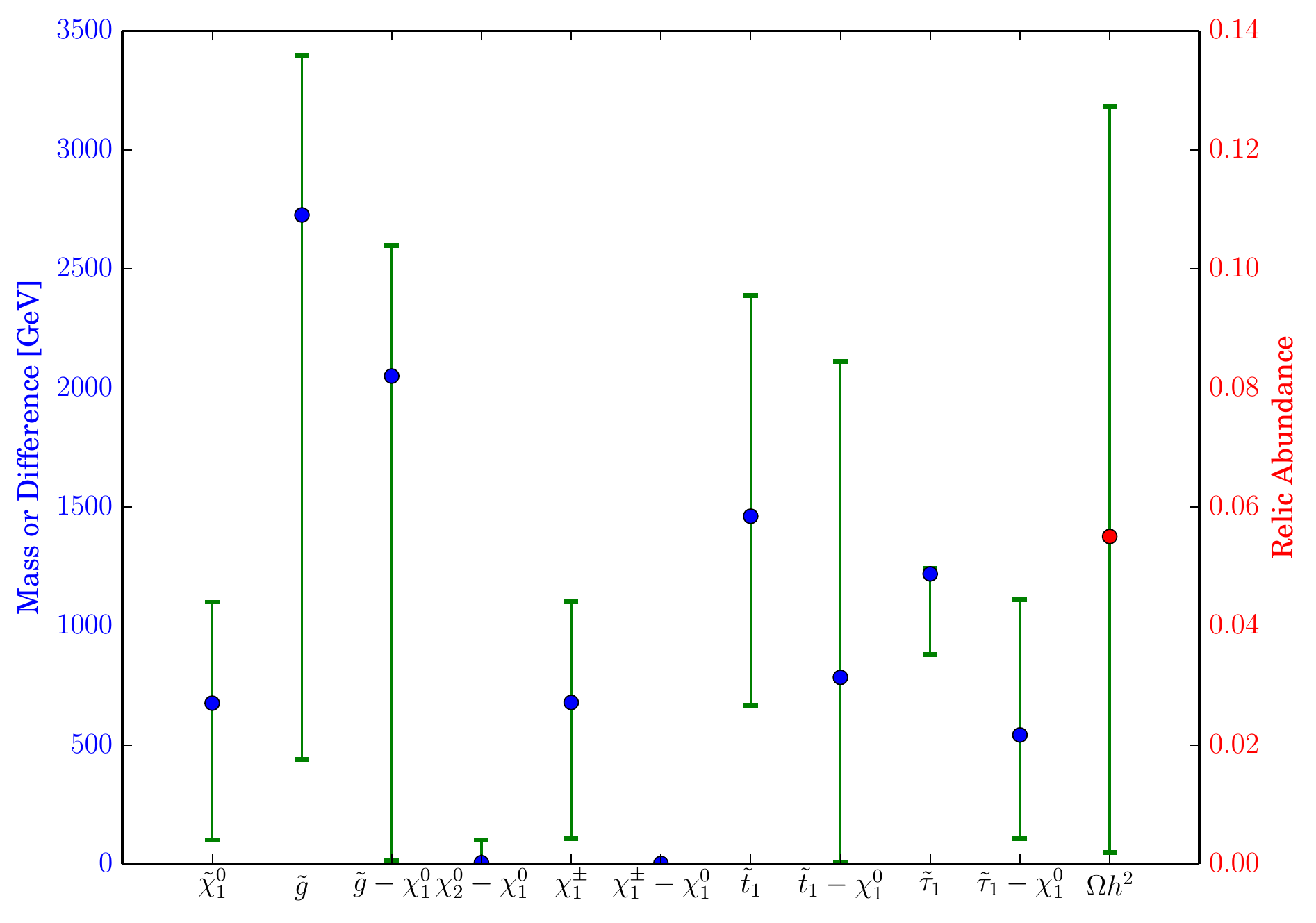}
\includegraphics[width=0.45\textwidth]{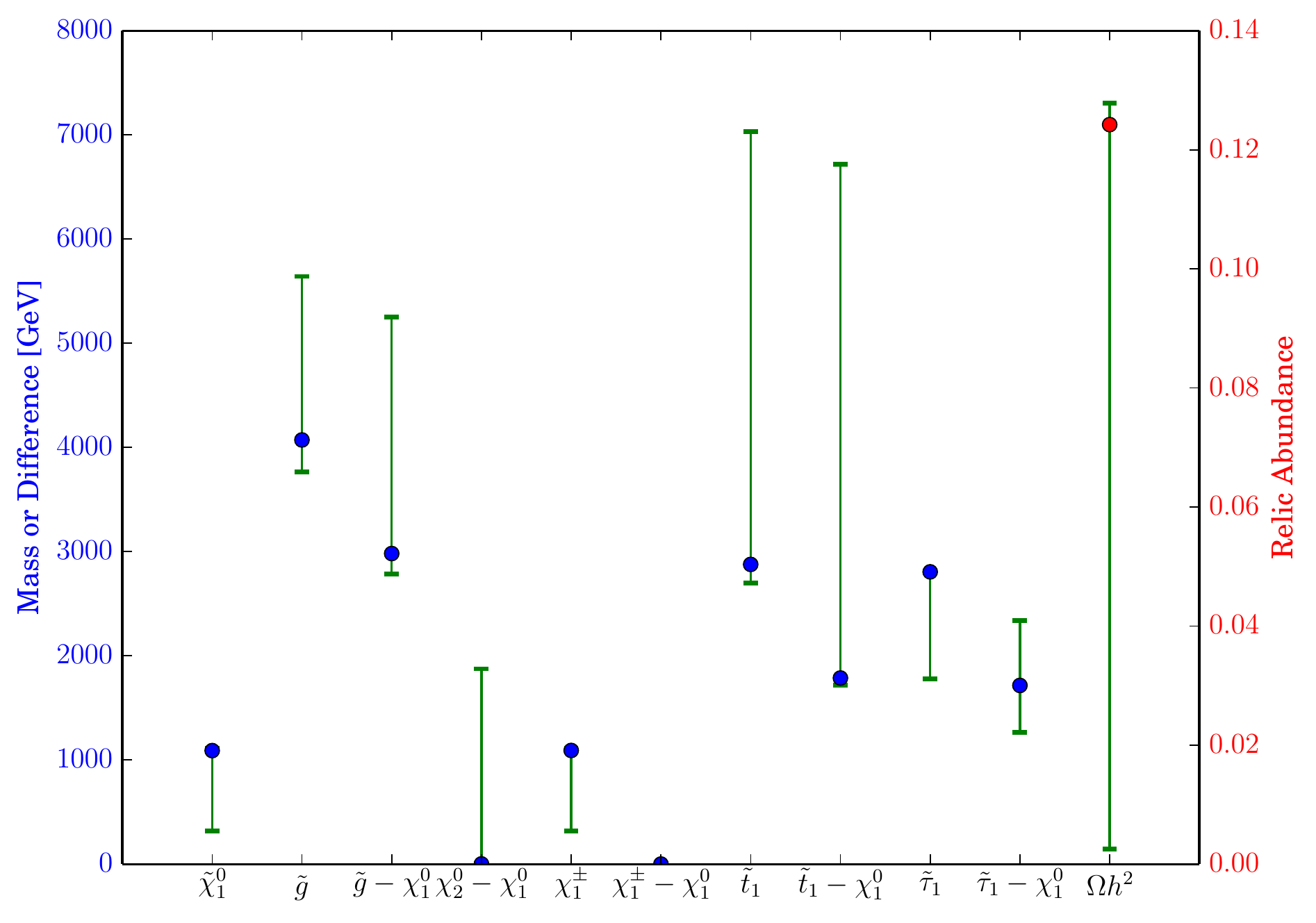}
\caption{Low energy mass ranges for point~5 (left) and point 8 (right) for the quantities in Table~\ref{table:DMMKKLT}. The dots represent the values for the corresponding KKLT base point with zero messengers, from Table~\ref{table:KKLTBenchmarks}.}
\label{plot:n050}
\end{center}
\end{figure}
%%%%%%%%%%%%%%%%%%%%%%%%%%%%%%%%%%%%%%%%%%%%%%%%%%%%%%

As a general rule, addition of deflected mirage mediation results in a more compressed superpartner spectrum, though there are variations depending on the KKLT base point. Figure~\ref{plot:n050} depicts the minimum and maximum values of Table~\ref{table:DMMKKLT} for point~5 (left panel), and point~8 (right panel). Consider first the DMM ensemble based on point~5. Here we see a degenerate system of electroweak gauginos which is not significantly affected by the DMM deformation. For the base point in Table~\ref{table:KKLTBenchmarks}, the other superpartners (particularly those carrying color) are not close to the LSP in mass. The addition of gauge messengers, for positive $\alpha_g$, has the potential to drive these masses down signficantly, potentially yielding a rich diversity of particles at, or just around, the TeV~scale. Similar behavior is seen with the DMM ensemble based on points~3, 4~and~7. Yet for base point~8, the KKLT point is already near the low end of the ranges for gluino and stop masses. The DMM addition can only increase these masses (and decrease the LSP mass), meaning that here the gauge messengers generally {\em reduce} the compression in the spectrum.
Naturally, these effects have profound implications for superpartner searches at the LHC~\cite{Krippendorf:2013dqa,Anandakrishnan:2014nea}, which will be our focus in Section~\ref{sec:collider}.

%%%%%%%%%%%%%%%%% Point 2: scatter plots of alphas, alpha vs m0 %%%%%%%%%%%%%%%%%%
%
\begin{figure}[t]
\begin{center}
\includegraphics[width=0.45\textwidth]{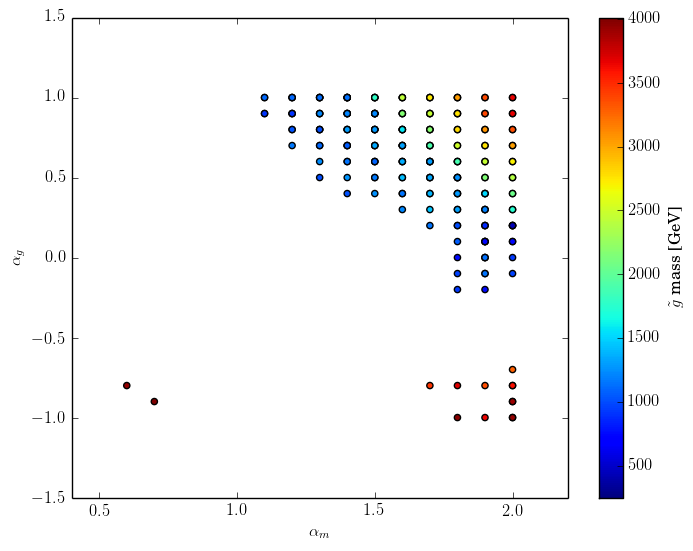}
\includegraphics[width=0.45\textwidth]{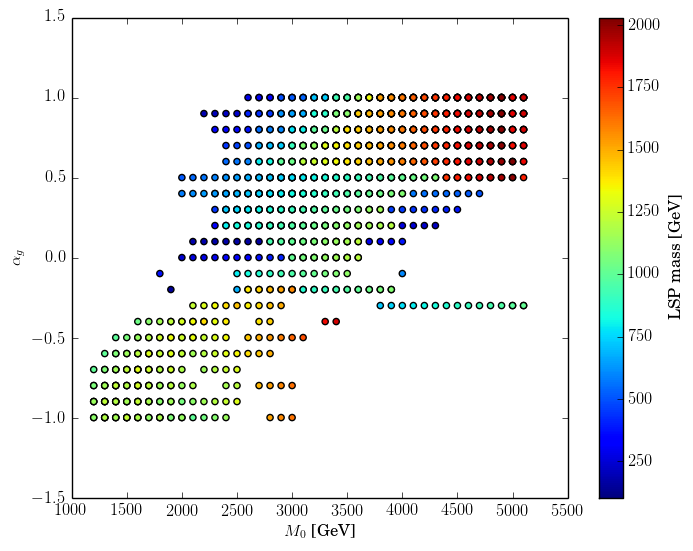}
\caption{Allowed parameter space for DMM perturbations on point~2 of Table~\ref{table:KKLTBenchmarks}. In both cases we assume $(n_M,n_H) = (0,0)$, $\tan\beta=9$, and fix either $M_0 = 2900\,{\rm GeV}$ (left panel) or $\alpha_m = 1.8$ (right panel), as is the case for the KKLT base point~2 in Table~\ref{table:KKLTBenchmarks}. The allowed region in the $(\alpha_m,\alpha_g)$ plane (left) and $(M_0,\alpha_g)$ planes are displayed for $N=3$.}
\label{plot:point2scan}
\end{center}
\end{figure}
%%%%%%%%%%%%%%%%%%%%%%%%%%%%%%%%%%%%%%%%%%%%%%%%%%%%%%

Before proceeding to the~LHC implications of deflected mirage mediation, it is instructive to consider how the inclusion of gauge messengers can affect the space of viable KKLT base points themselves. We focus here on the particular case of base point~2 from Table~\ref{table:KKLTBenchmarks}. 
Figure~\ref{plot:point2scan} shows the effect of adding $N=3$ messenger multiplets, over all messenger mass scales, to this point. In the left panel, $\alpha_m$ is allowed to vary away from the original value of $\alpha_m = 1.8$, while keeping $M_0 = 2900\,{\rm GeV}$ fixed. In the right panel, $M_0$ is allowed to vary while $\alpha_m$ is held fixed.  Our KKLT base point is clearly part of the $\alpha_g=0$ line in the left panel, but the inclusion of $\alpha_g \neq 0$ allows other $\alpha_m$ values, including the possibility of very light gluinos for $\alpha_g \to 1$ near the KKLT limit of $\alpha_m=1$. 

In the left panel  of Figure~\ref{plot:point2scan} we see three distinct regions, two points where $\alpha_m\sim0.5$ and $\alpha_g\sim-1$, an area with $\alpha_m\sim1.8$ and $\alpha_g<-0.5$, and finally a large region with $\alpha_m>1$ and $\alpha_g>-0.5$. The first region comes about due to points with a bino-like LSP and an intermediate messenger scale $M_{\rm mess} \sim 10^{9}\,{\rm GeV}$, driving the gluino mass down and softening the running of the stop so that the Higgs mass for these points is boosted by a highly mixed, and light, stop. The other region, below $\alpha_g<-0.5$, are points with light Higgsino LSPs and mixed stops as well. The gap around $\alpha_g \simeq -0.5$ consists of points where the stop is either very light or the LSP. Here the anomaly mediated contributions are large, leading to a light stop mass in the UV, while the messenger-scale corrections are too small to drive the neutralino mass below that of the stop.  
The upper region consists of points with either a wino LSP or Higgsino LSP and heavy stops. As $\alpha_g$ decreases, $\alpha_m$ needs to increase to compensate to make the stop heavy and thereby obtain the correct Higgs mass. 

In the right panel of Figure~\ref{plot:point2scan}, we see that varying $\alpha_g$ allows for a large range of overall mass scales $1\,{\rm TeV} \leq M_0 \leq 5 \,{\rm TeV}$. However, the resulting mass of the superpartners is not simply correlated with this quantity, once the phenomenological constraints are imposed on the parameter space. This is evidenced by the color in the right panel of Figure~\ref{plot:point2scan}, which gives the LSP mass in GeV. Larger values of $M_0$ tend to require a smaller value of the messenger scale, and larger values of $\alpha_g$, to get the correct Higgs mass. Thus, for a fixed value of $\alpha_g$, the points in the figure at different $M_0$ values tend to have differing messenger mass scales.  For positive $\alpha_g$, the largest messenger scales are at the far right of the plot (large $M_0$). For negative $\alpha_g$ this relationship is inverted. Above $\alpha_g=0.5$, the LSP is exclusively wino-like, as the deflection is now large enough to push the wino mass to small values. This is significant, as wino-like LSPs were not found in the pure KKLT scenario studied in~\cite{Kaufman:2013pya}.

\section{LHC Implications of Deflected Mirage Mediation}
\label{sec:collider}

The discovery of the Higgs boson at the LHC in 2012 and ever-improving bounds on the dark matter relic density from the PLANCK experiment have placed considerable constraints on the form that any new model of physics might take. As the LHC begins running at $13$~TeV, and later $14$~TeV, these constraints are expected to further tighten. As we enter into this new era, we are particularly interested in how DMM will fare in the coming years. 
For pure mirage mediation, embedded in the KKLT framework for Type~IIB string theory, previous work~\cite{Kaufman:2013oaa} demonstrated that the heavy mass spectra left KKLT undiscoverable at $\sqrt{s} =8\TeV$, with rather dim discovery prospects at $\sqrt{s}=14\TeV$. Further, direct detection of dark matter was left nearly impossible.
However, with DMM we have seen that the addition of a small, fixed number of vector-like messengers can affect the running of these masses at some scale between the electroweak and the Planck. As was shown in the previous section, this ultimately results in lighter superpartners which could be within the reach of the LHC for detection in the near future. In the following section we will determine those portions of the parameter space which have been ruled out by direct searches at $\sqrt{s} = 8\TeV$, and evaluate the extent to which DMM modifications can enhance accessibility at $\sqrt{s} = 14\TeV$. 
%We will then follow by discussing the extent to which this model could present itself at dark matter direct detection experiments, and discuss the implications of a possible $100\TeV$ collider in the future.

\subsection{Benchmark Points}

\begin{table}[t]
\begin{center}
\begin{tabular}{|c|cc|ccccc|ccc|cc|ccc|cc|}
\hline
 &  \multicolumn{2}{c|}{Point 1} & \multicolumn{5}{c|}{Point 2} & \multicolumn{3}{c|}{Point 3} & \multicolumn{2}{c|}{Point 5} & \multicolumn{3}{c|}{Point 6} &  \multicolumn{2}{c|}{Point 7} \\
Quantity & 1.0 & 1.1 & 2.0 & 2.1 & 2.2 & 2.3 & 2.4 & 3.0 & 3.1 & 3.2 & 5.0 & 5.1 & 6.0 & 6.1 & 6.2 & 7.0 & 7.1 \\
\hline
$(n_M,n_H)$ & \multicolumn{2}{c|}{$(0,0)$} & \multicolumn{5}{c|}{$(0,0)$}  & \multicolumn{3}{c|}{$(0,0.5)$} & \multicolumn{2}{c|}{$(0.5,0)$} & \multicolumn{3}{c|}{$(0.5,0.5)$} & \multicolumn{2}{c|}{$(0.5,0.5)$}\\
$M_0$ & \multicolumn{2}{c|}{$1900$} & \multicolumn{5}{c|}{$2900$} & \multicolumn{3}{c|}{$1950$} & \multicolumn{2}{c|}{$2000$} & \multicolumn{3}{c|}{$1800$} & \multicolumn{2}{c|}{$3200$} \\
$\alpha_m$ & \multicolumn{2}{c|}{$1.05$} & \multicolumn{5}{c|}{$1.80$} & \multicolumn{3}{c|}{$1.65$} & \multicolumn{2}{c|}{$1.25$} & \multicolumn{3}{c|}{$0.7$} & \multicolumn{2}{c|}{$1.45$} \\
$\tan\beta$ & \multicolumn{2}{c|}{$9$} & \multicolumn{5}{c|}{$9$} & \multicolumn{3}{c|}{$27$} & \multicolumn{2}{c|}{$28$} & \multicolumn{3}{c|}{$9$} & \multicolumn{2}{c|}{$7$} \\ \hline
$\alpha_g$ & 0 & $0.55$ & 0 & $0.35$ & $0.10$ & $0.2$ & $0.1$ & 0 & $0.15$ & $-0.90$ & 0 & $0.2$ & 0 & $0.3$ & $0.1$ & $0$ & $-0.35$ \\
$M_{\rm mess}$ & -- & $10^5$ & -- & $10^{14}$ & $10^5$ & $10^6$ & $10^{12}$ & -- & $10^{14}$ & $10^6$ & -- & $10^9$ & -- & $10^5$ & $10^5$ & -- & $10^5$  \\
$N$ & 0 & $2$ & 0 & $3$ & $3$ & $3$ & $4$ & 0 & $5$ & $3$ & 0 & $3$ & 0 & $3$ & $3$ & 0 & $3$ \\ \hline \hline
$m_{\tilde{g}}$ & 2873 & 1002 & 3084 & 1448 & 1061 & 1010 & 1065 & 2264 & 1074 & 1901 & 2727 & 1013 & 3055 & 713 & 902 & 3924 & 1228 \\
$m_{\tilde{t}_1}$ & 1434 & 1265 & 1554 & 1061 & 2582 & 2536 & 1008 & 1500 & 1062 & 1109 & 1461 & 668 & 1978 & 1227 & 1171 & 2478 & 1630 \\ \hline
$m_{\tilde{\chi}^0_1}$ & 1406 & 986 & 1547 & 836 & 147 & 942 & 727 & 1415 & 1042 & 1101 & 676 & 661 & 1150 & 654 & 696 & 974 & 1067 \\
B\% & 99.8\% & 0.4\% & 0.1\% & 0.1\% & 0.0\% & 0.1\% & 0.1\% & 0.9\% & 0.5\% & 0.3\% & 0.1\% & 0.4\% & 99.6\% & 5.6\% & 98.7\% & 0.0\% & 0.1\% \\
H\% & 0.2\% & 0.4\% & 99.7\% & 99.5\% & 99.7\% & 97.6\% & 99.7\% & 97.3\% & 96.5\% & 98.8\% & 99.6\% & 98.2\% & 0.38\% & 72.6\% & 0.3\% & 99.9\% & 99.3\% \\
$\Omega h^2$ & 0.062 & 0.028 & 0.077 & 0.070 & 0.003 & 0.091 & 0.057 & 0.124 & 0.044 & 0.041 & 0.055 & 0.018 & 0.069 & 0.019 & 0.113 & 0.106 & 0.107 \\ \hline
\end{tabular}
\caption{Benchmarks for LHC study of DMM parameter space. A subset of the KKLT base points in Table~\ref{table:KKLTBenchmarks} is here reproduced, together with one or more perturbations that involve gauge-charged messengers. The collection represents a variety of input parameters, LSP type, thermal relic densities, and mass scales. These example parameter points will be the focus of our detailed study of LHC~phenomenology to follow.\label{table:Partners}}
\end{center}
\end{table}%

The DMM framework leaves us with a large number of possible input parameters consistent with the constraints we have placed. While we are interested in the detection prospects of the entirety of the remaining parameter space, we can gain a sense of the reach that an experiment would have by considering the extrema of the parameter space, and using these benchmarks to evaluate whether or not a given subsection of the parameter space would be accessible for a given experiment. In~\cite{Kaufman:2013oaa} it was found that for the KKLT framework, roughly a dozen benchmark points could be chosen to give a sense of the discoverability of the parameter space at large. Here too we will consider a small number of benchmark points that are representative of the DMM framework.

We will consider each of the benchmark points discussed in Table~\ref{table:KKLTBenchmarks}, around which we performed a scan in the DMM parameter space. For each grid of DMM points, we consider the following: the point with the lightest LSP, the point with the lightest gluino, and the point with the lightest stop $\tilde{t}_1$. We further isolate only the points where $N=3$, and again choose the same trio of points. We do not impose a lower bound on the gluino mass in this exercise, and will often find cases with gluino masses below the often-quoted bound of about $1200\GeV$. In fact, as we shall see below, many of these cases would indeed have been discovered at the previous LHC run, while still others would have escaped detection due to compression between the gluino mass and that of the LSP. 

The points that we have chosen to present in detail represent a subset of the entirety of the DMM parameter space, and are collected in Table~\ref{table:Partners}. We choose this set to gain a fuller understanding of the amount of DMM parameter space that the LHC can probe, while exhibiting a large range in the input parameters, LSP type, mass spectra, and mass scales we consider. For example, point~1.1 in Table~\ref{table:Partners} represents a case described in the previous section, near the lower range of the vertical bars in the left panel of Figure~\ref{plot:n00}. The KKLT base point (1.0) has a primarily bino-like LSP, while the DMM perturbation produces a primarily wino-like LSP, with a significantly smaller mass. All superpartner masses are reduced by the gauge mediation in this case, with a highly-compressed spectrum emerging. 

The four perturbations on KKLT base point~2.0 all involve a relatively small and positive $\alpha_g$, and at least three messenger fields, thus reducing the masses of the gauginos (the LSP mass and gluino mass are given explicitly in Table~\ref{table:Partners}). Yet the effect on the stop mass depends crucially on the messenger mass scale, $M_{\rm mess}$. Larger messenger scales means fewer decades in energy for the messenger-corrected renormalization group equations to operate, and consequently lighter stops. This is the complement to the discussion in the previous section regarding the right panel of Figure~\ref{plot:point2scan}. As we will see in the following section, the discovery prospects for points 2.1~-~2.4 are quite different, despite the roughly similar key masses. 

The two perturbations on KKLT base point~3.0 show two very different types of compressed spectra that can emerge at opposite ends of the $\alpha_g$ parameter space. Cases with large numbers of messengers, positive $\alpha_g$, and high messenger scale tend to exhibit the largest amount of compression, while negative $\alpha_g$ has less impact on the gluino mass, while still compressing the stop and LSP masses. The stop mass $m_{\tilde{t}_1}$ and LSP mass $m_{\tilde{\chi}^0_1}$ are roughly similar for points~3.1 and~3.2, yet we will see that the former point will be discovered within the first 40~fb$^{-1}$ at $\sqrt{s} = 14\TeV$, while the latter will require as much as 240~fb$^{-1}$ for discovery.

We round out our benchmarks with a perturbation each on KKLT base points~5.0 and~7.0, and two for base point~6.0. This latter case is particularly interesting, in that the KKLT base point is in the region identified in~\cite{Kaufman:2013oaa} as giving purely bino-like LSP neutralinos. Indeed, case~6.2 is such a point, with roughly the correct thermal relic density. But the pertrubation in~6.1, with only slightly larger $\alpha_g$, and identical $M_{\rm mess}$ and $N_{\rm mess}$, gives a mixed wavefunction LSP and much lighter gluino. As we will discuss in the next section, both points~6.1 and~6.2 would have escaped detection at the previous LHC~run, but both are prime candidates for discovery in the first 5~fb$^{-1}$ in the upcoming run.

\subsection{Relevant LHC Searches}

In addressing, more specifically, the issue of detection at the LHC, we hope to identify some commonalities among these points, and the larger set of~60 benchmarks from which they are chosen,  that will guide our search strategies. 
%----- Simulation ---------------
In order to simulate the LHC signature for each of these benchmark points, we take the electroweak-scale SLHA file generated by SoftSUSY (as discussed previously), then generate the full decay table with SUSY-HIT~\cite{Djouadi:2006bz}. For several of the benchmark points, the mass spectrum features a neutralino LSP with a stop~NLSP only slightly heavier. For these points, the decay $\tilde{t}_1\rightarrow t \tilde{\chi}_1^0$ is highly suppressed, as is the decay $\tilde{t}_1\rightarrow b W \tilde{\chi}_1^0$, which are the dominant decay channels when $\Delta m(\tilde{t}_1,\tilde{\chi}_1^0)>m_t$ and $\Delta m(\tilde{t}_1,\tilde{\chi}_1^0)>m_W+m_b$, respectively. With such a small mass gap, the only decay processes allowed are $\tilde{t}_1\rightarrow c\tilde{\chi}_1^0$ and $\tilde{t}_1\rightarrow b f \bar{f} \tilde{\chi}_1^0$. These processes require additional calculations separate from the main SUSY-HIT routines~\cite{Grober:2014aha}.
Once these decay tables are generated, we use MadGraph5\_aMC@NLO 2.2.2~\cite{Alwall:2014hca} to simulate production of all $\rm{p} \rm{p} \to \tilde{X}\tilde{X}$ processes, where $\tilde{X}$ represents any supersymmetric particle. We then use MadEvent to generate 10,000~events for each parameter point, followed by PYTHIA 6.4 to perform the showering and hadronization. Detector simulation is performed in Delphes-3.1.2 using the default ATLAS detector card~\cite{deFavereau:2013fsa}. In total, we generated events for~60 parameter points, including those of Table~\ref{table:Partners}.

When considering the LHC implications of DMM, we consider the results published by the ATLAS and CMS collaborations. Both have conducted many searches for possible SUSY signatures in the $\sqrt{s}=8\TeV$ data they have collected. To date, however, no signal above background expectations has been found by either of the two collaborations. For the sake of simplicity, we will consider only the ATLAS search results; the searches conducted by ATLAS tend to utilize geometric cuts in their signal region definitions, which are better suited to simple computer simulation. To date, ATLAS has published 32 searches using the full $\sqrt{s}=8\TeV$ data set, as well as a number of summary documents. By considering the properties of the 60~DMM points generated, we can focus on a small number of these searches to target the event topologies most likely to be produced. 

For example, of the 60~benchmark points for which we performed simulations at $\sqrt{s}=8\TeV$, we find that lepton production is generally rare for the DMM points, though it can be substantial for the associated KKLT~base points. Before considering lepton $p_T$ and angular distribution, we find that across the benchmarks, events with leptons generally make up less than 10\% of the total number of events; events with two leptons or more are even rarer, typically making up no more than 5\% of the total number of events. Conversely, 51 of the 60 benchmark points have zero reconstructed leptons for at least 80\% of the events. At $\sqrt{s}=14\TeV$, this property persists. Only in two cases do signatures with leptons become the predominant topology, and in only two others does it surpass 33\% of the total event count. At $\sqrt{s}=14\TeV$, these points have a total SUSY production cross section of $\mathcal{O}(10^{-4})\rm{~fb}$ and $\mathcal{O}(10^{-2})\rm{~fb}$, respectively. These cross sections are sufficiently low that, despite their high lepton production rates, they will not result in a significant number of leptons produced at the LHC during the 14~TeV run. With this in mind, we can safely consider only searches that contain a lepton veto. 

Jet multiplicities tend to be relatively low for these 60~benchmarks, with two-thirds of the cases studied having a peak jet multiplicity of $N_{\rm jet} \leq 5$. A small subset of the remainder have a broad distribution of jet multiplicities, peaking at $N_{\rm jet} = 7~-~8$, with long tails that extend to very large multiplicities. But the vast majority of our cases will be visible first in the low jet-multiplicity channels. As most of the cases that are accessible at the 8~TeV run involve light stops, it is not surprising that a large fraction of the events we simulated have one or more b-tagged jets. Across the 60~cases studied, just under 40\% of events contain at least one b-tagged jet, with several of the benchmarks exhibiting twice that fraction of events with b-tagged jets.

% ------------- TABLE: BASIC SIGNAL PROPERTIES ---------------
\begin{table}
\begin{center}
\begin{tabular}{|c||c|c||c|c|c|}\hline
\parbox{1in}{Benchmark}	& \parbox{1in}{$\sigma_{8\TeV}~\rm{(fb)}$}		& \parbox{1in}{$\sigma_{14\TeV}~\rm{(fb)}$}	 & \parbox{0.8in}{\% Lepton} 	& \parbox{0.8in}{Peak $N_{\rm jets}$} & \parbox{0.8in}{\% B-Jets} \\\hline\hline
1.0			  & $5.1\times10^{-3}$	& $0.67$ & 3.6\% & 3 & 27.8\% \\
1.1			  & $10.1$		& $2.6\times10^{2}$	 & 0\% & 3 & 4.6\% \\ \hline
2.0			 & $1.8\times10^{-3}$ & $0.31$	& 0\%	& 2 & 23.0\% \\
2.1			& $0.3$ & $22.9$	& 21.1\% 	& 8 & 85.3\% \\
2.2			& $6.0$	 & $1.7\times10^{2}$	& 21.9\%	& 8 & 87.5\% \\
2.3			& $9.9$	 & $2.4\times10^{2}$ & 0.3\%	& 4 & 5.1\% \\
2.4			& $5.6$	 & $1.8\times10^{2}$	 & 10.5\% 	& 6 & 53.6\% \\ \hline
3.0			& $1.3\times10^{-2}$	 & $3.9$	 & 18.6\%	 & 6 & 54.5\% \\
3.1			& $7.7$		& $2.8\times10^{2}$	 & 0.4\%	 & 4 & 6.2\% \\
3.2			& $0.7$	& $35.8$	& 6.7\% 	& 4 & 31.7\% \\ \hline
4.0			& $3.8\times10^{-3}$ & $2.1$ & 14.9\%	& 5 & 43.1\% \\ \hline
5.0			& $6.7\times10^{-3}$ & $1.8$ & 27.0\%	& 5 & 42.5\% \\
5.1			& $20.0$	 & $4.6\times10^{2}$ & 0\%	& 6 & 57.7\% \\ \hline
6.0			& $1.2\times10^{-4}$ & $0.32$	& 12.8\%		& 4 & 32.1\% \\
6.1			& $2.0\times10^{2}$  & $2.6\times10^{3}$	& 0\%		& 4 & 19.1\% \\
6.2			& $26.9$	& $5.9\times10^{2}$	 & 4.1\% 	& 5 & 36.3\% \\ \hline
7.0			& $7.7\times10^{-7}$ & $3.1\times10^{-2}$	& 41.3\% 	& 4 & 30.6\% \\
7.1			& $1.4$	 & $86.1$ & 0.1\%	& 4 & 7.9\% \\ \hline
8.0			& $3.5\times 10^{-6}$	& $7.8\times 10^{-2}$	& 0.1\%	 & 3 & 9.7\% \\
9.0			& $7.5\times 10^{-15}$	& $4.9\times 10^{-6}$	& 26.6\% 	& 4 & 58.0\% \\
10.0			& $2.4\times 10^{-12}$	& $ 1.9\times 10^{-4}$	& 76.7\%	 & 4 & 17.2\% \\ \hline
\end{tabular}
\caption{This table contains the SUSY production cross sections at $\sqrt{s}=8\TeV$ and $\sqrt{s}=14\TeV$, as well as the percentage of events containing at least one high-$p_T$ lepton, the peak in the jet multiplicity distribution, and the percentage of events containing at least one b-tagged jet. Note that this is simply the number of jets whose $p_T>20~\rm{GeV}$, and does not include other quality requirements placed on jets. Note that the higest lepton multiplicities occur for the KKLT base points, whose cross-sections are well below the femptobarn scale.\label{table:sigprops}}
\end{center}
\end{table}

We summarize the gross LHC phenomenology of some of our benchmark points in Table~\ref{table:sigprops}, which includes all ten KKLT base points from Table~\ref{table:KKLTBenchmarks}, as well as the perturbations listed in Table~\ref{table:Partners}. Displayed are the overall SUSY production cross sections at both $\sqrt{s}=8\TeV$ and $\sqrt{s}=14\TeV$, the percentage of events containing at least one high-$p_T$ lepton, the peak in the jet multiplicity distribution, and the percentage of events containing at least one b-tagged jet. Given the broad features of the benchmarks, we have chosen to pursue the general-purpose ATLAS SUSY~searches which involve low jet and lepton multiplicity. More specifically, we will consider the general (low-multiplicity) jets plus missing transverse energy ($\MET$) search~\cite{Aad:2014wea}, the so-called `monojet' signatures of the stop search~\cite{Aad:2014nra} for small $\Delta m(\tilde{t}_1,\tilde{\chi}_1^0)$, and the dedicated stop searches of~\cite{Aad:2014bva} and~\cite{Aad:2014lra}, which require b-tagged jets, for large $\Delta m(\tilde{t}_1,\tilde{\chi}_1^0)$. Though we expect leptonic signatures to be sub-dominant, we nevertheless also simulate the signatures of the one-lepton search of~\cite{Aad:2014lra}, which requires at least three b-tagged jets in the final state, as well as the single hard lepton plus multijets searches of~\cite{Aad:2015mia}. 

Each of the search strategies utilized by ATLAS divides the search into two parts: object reconstruction and event selection. The object reconstruction sets requirements for each object in an event, typically the jets, leptons, photons, and missing energy. For the ATLAS searches conducted at $8$~TeV, jets are reconstructed using the anti-$k_T$ algorithm with a radius parameter of~$0.4$. Jets are required to be isolated from leptons by calculating $\Delta R\equiv\sqrt{\Delta\eta^2+\Delta\phi^2}$, and demanding $\Delta R>0.2$. If $\Delta R<0.2$ between any jet candidate and any electron, the jet is discarded. For any surviving jet candidates, if $\Delta R<0.4$ between a jet candidate and any leptons, the lepton is discarded. For remaining jet and lepton candidates, further requirements are placed on $|\eta|$ and $p_T$ that vary from one search to another. Further isolation requirements are placed on each of the jet and lepton candidates. Finally, the missing transverse energy, $\slashed{E}_T$, is calculated to be the negative of the vector sum of the $p_T$ of all reconstructed objects with $|\eta|<4.9$, not belonging to other reconstructed objects. 
%
%All of this is built into Delphes by using the ATLAS detector card; once this reconstruction takes place, we can simply apply the event selection criteria for each signal region definition for each of the searches that we consider.

\subsubsection{Low multiplicity jets plus missing transverse energy}

\begin{table}[htdp]
%\vspace{-1mm}
  \footnotesize
  \begin{center}
    \begin{tabular}{|l||c |c|c| c|c|c|c|c|c|c|c|c|c|}
      \hline
      \multirow{2}{*}{Requirement}      &\multicolumn{13}{|c|}{Signal Region} \\
  \cline{2-14}
  										& \parbox{0.35in}{\bf 2jl} 	& \parbox{0.35in}{\bf 2jm}	& \parbox{0.35in}{\bf 2jt} 	& \parbox{0.35in}{\bf 3j}	& \parbox{0.35in}{\bf 4jl-} 	& \parbox{0.35in}{\bf 4jl} 	& \parbox{0.35in}{\bf 4jm} 	& \parbox{0.35in}{\bf 4jt} 	& \parbox{0.35in}{\bf 5j} 	& \parbox{0.35in}{\bf 6jl} 	& \parbox{0.35in}{\bf 6jm} 	& \parbox{0.35in}{\bf 6jt} 	& \parbox{0.35in}{\bf 6jt+} \\ \hline \hline
$\slashed{E}_T/\sqrt{H_{\rm T}}$ [GeV$^{1/2}$]  	& 8 &\multicolumn{2}{|c|}{15} &\multicolumn{1}{|c|}{--} &\multicolumn{2}{|c|}{10} &\multicolumn{7}{|c|}{--}  \\ \hline
$\slashed{E}_T/m_{\rm eff}(N)$ 					&\multicolumn{3}{|c|}{--} 		& 0.3 	&\multicolumn{2}{|c|}{--} & 0.4 	&0.25 	&\multicolumn{3}{|c|}{0.2} 		& 0.25 	& 0.15 \\\hline
$m_{\rm eff}({\rm incl.})$ [TeV]	& 0.8 	&1.2 		&1.6		&2.2		&0.7		&1.0	 	&1.3	 	& 2.2	 	&1.2	 	&0.9	 	&1.2 		&1.5	 	&1.7\\ \hline\hline
Observed Events							& 12315	& 715	& 33		& 7		& 2169	& 608	& 24		& 0		& 121	& 121	& 39		& 5		& 6\\\hline
$S_{95}^{\rm obs}$								&  1200	& 90		& 38		& 8.2		& 270	& 91		& 10		& 3.1		& 35		& 39		& 25		& 6.6		& 7.9\\\hline
%\hline
\end{tabular}
\caption{Signal region definitions, observed number of events, and $S_{95}^{\rm obs}$ for the low-multiplicity jets plus $\MET$ search of~\cite{Aad:2014wea}. The numbers in the first three rows represent the minimum value for the kinematic quantity in the first column. Further description of the signal characteristics is given in the text. All data corresponds to $20.3\,\rm{ fb}^{-1}$ of integrated luminosity. \label{table:multijet}}
  \end{center}
\end{table}

The ATLAS low-multiplicity jets plus missing transverse energy search contains a number of signal regions, each requiring between~2 and 6~jets. This is typical in events with production of $\tilde{q}\tilde{q}$, $\tilde{q}\tilde{g}$ and $\tilde{g}\tilde{g}$, where the decay $\tilde{g}\rightarrow q\bar{q}\tilde{\chi}^0_1$ produces two jets, and the decay $\tilde{q}\rightarrow q\tilde{\chi}^0_1$ produces a single jet. It thus represents a very general search that fits the gross phenomenology of a wide range of SUSY models. 

The most recent iteration of this search was published in May of 2014~\cite{Aad:2014wea}, and extends the reach of possible SUSY production beyond previous searches. This search defines 15 signal regions, 13 of which were studied in this work and are defined in Table~\ref{table:multijet}.\footnote{Two of the signal regions involved attempting to identify hadronically-decaying $W$-bosons via a particular jet-pairing algorithm, and will not be included in our study.}
For each signal region, $\slashed{E}_T>150$~GeV is required, as is at least one jet with $p_T>130$~GeV. A lepton veto is placed on all events containing a single electron or muon with $p_T>10$~GeV. Between two and six jets are required for these signal regions, with each additional jet requiring $p_T>60$~GeV. The first three jets are required to be separated from the reconstructed $\slashed{E}_T$ direction with a minimum $\Delta\phi(Jet,\slashed{E}_T)>0.4$, while any additional jets must be separated by $\Delta\phi(Jet,\slashed{E}_T)>0.2$. Signal regions are then defined using the ratio $\slashed{E}_T/\sqrt{H_T}$, $\slashed{E}_T/m_{\rm eff}(N)$, and $\slashed{E}_T/m_{\rm eff}({\rm incl.})$ where $H_T$ is the sum of the $p_T$ of all jets with $p_T>40$~GeV, $m_{\rm eff}(N)$ the scalar sum of $\slashed{E}_T$ and the $N$~hardest jets, and $m_{\rm eff}({\rm incl.})$ is the scalar sum of $\slashed{E}_T$ and all jets with $p_T>40$~GeV. The signal regions are named by the number of jets, and a criterion `loose', `medium', or `tight', depending on the values of each of these discriminants. 

\subsubsection{Monojet signatures for light stops}

\begin{table}[htdp]
\begin{center}
\begin{tabular}{|l||c|c|c|}\hline
\multirow{2}{*}{\parbox{1.5in}{Requirement}}      &\multicolumn{3}{|c|}{Signal Region} \\
  \cline{2-4}
	& \parbox{1in}{M1}	& \parbox{1in}{M2}	& \parbox{1in}{M3} \\ \hline \hline
$p_T(Jet,1)$ [GeV] & 280 & 340 & 450 \\ \hline
$\slashed{E}_T$ [GeV] & 220 & 340 & 450 \\ \hline \hline
Observed Events	& 33054	& 8606	& 1776 \\ \hline
$S_{95}^{\rm obs}$		& 1951	& 575	& 195 \\ \hline
\end{tabular}
\caption{Signal region definitions, observed number of events, and $S_{95}^{\rm obs}$ for the the three monojet-like searches of~\cite{Aad:2014nra}. The numbers in the first three rows represent the minimum value for the kinematic quantity in the first column. Further description of the signal characteristics is given in the text. All data corresponds to $20.3\,\rm{ fb}^{-1}$ of integrated luminosity. \label{table:monojets}}
\end{center}
\end{table}

For some SUSY spectra, particularly those with small mass gaps between $SU(3)$-charged superpartners and the LSP, production in association with a single hard jet is of particular interest. The most recent relevant search for such ``monojet'' topologies was published by ATLAS in July of 2014~\cite{Aad:2014nra} with an integrated luminosity of $20.3\rm{~fb}^{-1}$. This search was designed to target direct stop production via the two-body decay $\tilde{t}\rightarrow c\tilde{\chi}^0_1$, as well as the 4-body decay $\tilde{t}\rightarrow bf\bar{f}\tilde{\chi}^0_1$ for compressed spectra.\footnote{Additionally, there are two signal regions that require charm-tagged jets; we will not be considering these, as Delphes does not incorporate charm-tagging.} This is of particular interest for points which feature a heavy LSP, a stop NSLP, and the remaining superpartners sufficiently heavy as to be effectively integrated out. Each event is required to have a reconstructed primary vertex with at least~5 associated tracks. Further, each event is required to have $\slashed{E}_T>150$~GeV, and at least one jet with $p_T>150$~GeV and $|\eta|<2.8$. To eliminate multiple jets, a maximum of 2 additional jets are permitted with $p_T>30$~GeV; events with additional hard jets are rejected. Each of these jets must have a minimum $\Delta\Phi(Jet,\slashed{E}_T)>0.4$. Events with reconstructed electrons or muons are also rejected.
Three signal regions are then defined by additional requirements on the $p_T$ of the hardest jet and the $\slashed{E}_T$. Signal~M1 is defined to have $p_T(Jet,1)>280$~GeV and $\slashed{E}_T>220$~GeV; M2~is defined to have $p_T(Jet,1)>340$~GeV and $\slashed{E}_T>340$~GeV; and~M3 is defined to have $p_T(Jet,1)>450$~GeV and $\slashed{E}_T>450$~GeV. The number of observed events for each signal region is listed in~Table\ref{table:monojets}.

\subsubsection{B-tagged jets and missing transverse energy with a lepton veto}

\begin{table}[h]
\begin{center}
\begin{tabular}{|c||c|c|c|c||c|c|c|}\hline
\multirow{2}{*}{\parbox{1.5in}{Requirement}}      &\multicolumn{7}{|c|}{Signal Region} \\
  \cline{2-8}
							& \parbox{0.7in}{SRA1}	& \parbox{0.7in}{SRA2} 	& \parbox{0.7in}{SRA3}	& \parbox{0.7in}{SRA4}  & \parbox{0.7in}{SRC1}  & \parbox{0.7in}{SRC2}  & \parbox{0.7in}{SRC3} \\ \hline
$m_{bjj}^0$ [GeV] & \multicolumn{2}{c|}{$<225$}	& \multicolumn{2}{c||}{$[50,250]$} & -- & -- & -- \\ \hline
$m_{bjj}^1$ [GeV] & \multicolumn{2}{c|}{$<225$}	& \multicolumn{2}{c||}{$[50,400]$} & -- & -- & --\\ \hline
${\rm min}[m_T(Jet_i,\slashed{E}_T)] $	[GeV] & \multicolumn{2}{c|}{--}		& \multicolumn{2}{c||}{$>50$}  & -- & -- & --\\ \hline
$m_T^{b,{\rm min}}$ [GeV] & \multicolumn{4}{c||}{$>175$}	 & $>185$ & \multicolumn{2}{c|}{$>200$}  \\ \hline
$m_T^{b,{\rm max}}$ [GeV] & \multicolumn{4}{c||}{--}	 & $>205$ & $>290$ & $>325$ \\ \hline
$\slashed{E}_T$	 [GeV]			& $>150$	& $>250$			& $>300$	& $>350$ & \multicolumn{2}{c|}{$>160$}  & $>210$ \\ \hline \hline
Observed Events	& 11	& 4	& 5 & 4 & 59 & 30 & 15 \\ \hline
$S_{95}^{\rm obs}$		& 6.6 & 5.7 & 6.7 & 6.5 & 15.7 & 12.4 & 8.0 \\ \hline
\end{tabular}
\caption{Signal region definitions, observed number of events, and $S_{95}^{\rm obs}$ for the two classes of b-tagged jets plus $\slashed{E}_T$ searches of~\cite{Aad:2014bva}. The numbers in the first six rows represent the minimum, maximum, or allowed range of values for the kinematic quantity in the first column. Further description of the signal characteristics is given in the text. All data corresponds to $20.1\,\rm{ fb}^{-1}$ of integrated luminosity. \label{table:bjets}}
\end{center}
\end{table}

In this category we will discuss two separate ATLAS publications, each representing a stop search via b-tagged jets. The first analysis focuses on stop production where the stop decays via $\tilde{t}_1\rightarrow t\tilde{\chi}_1^0$ or $\tilde{t}_1\rightarrow b\tilde{\chi}_1^\pm \rightarrow bW^{(*)}\tilde{\chi}_1^0$, where the $W$ is assumed to decay hadronically~\cite{Aad:2014bva}. In either case, the result will be an LSP, a b-tagged jet and 2 non-b-tagged jets per $\tilde{t}$ produced. For all signal regions, a minimum of $\slashed{E}_T>150~\rm{GeV}$ is required, a minimum of six jets, two of which must be b-tagged, and no reconstructed leptons (electrons or muons). The two highest $p_T$ jets must have an energy of at least $80~\rm{GeV}$, with remaining jets satisfying $p_T>35~\rm{GeV}$, and the three highest $p_T$ jets must be separated from the missing energy by at least $\Delta\phi>\pi/5$. A further requirement is placed on the b-tagged jet closest in angle to the missing energy. The transverse mass, defined as
\begin{equation}
m_T^{b,{\rm min}}=\sqrt{2p_T^b\slashed{E}_T\left[ 1-\cos\Delta\phi(p_T^b,\slashed{E}_T)\right]} \label{mTbmin}
\end{equation}
must have a minimum of $m_T^{b,{\rm min}}>175~\rm{GeV}$.

The search is then divided into three subsections. We will consider only the first (SRA) and third (SRC). For the first (SRA), the two jets with the highest b-tag weight are selected, then of the remaining jets, the two closest in the $\eta-\phi$ plane are combined to form a W candidate, which is then combined with the first b-tagged jet to form a top candidate with mass $m_{bjj}^0$. A second W candidate is formed by repeating the procedure with the remaining jets. Lastly, the value ${\rm min}[m_T(Jet_i,\slashed{E}_T)] $ is calculated as the minimum $m_T$ of each of the signal jets and the missing energy.  With all of these quantities, the signal regions are defined as in the first four columns of Table~\ref{table:bjets}.

The second subsection that we consider (SRC) focuses on the specific case when one of the stops decays via $\tilde{t}_1\rightarrow b\tilde{\chi}_1^\pm\, ; \quad \tilde{\chi}_1^\pm\rightarrow W^{(*)}\tilde{\chi}_1^0$. Only five jets are now required, and a minimum of $\Delta\phi>0.2\pi$ between the two hardest b-tagged jets is required. The $m_T$ is further constrained: for SRC1, $m_T^{b,{\rm min}}>185\GeV$ while for SRC2 and SRC3, $m_T^{b,{\rm min}}>200\GeV$. A further quantity, $m_T^{b,{\rm max}}$, is computed similarly to $m_T^{b,min}$ in~(\ref{mTbmin}), but now with the b-tagged jet being that with the largest $\Delta\phi$ from $\slashed{E}_T$. For SRC1, $m_T^{b,{\rm max}}>205\GeV$; for SRC2, $m_T^{b,{\rm max}}>290\GeV$; and for SRC3, $m_T^{b,{\rm max}}>325\GeV$. These values are collected in the final three columns of Table~\ref{table:bjets}.

\begin{table}[h]
\begin{center}
\begin{tabular}{|c||c|c|c||c|c|c|}\hline
\multirow{2}{*}{\parbox{1.5in}{Requirement}}      &\multicolumn{6}{|c|}{Signal Region} \\
  \cline{2-7}
							& \parbox{0.7in}{SR-4jA}	& \parbox{0.7in}{SR-4jB} 	& \parbox{0.7in}{SR-4jC}	&  \parbox{0.7in}{SR-7jA}  & \parbox{0.7in}{SR-7jB}  & \parbox{0.7in}{SR-7jC} \\ \hline
$\Delta \phi_{\rm min}^{4j}$ & \multicolumn{6}{c|}{$0.5$} \\ \hline
$\slashed{E}_T/m_{\rm eff}^{4j}$ & \multicolumn{6}{c|}{$0.2$} \\ \hline
$m^{4j}_{\rm eff}$ [GeV] & $1300$ & $1100$ & $1000$ & -- & -- & -- \\ \hline
$\slashed{E}_T/\sqrt{H_T^{4j}}$ [GeV$^{1/2}$] &  -- & -- & $16$ & -- & -- & -- \\ \hline
$m^{incl}_{\rm eff}$ [GeV] & -- & -- & -- & $1000$ & $1000$ & $1500$ \\ \hline
$\slashed{E}_T$	 [GeV]			& $250$	& $350$			& $400$	& $200$ & $350$  & $250$ \\ \hline \hline
Observed Events	& 2	& 3	& 1 & 21 & 3 & 1  \\ \hline
$S_{95}^{\rm obs}$		& 5.2 & 6.5 & 3.9 & 13.9 & 6.1 & 4.2 \\ \hline
\end{tabular}
\caption{Signal region definitions, observed number of events, and $S_{95}^{\rm obs}$ for the two classes of b-tagged jets plus $\slashed{E}_T$ searches of~\cite{Aad:2014lra}. The numbers in the first six rows represent the minimum value for the kinematic quantity in the first column. Further description of the signal characteristics is given in the text. All data corresponds to $20.1\,\rm{ fb}^{-1}$ of integrated luminosity. \label{table:bjets2}}
\end{center}
\end{table}

The second analysis we include is a more general search that targets third generation squark production and/or gluino production. Like the previous analysis, it relies heavily on the presence of b-tagged jets and requires large $\slashed{E}_T$. For this study we will focus on the channels which impose a veto on electrons with $p_T>20~\rm{GeV}$ and muons with $p_T>10~\rm{GeV}$. Events are separated into two categories. The first category (4jA,B,C) requires at least four jets, all four of which having a minimum jet $p_T > 50\,{\rm GeV}$. The second category (7jA,B,C) requires at least seven jets, all seven of which having a minimum jet $p_T > 30\,{\rm GeV}$. All channels require the leading jet to satisfy $p_T(Jet,1) \geq 90\,{\rm GeV}$, and require the presence of at least three b-tagged jets, all with $p_T > 30\,{\rm GeV}$. 

Additional global cuts for all signal regions are given in terms of $H_T^{4j}$, defined as the scalar sum of the $p_T$ of the four hardest jets in the event, $m_{\rm eff}^{4j}$, defined as the scalar sum of $\slashed{E}_T$ and the $p_T$ of the four hardest jets in the event, $\Delta \phi_{\rm min}^{4j}$, defined as the minimum azimuthal separation between any of the four leading jets and the direction of the missing transverse energy, and $m_{\rm eff}^{\rm incl}$ defined as the scalar sum of $\slashed{E}_T$ and the sum of the $p_T$ of all jets in the event with $p_T> 30\,{\rm GeV}$. These cuts are given in Table~\ref{table:bjets2}, along with the observed number of events, and $S_{95}^{\rm obs}$ for the six signal regions.

%------------------------------------
\subsection{Results and Case Studies}

We begin this section by considering the reach of the already completed searches at $\sqrt{s}=8\TeV$, described in the previous subsection, on the parameter space of the DMM framework. We then investigate a number of case studies motivated by the points in Table~\ref{table:Partners}, which are representative of the LHC~phenomenology of the model space as a whole.

As stated previously, none of the parameter combinations in Tables~\ref{table:Partners} and~\ref{table:sigprops}
would have given a significant excess over background at the previous LHC run. This is despite some relatively large cross-sections at $\sqrt{s} = 8\TeV$ (see Table~\ref{table:sigprops}).
In general, we can determine the likelihood that a given parameter point would have produced a detectable signal above background by comparing the simulated number of signal events at $\sqrt{s}=8\TeV$ versus the reported $S_{95}^{\rm obs}$ value, which gives the 95\% confidence-level upper bound on the number of signal events compatible with the ATLAS observations. Thus, for example, our simulation of point~6.1 (the benchmark point with the largest production cross-section) suggests an overall production of 4060~events in all channels after 20.3~fb$^{-1}$ of data-taking, but the compression between the gluino mass and LSP mass translates into only 282~events in the 2jl~channel, 34~events in the 2jt~channel, and 39~events in the 4jl~channel, all of which are below the reported $S_{95}^{\rm obs}$ values of 1200, 38~and~91 events, respectively (see Table~\ref{table:multijet}). We estimate the greatest excess would have been in the 2jt~channel, but the signal significance would be only $0.7\sigma$ in this channel.

We find that while none of the KKLT base points would have been discovered thus far, a fair fraction of the DMM parameter space involving stop or gluinos with masses at or below 1~TeV would now be ruled out. In general, we find the best channels at $\sqrt{s}=8\TeV$ are in the low-multiplicity jets plus $\slashed{E}_T$ search, with a reach of $m_{\tilde{g}}\lappeq 600\GeV$ for small mass gaps between the gluino and the LSP $\Delta m(\tilde{g},\tilde{\chi}_1^0)\lappeq 50\,{\rm GeV}$, and $m_{\tilde{g}}\lappeq 900\GeV$ for mass gaps of greater than 50-100~GeV. We find the reach in the lightest stop mass to be about 100~GeV less for the two different mass gaps between the stop and the LSP. However, DMM~corrections tend to move the masses of $SU(3)$-charged objects in a correlated way, tending to compress both the gluino and the stop toward the lightest neutralino mass. In extreme cases, a nearly degenerate trio of masses $(m_{\tilde{\chi}^0_1},\,m_{\tilde{t}_1},\, m_{\tilde{g}})$ would have escaped detection, even at a mass scale of 500-600~GeV. Such outcomes are rare in the DMM landscape, but not impossible (see point~3.1 of Table~\ref{table:Partners}, to be discussed in more detail below). In terms of theoretical parameters, the modular weight combinations most likely to produce spectra detectable at $\sqrt{s}=8\TeV$ are the $(n_M,n_H)= (0,\,0)$, $(0,\,0.5)$ and $(0.5,\,0)$ cases, with low messenger scale and $\alpha_g >0$.

To discuss detection in the near future, we will use the signal definitions at $\sqrt{s}=8\TeV$ described above as a first approximation to what will be done at 14~TeV. Signal significance is estimated by calculating the background counts using the pre-generated Snowmass~2013 published backgrounds at $\sqrt{s} = 14\,{\rm TeV}$~\cite{Avetisyan:2013onh}, which were generated in a manner identical to that in which our signal files were produced. In general, we find that the loosest possible cuts tends to preferentially populate the signal relative to background, as most of the parameter space within reach at the~LHC will feature a compressed spectrum allowing less phase space for hard jets and leptons. We thus do not attempt to optimize beyond the $\sqrt{s}=8\TeV$ criteria by tightening the requirements on various distributions, though we will consider a modified monojet, or `lopsided jet', signature in what follows.

% ------------- TABLE: DISCOVERY CHANNELS and LMIN---------------
\begin{table}
\begin{center}
\begin{tabular}{|c||c|c||c|c||c|c|c|}
\multicolumn{3}{c}{  } & \multicolumn{2}{c}{Monojet $L_{\rm min}$} & \multicolumn{3}{c}{B-tagged jet $L_{\rm min}$} \\ \hline
\parbox{0.8in}{Benchmark}	& \parbox{0.8in}{Overall $L_{\rm min}$}		& \parbox{0.8in}{Best Channel}	 & \parbox{0.8in}{ 2jl (opt)}  & \parbox{0.8in}{monojet M2}	& \parbox{0.8in}{B-tag SRA2} & \parbox{0.8in}{B-tag 4jB} & \parbox{0.8in}{B-tag 7jC} \\ \hline\hline
%1.0			  & --	& -- & -- & -- & -- & -- & -- & --\\
1.1	  		& $80$ & \textbf{2jt} & $159$ & $303$ & -- & -- & --  \\ \hline
%2.0			 & -- & -- & -- & -- & -- \\
2.1			& $94$ & \textbf{6jt+} & 	-- & -- & --  & $249$ & $133$ \\
2.2			& $1.9$	 & \textbf{6jt+} & $64$ & -- & $123$ & $4$ & $2$ \\
2.3			& $141$ & \textbf{2jl} & $185$ & $2909$ & -- & -- & -- \\
2.4			& $22$	 & \textbf{5j} & $46$ & -- & $216$ & $97$ & $100$\\ \hline
%3.0			& --	 & -- & -- & -- & -- \\
3.1			& $22$ & \textbf{2jm} & $29$ & $96$  & -- & -- & -- \\
3.2			& $240$ & \textbf{2jt}	& $574$ & -- & -- & -- & -- \\ \hline
%4.0			& -- & -- & -- & -- & -- \\ \hline
%5.0			& -- & -- & -- & -- & -- \\
5.1			& $2.9$ & \textbf{5j} & $11$ & $2466$  & $55$  &  $5$ &  $7$ \\ \hline
%6.0			& -- & -- & -- & -- & -- \\
6.1			& $1.7$  & \textbf{3j} & $2.3$ & $32$ & $90$ & -- & $929$ \\
6.2			& $4.6$ & \textbf{4jl} & $12$ & -- & $69$ & $70$ & $53$ \\ \hline
%7.0			& -- & -- & -- & -- & -- \\
7.1			& $65$ & \textbf{4jl} & $168$ & -- & -- &  -- &  -- \\ \hline
%8.0			& -- & -- & -- & -- & -- \\
%9.0			& -- & -- & -- & -- & -- \\
%10.0			& -- & -- & -- & -- & -- \\ \hline
%
\end{tabular}
\caption{Minimum integrated luminosity $L_{\rm min}$ (in fb$^{-1}$) to achieve a $5\sigma$~signal significance in a given channel, at $\sqrt{s}=14\TeV$. In all cases, the strongest signal will be in the low-multiplicity jets plus $\MET$ search. The overall $L_{\rm min}$ therefore reflects the strongest of the 13~channels in Table~\ref{table:multijet}, given in the third column. The value of $L_{\rm min}$ to achieve the same signal significance in certain sub-dominant channels is also given, for reference. Columns 4~and~5 represent monojet-like channels, while the final three columns represent various signatures that involve b-tagged jets. Signal region descriptions are given in the text. Note that none of the KKLT base points (X.0) will yield a $5\sigma$ excess after 3000~fb$^{-1}$ of integrated luminosity.\label{table:Lmin}}
\end{center}
\end{table}

A summary of the results of our simulations at $\sqrt{s}=14\TeV$ is given in Table~\ref{table:Lmin} for the DMM perturbations shown in Table~\ref{table:Partners}. We show results for the quantity $L_{\rm min}$, defined as the minimum amount of integrated luminosity needed to achieve a $5\sigma$ ($S/\sqrt{B} = 5$) signal significance in that particular channel. The most effective discovery channels are uniformly found to be from the low-multiplicity jets plus $\slashed{E}_T$ search. The first two columns give the best signal region from Table~\ref{table:multijet} with the corresponding $L_{\rm min}$ value. The next two columns give the corresponding $L_{\rm min}$ from the best signal region of the monojet search in Table~\ref{table:monojets}, as well as a monojet-oriented perturbation on the 2jl channel introduced in~\cite{Kaufman:2015nda} and to be discussed below. Finally, the last three columns indicate the $L_{\rm min}$ value for the best signal regions involving multiple b-tagged jets. If a $5\sigma$ excess is not expected within 3000~fb$^{-1}$ the entry is left empty. Note that none of the KKLT base points (X.0) will yield a $5\sigma$ excess after 3000~fb$^{-1}$ of integrated luminosity. We will now discuss the details behind many of the numbers in Table~\ref{table:Lmin} via a sequence of case studies.

\subsubsection{Case Study 1: Signatures involving b-tagged jets; Points 2.2, 6.1 and 6.2}

Let us begin our study of comparative LHC signatures of DMM parameter points by considering the case of point~2.2 versus~6.1. These two points share common gauge mediation parameters: $N=3$, $\Lambda_{\rm mess} = 10^5\,{\rm GeV}$ and $\alpha_g$ positive, but relatively small. Point~2.2 has a relatively large large mirage parameter, $\alpha_m=1.8$, while that of point~6.1 has $\alpha_m =0.7$. Both involve universal scalar masses (to leading order) at the unification scale, with modular weights $(n_m, n_H)$ equal to $(0,0)$ and $(1/2,1/2)$, respectively.

Point~2.2, with vanishing modular weights, produces a heavier stop mass than that of point 6.1 as indicated in Table~\ref{table:Partners}: $m_{\tilde{t}_1} = 2582\,{\rm GeV}$ versus $1227\,{\rm GeV}$. More relevant to LHC~phenomenology is the differences in the gaugino sector. Point~2.2 has a gluino mass of  $m_{\tilde{g}} = 1061\,{\rm GeV}$ and LSP mass of  $m_{\tilde{\chi}^0_1} = 147\,{\rm GeV}$, while point~6.1 has a gluino mass of 713~GeV and LSP mass of 654~GeV. As a consequence, at $\sqrt{s}=14\,{\rm TeV}$, the cross-section for overall superpartner production for point~6.1 is 2.6~pb, versus 169~fb for point~2.2. The difference is entirely accounted for by the gluino mass, as gluino pair production represents over 94\% of the total production cross-section in both cases. Despite the huge disparity in total production cross-sections, the two points have nearly identical values of the overall $L_{\rm min}$ value needed for discovery (just under 2~fb$^{-1}$).

%%%%%%%%%%%%%%%%% Effective Mass and Njet %%%%%%%%%%%%%%%%%%
%
\begin{figure}[t]
\begin{center}
\includegraphics[width=0.48\textwidth]{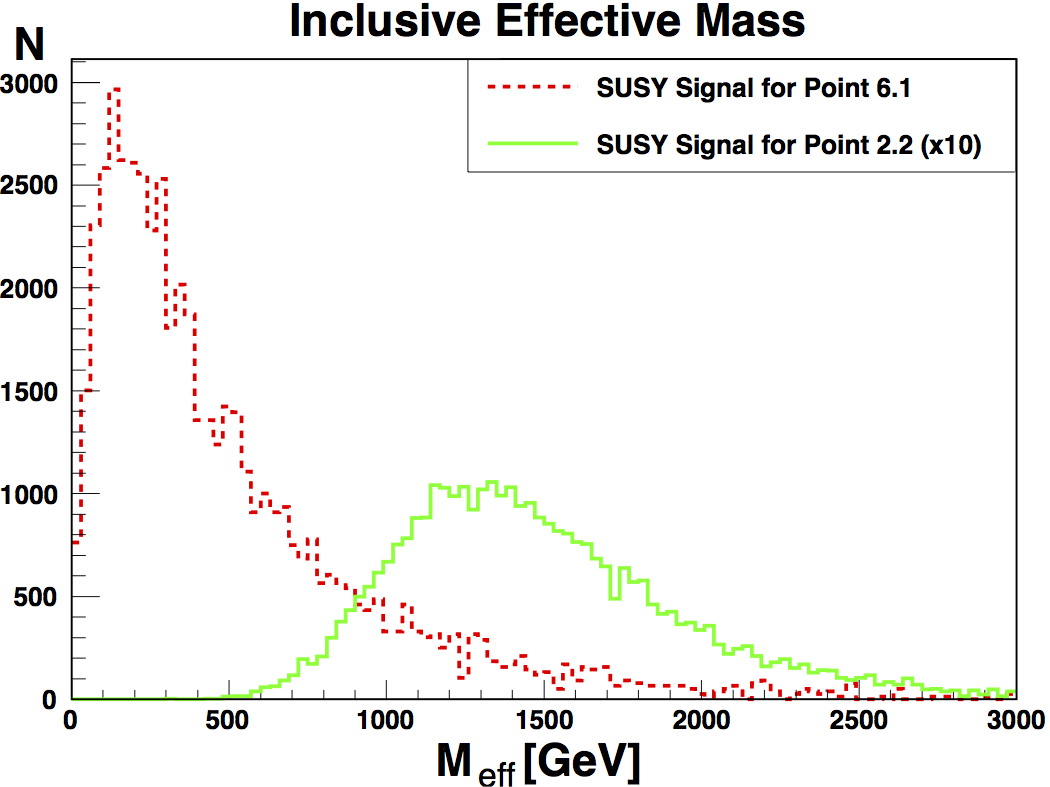}
\includegraphics[width=0.48\textwidth]{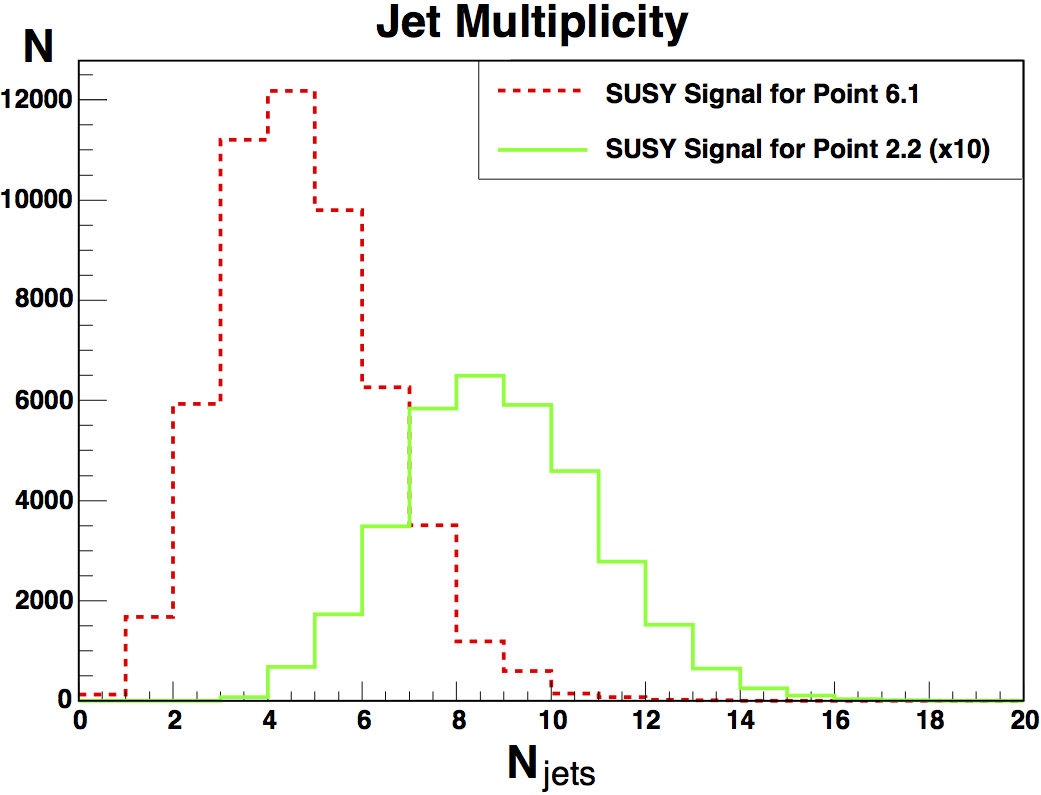}
\caption{Effective mass and total jet multiplicity for points~2.2 and~6.1. Left panel gives the effective mass distribution $m_{\rm eff}$ for points~2.2 and~6.1, while the right panel gives the total jet multiplicity $N_{\rm jets}$ for the same two points. Both plots are normalized to 20~fb$^{-1}$ at $\sqrt{s}=14\,{\rm TeV}$, but with the distributions for point~2.2 multiplied by a factor of~ten to allow for greater readability.}
\label{plot:Meff2and6}
\end{center}
\end{figure}
%%%%%%%%%%%%%%%%%%%%%%%%%%%%%%%%%%%%%%%%%%%%%%%%%%%%%%

It is not difficult to understand why this is. Despite the light gluino, point~6.1 has a mass difference between the gluino and the LSP of roughly 60~GeV, while point~2.2 has a spectrum not dramatically different from the so-called ``simplified'' models that are often used as benchmarks in the interpretation of LHC search results. 
Indeed, in the left panel of Figure~\ref{plot:Meff2and6} we see the effective mass distribution at $\sqrt{s}=14\,{\rm TeV}$. Despite the much larger signal size for point~6.1 (note that distributions for point~2.2 are multiplied by a factor of ten for readability), the small mass gap between the gluino and the LSP translates into an effective mass distribution heavily weighted towards those bins below the cut value of 700-800~GeV which defines the 2jl and 4jl signal regions. Table~\ref{table:Lmin} indicates that within the first 2~fb$^{-1}$ at $\sqrt{s}=14\,{\rm TeV}$ both models would be ``discovered'' at the LHC, in the six-jet channel for point~2.2, and in the three-jet channel for point~6.1. The large mass gap between the gluino and LSP in the case of point~2.2 allows more phase space to generate jets with $p_T>40\GeV$, as indicated by the jet multiplicity distributions shown in the right panel of Figure~\ref{plot:Meff2and6}.

%%%%%%%%%%%%%%%%% pT (jet4) and bjet multiplicity %%%%%%%%%%%%%%%%%%
%
\begin{figure}[th]
\begin{center}
\includegraphics[width=0.48\textwidth]{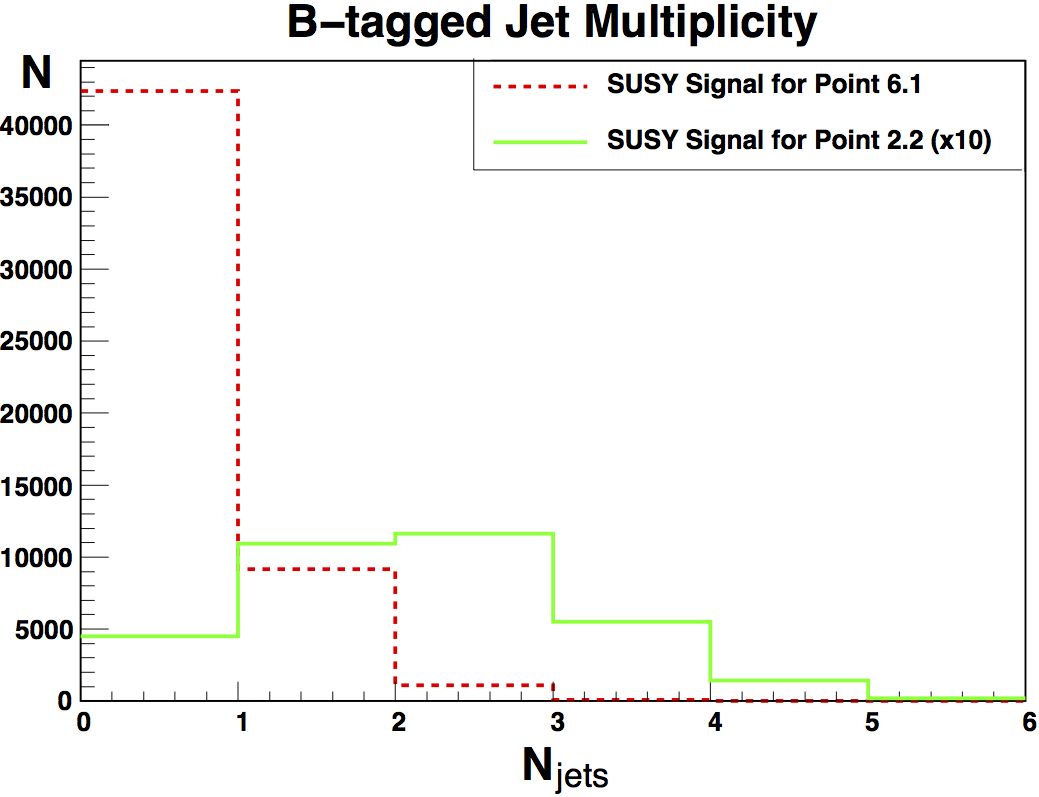}
\includegraphics[width=0.48\textwidth]{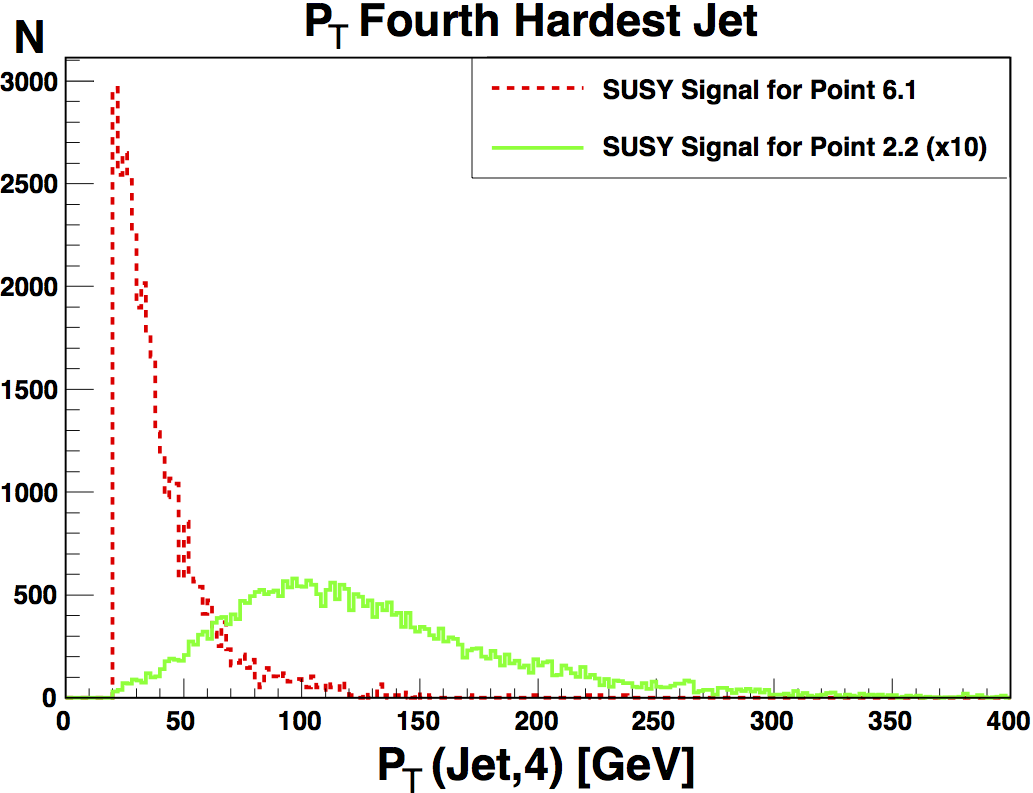}
\caption{Multiplicity of b-tagged jets and fourth jet $p_T$ for points~2.2 and~6.1. Left panel gives the number of events with one or more b-tagged jets. The right panel gives the $p_T(Jet,4)$ of the fourth hardest jet in the event, a key discriminant in the b-tagged searches described in Table~\ref{table:bjets2}. Both distributions are generated prior to the imposition of any other signal region cuts, and are normalized to 20~fb$^{-1}$ of data. The signal distribution for point~2.2 is multiplied by a factor of~ten to allow for greater readability.}
\label{plot:bjet}
\end{center}
\end{figure}
%%%%%%%%%%%%%%%%%%%%%%%%%%%%%%%%%%%%%%%%%%%%%%%%%%%%%%

It is instructive to consider sub-dominant channels, as these signals will provide corroboration of new physics and a powerful descriminant between potential models of this new physics. In the case of points~2.2 and~6.1 all of the signal regions in the general class of jets plus missing transverse energy will show signals in the first 30-40~fb$^{-1}$. Thus we consider those signal regions that are defined as having one or more b-tagged jets. Here we find the corroborating signals will arise nearly immediately for point~2.2, while requiring approximately 100~fb$^{-1}$ for point~6.1, despite the much larger overall production cross-section. The precise values of $L_{\rm min}$ for the particular signatures~SRA2 of Table~\ref{table:bjets}, and~4jB and~7jC of Table~\ref{table:bjets2}, are given in the last three columns of Table~\ref{table:Lmin}.

As mentioned previously, both points are dominated by gluino pair production. For point~2.2, the gluino decays to $\tilde{\chi}_{1,2}^0 t\bar{t}$ 40\% of the time, and $\tilde{\chi}_1^{\pm}bt$ 54\% of the time, thus assuring at least four genuine b-jets in the vast majority of signal events, with up to four leptons possible from the $W^{\pm}$ bosons produced in the top decays. The high probability of b-tagged jets and leptons in the final state is reflected in Table~\ref{table:sigprops}. The smaller mass gap in the case of point~6.1 eliminates the possibility of top pairs from the gluino decay, all but eliminating the prospect of high-$p_T$ leptons in the final state.\footnote{Thus we expect no signal in the most promising single-lepton channel for point~6.1, even after 3000~fb$^{-1}$ of integrated luminosity.}
Furthermore, the gluino decays into $\tilde{\chi}_{1,2}^0 b\bar{b}$ only 14\% of the time, decaying to lighter flavors for the vast majority of events. Thus we expect, on average, fewer b-tagged jets per event for point~6.1, which is only partially mitigated by the much higher cross-section for this point relative to point~2.2. This is displayed in the left panel of Figure~\ref{plot:bjet}, where the multiplicity of b-tagged jets is given for both points. 
Of the b-jet searches in Table~\ref{table:bjets}, the signal region SRA2 tends to be the most promising of the cases that require only two b-tagged jets. Here we find comparable values of $L_{\rm min} \simeq \order(100\,{\rm fb}^{-1})$ in this channel, despite the great disparity in production cross-sections.

This outcome is somewhat disappointing, given that point~6.1 still produces ample events with at least one b-tagged jet. The signatures of Table~\ref{table:bjets} have the advantage of only requiring two b-jets, but they simultaneously require a minimum of six jets overall, a property satisfied by less than a quarter of the event sample for point~6.1. Given the low overall jet multiplicity, we might have hoped that the four-jet channels of Table~\ref{table:bjets} would be particularly effective, but note that the signal region definition calls for $p_T(Jet,4)\geq 50\GeV$, while the small mass difference $\Delta m(\tilde{g},\tilde{\chi}_1^0)$ implies that such sub-leading jets will generally have very low $p_T$, often only slightly above the threshold $p_T$ to be classified as a jet in the initial event reconstruction (see the right panel of Figure~\ref{plot:bjet}). Furthermore, only a negligible fraction of the events for point~6.1 have three b-tagged jets -- a fraction roughly consistent with the mis-tagging rate for assigning a b-tag to a light quark jet in Delphes. By contrast, point~2.2, with its substantial mass differences between the gluino and the light electroweak gauginos, seems ideally suited to these general-purpose b-jet searches, and should produce signals in the 4j~and~7j channels of Table~\ref{table:bjets2} roughly simultaneous with the initial discovery in the 6jt+~channel.

Let us now extend this discussion to include point~6.2 from Table~\ref{table:Partners}. Point~6.2 shares all KKLT base point parameters with point~6.1, and differs in the gauge mediation sector only in the value of $\alpha_g$, which is slightly smaller. The spectrum of point~6.2 therefore is, not surprisingly, rather similar to that of point~6.1, though the gaugino sector is slightly different. Note that the LSP of point~6.2 is almost entirely bino-like, with a thermal relic density consistent with WMAP/PLANCK, while that of point~6.1 is a mixed state that is mostly Higgsino-like with a relic density that is smaller by an order of magnitude. The gluino mass for point~6.2 is $m_{\tilde{g}} = 902\GeV$, and the cross-section for superpartner production is therefore intermediate between points~2.5 and~6.1 (though now gluino pair production constitutes only 84\% of the total production cross-section, with associated production of a gluino with a light-flavor squark representing 15\% of the total). Gluinos decay via various three-body decays involving top and bottom quarks, suggesting relatively high jet multiplicity with many b-tagged jets in the final state.

We might, therefore, expect that this point would have an $L_{\rm min}$ in the low-multiplicity jets plus $\slashed{E}_T$ channel that is similar to the other two points. And indeed, Table~\ref{table:Lmin} shows that the best discovery channel is 4jl for this point, with $L_{\rm min} = 4.6\,{\rm fb}^{-1}$. It is noteworthy that this point has the lowest $L_{\rm min}$ of the trio in the sub-dominant b-jet channel SRA2 and the monojet channel M2. The branching fractions of the gluino and lightest stop are nearly identical for points~6.1 and~6.2, but the increased signal in the three b-jet channels of Table~\ref{table:Lmin} is entirely due to the large mass gaps between the gluino and the light electroweak gauginos. We note that all b-jet channels studied are equally effective in this case. 

\subsubsection{Case Study 2: `Optimized' monojet signatures; Points 3.1, 3.2 and 5.1}

Of the previous trio of points, we might note a curious fact about the data presented in Table~\ref{table:Lmin}: point~6.1 is the only one of the three that yields a $5\sigma$ excess in the most advantageous `monojet' channel, signal region M2 from Table~\ref{table:monojets}. In fact, it yields a signal in this channel well before the corroborating b-jet channels studied in the previous subsection. This is particularly odd, in that gluino pair-production dominates the signal, with gluinos decaying universally via three-body decays involving two quarks and an electroweak gaugino. In short, there is no reason to anticipate this particular model would be a natural candidate for a monojet signal at all. 
Clearly, then, these signatures are not adequately addressing the topologies they were designed to attack (at least within our model framework). This begs an obvious question -- just how `monojet-like' are the events captured by the so-called monojet signature M2? As it happens, this signal region is not really a monojet search at all, but rather a skewed two- and three-jet search in which stringent demands are placed on the leading jet; see the conditions outlined in Table~\ref{table:monojets}. Thus, point~6.1 produces a signal in this channel primarily because it produces no leptons, has a long tail in the $p_T$ distribution of the hardest jet, and (crucially) has a jet multiplicity skewed towards small numbers of jets. In other words, the monojet channel here simply captures the same sorts of events that appear in the two and three-jet channels of Table~\ref{table:multijet}.

To get a clearer picture of this phenomenon, let us now consider points~3.1, 3.2 and~5.1 from Table~\ref{table:Partners}. Point~3.1 is the most compressed model in Table~\ref{table:Partners}. The heaviest squarks in this case have masses of roughly 1760~GeV, while the gluino and lightest stop are nearly degenerate with the LSP at just above 1~TeV. Thus, the entire spectrum is compressed and all cascade decays will involve soft outgoing particles. We might therefore expect this point to be one for which the monojet-like search strategies would be most effective.

%%%%%%%%%%%%%%%%% Trio: Jet multiplicity %%%%%%%%%%%%%%%%%%
%
\begin{figure}[t]
\begin{center}
\includegraphics[width=0.48\textwidth]{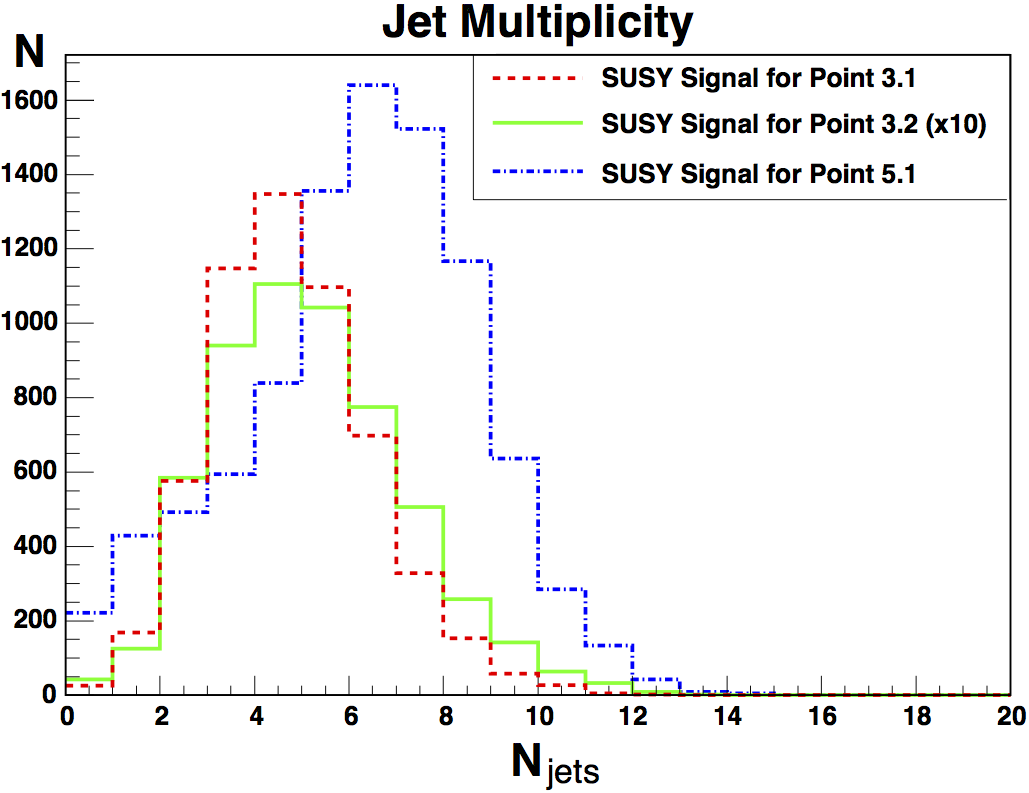}
\caption{Total jet multiplicity for points~3.1, 3.2 and~5.1. Distributions give the total number of reconstructed jets, prior to any event selection cuts, normalized to 20~fb$^{-1}$ of data. Red (dotted) distribution is for point~3.1, green (solid) distribution is for point~3.2, and blue (dot-dashed) distribution is for point~5.1. The signal distribution for point~3.2 is multiplied by a factor of~ten to allow for greater readability.}
\label{plot:trio1}
\end{center}
\end{figure}
%%%%%%%%%%%%%%%%%%%%%%%%%%%%%%%%%%%%%%%%%%%%%%%%%%%%%%

Production is roughly evenly split between gluino pair production and (light-flavored) squark production in association with a gluino. These light-flavored squarks decay back to a light quark and a gluino 95-99\% of the time, while the gluino decays 100\% of the time to a gluon and an LSP. So we can expect a small number of very soft jets, with the potential that one jet will have larger $p_T$ from the decay of a light squark (or from initial state radiation). Indeed, the peak in the jet multiplicity distribution for point~3.1 is at four jets (see Figure~\ref{plot:trio1}), with less than 1\% of all events containing a high-$p_T$, isolated lepton. But the best discovery channel for this point is the two-jet channel with relatively mild kinematic requirements (2jm), with an associated $L_{\rm min}$ of 22~fb$^{-1}$. In contrast, the best monojet-based search is~M2, with a comparatively large $L_{\rm min} = 96\,{\rm fb}^{-1}$.  The failure of the monojet signal to be competitive is mostly due to the veto placed on events with a fourth jet satisfying $p_T(Jet,4)>30\,{\rm GeV}$, which effectively vetos {\em all} events with $N_{\rm jets} > 3$. This is a sizeable fraction of the total event sample, as Figure~\ref{plot:trio1} indicates. 

%%%%%%%%%%%%%%%%% Trio: Leading Jet pT and  %%%%%%%%%%%%%%%%%%
%
\begin{figure}[t]
\begin{center}
\includegraphics[width=0.48\textwidth]{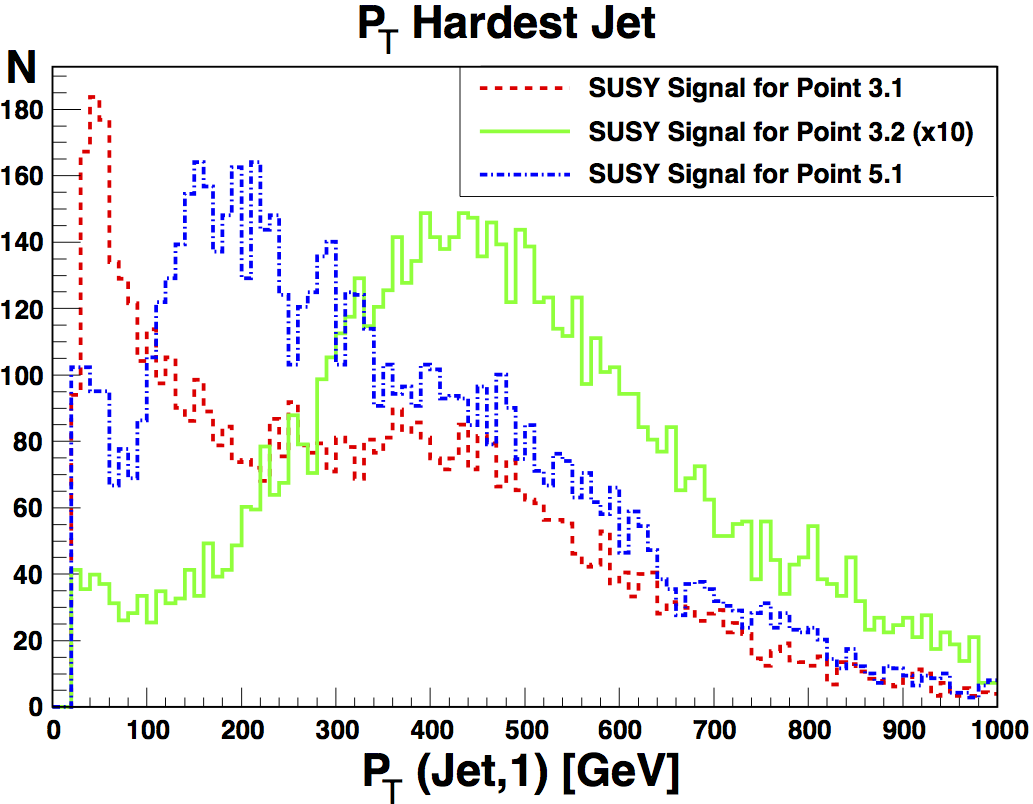}
\includegraphics[width=0.48\textwidth]{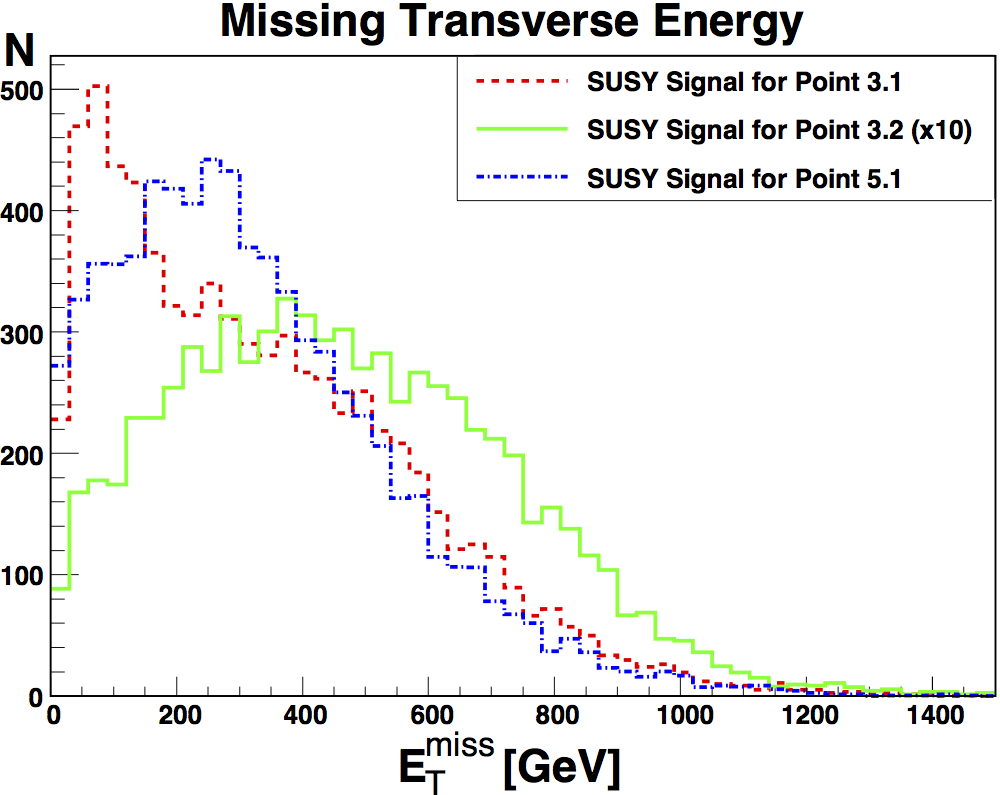}
\caption{Transverse momentum of leading jet (left) and missing transverse energy (right) for points~3.1, 3.2 and~5.1. Distributions are constructed prior to any event selection cuts and normalized to 20~fb$^{-1}$ of data. Red (dotted) distributions are for point~3.1, green (solid) distributions are for point~3.2, and blue (dot-dashed) distributions are for point~5.1. The signal distributions for point~3.2 are multiplied by a factor of~ten to allow for greater readability.}
\label{plot:trio2}
\end{center}
\end{figure}
%%%%%%%%%%%%%%%%%%%%%%%%%%%%%%%%%%%%%%%%%%%%%%%%%%%%%%

It would seem, therefore, that there is potential for improvement in the choice of cuts for monojet-like signatures, just as there were for b-jet searches in the previous case study. Here we consider an `optimized' two-jet signature, originally introduced in the recent work by the authors~\cite{Kaufman:2015nda} in the context of searching for light stops in minimal supergravity models. This signature is a simple modification of the 2jl~signature of Table~\ref{table:multijet}, in which the separation cuts between the direction of the $\slashed{E}_T$ and that of the two hardest jets are significantly increased from $\Delta \phi(Jet,\slashed{E}_T) > 0.4$ to $\Delta \phi(Jet1,\slashed{E}_T) > \pi/2$ and $\Delta \phi(Jet2,\slashed{E}_T) > 1$ for the hardest and second-hardest jet, respectively. This preferentially selects signal events with a highly `lopsided' nature, more in keeping with the notion of soft cascade decay products recoiling against a single hard jet from initial state radiation. For point~3.1 we find that this optimized 2jl~signature far out-performs the traditional monojet signature~M2, becoming comparable in effectiveness to the discovery channel itself. A comparison of the distribution for $p_T(Jet,1)$ with that of the $\slashed{E}_T$ for this point (Figure~\ref{plot:trio2}) shows that a large portion of events will indeed be characterized by $p_T(Jet,1) \simeq \slashed{E}_T$, but that these events are in the lowest $p_T$ bins, often below the cutoff of 340~GeV imposed on both quantities by signature~M2.

These sorts of gains in $L_{\rm min}$ depend on the details of the SUSY model, so generalizations are difficult to state unequivocally. For example, consider point~3.2, which might appear to be an ideal candidate for the optimized, or `lopsided', two jet signature. The work in~\cite{Kaufman:2015nda} was motivated by the decay $\tilde{t}_1 \to \tilde{\chi}_1^0c$, and in the case of point~3.2 we have only an 8~GeV mass difference between the lightest stop and the LSP. Here the gluino mass is heavier ($m_{\tilde{g}} = 1901\,{\rm GeV}$) and thus gluino pair production accounts for a minuscule fraction of the total signal, which is instead composed primarily of light-flavored squark pairs (51\%) and (light-flavored) squark production in association with a gluino (32\%). Stop pairs account for only 9\% of the total SUSY production cross-section. Thus, the 2jl-optimized search is, in fact, optimal for only a small sub-component of the total production cross-section. The $L_{\rm min}$ value of 574~fb$^{-1}$ for 2jl(opt) is actually slightly worse than the original 2jl~signature, for which $L_{\rm min} = 500\,{\rm fb}^{-1}$. Again, this is because 91\% of the events for point~3.2 are {\em not} monojet like in nature at all. This is despite the fact that a large proportion of these events have $p_T(Jet,1) \simeq \slashed{E}_T$, as can be seen from the distributions in Figure~\ref{plot:trio2}. We note that these tend to be the events in the lowest $p_T$ bins, where the cuts of 340~GeV on both quantities from signature~M2 would eliminate most of the monojet-like sub-component of the signal.

%%%%%%%%%%%%%%%%% Trio: Jet multiplicity and  %%%%%%%%%%%%%%%%%%
%
\begin{figure}[th]
\begin{center}
\includegraphics[width=0.48\textwidth]{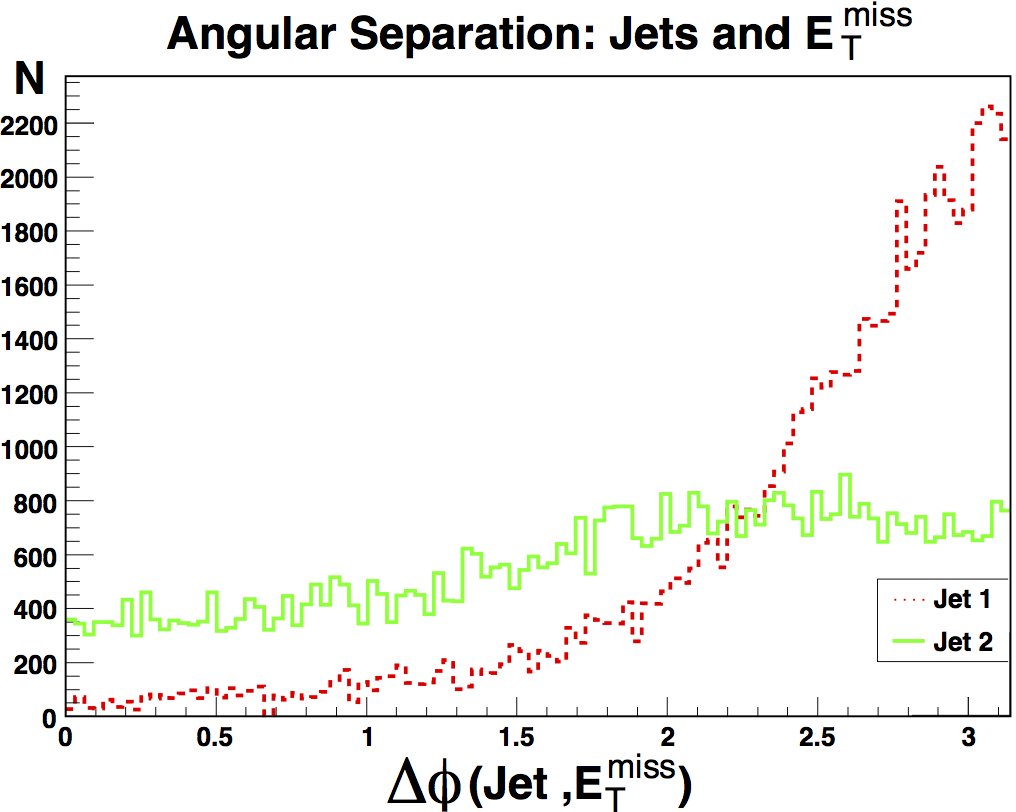}
\caption{Separation in azimuthal angle between leading jets and $\slashed{E}_T$ for point~5.1. The separation $\Delta\Phi$ between the hardest jet (red, dotted), and second-hardest jet (green, solid), in all events is shown. Both distributions are generated prior to the imposition of any other signal region cuts, and are normalized to 20~fb$^{-1}$ of data.}
\label{plot:trio3}
\end{center}
\end{figure}
%%%%%%%%%%%%%%%%%%%%%%%%%%%%%%%%%%%%%%%%%%%%%%%%%%%%%%

The improvement from `standard' monojet to `optimal' monojet is most dramatic for point~5.1, where the $L_{\rm min}$ for the monojet signal reduces by two orders of magnitude, making this channel extremely competitive with the five-jet `discovery' channel. Again, the $N_{\rm jet} \leq 3$ requirement effectively eliminates the `standard' monojet signal, even though a high proportion of these events really do have a lopsided kinematic profile. Conversely, the requirements on inclusive effective mass and $\slashed{E}_T$ listed in Table~\ref{table:multijet} are relatively easy to satisfy (see the distributions in Figure~\ref{plot:trio2}). The greatest impact is on the more restrictive separation requirements, $\Delta \phi(Jet1,\slashed{E}_T) > \pi/2$ and $\Delta \phi(Jet2,\slashed{E}_T) > 1$, which greatly enhance signal-to-background. We plot these quantities for point~5.1 in Figure~\ref{plot:trio3}, where the concentration of events near $\Delta \phi(Jet1,\slashed{E}_T) \to \pi$ is apparent. 

Table~\ref{table:Lmin} summarizes the effectiveness of the traditional monojet channel~M2 as well as the lopsided two-jet channel we call~2jl(opt). The lopsided two-jet channel is always superior in these cases, mostly due to the relaxation of the strict $N_{\rm jet} \leq 3$ requirement, and (to a lesser extent) the replacement of specific requirements on $p_T(Jet,1)$ with a strict requirement on  $\Delta \phi(Jet1,\slashed{E}_T)$ and $\Delta \phi(Jet2,\slashed{E}_T)$. In cases where the spectrum is indeed highly compressed, where the hardest jet arises from a prompt decay of a squark on one side of the event, or from ISR, the 2jl(opt) signature outperforms other two-jet signatures from Table~\ref{table:multijet}. When the model in question is not particularly monojet-like in the first case, this signature is slightly less effective than those in Table~\ref{table:multijet}.

\subsubsection{Case Study 3: Stop-gluino orderings; 2.1 vs. 7.1}

Our final case study is a comparison of points~2.1 and~7.1, in which we will study the impact of the gluino-stop mass hierarchy on detection prospects at the upcoming run of the LHC. Both points involve a certain degree of universality among the scalar masses, with point~2.1 having modular weights $(n_M, n_H)=(0,0)$ and $M_0 = 3200\GeV$, while point~7.1 has modular weights $(n_M, n_H)=(1/2,1/2)$ and $M_0 = 2900\GeV$. The values of $\alpha_m$ are 1.45 and 1.85, respectively. In pure mirage mediation, these values would imply a smaller mass gap between the gluino and neutralino LSP for point~7.1, but the introduction of gauge messengers inverts this. Both points involve three generations of messenger fields, with equal and opposite values of $\alpha_g$ ($\alpha_g = 0.35$ for point~2.1, and $\alpha_g = -0.35$ for point~7.1). Consequently, point~2.1 features the mass hierarchy $m_{\tilde{t}_1} < m_{\tilde{g}}$, with $\Delta m(\tilde{t}_1,m_{\tilde{\chi}_1^0}) = 225\GeV$, while point 7.1 exhibits $m_{\tilde{g}}< m_{\tilde{t}_1}$ with mass separation $\Delta m(\tilde{g}, m_{\tilde{\chi}_1^0}) = 161\GeV$.

Despite the different mass orderings of the lightest SU(3)-charged states, the overall mass scales between the two points are roughly the same. This allows us to address questions like the ultimate `reach' of the LHC at $\sqrt{s} = 14\TeV$ for this class of models. The overall production cross-section for supersymmetry is correlated to the gluino mass, and thus we find $\sigma_{\SUSY}^{2.1} = 22.9\,{\rm fb}$ while $\sigma_{\SUSY}^{7.1} = 86.1\,{\rm fb}$, despite the heavier LSP and stop. For both points, approximately 55\% of the total cross-section for superpartners is gluino pair production. For point~7.1 an additional 40\% of the total production cross-section involves associated production of a gluino with a light-flavored squark, while the light stop of point~2.1 reduces gluino/light squark production to 26\% of total events, with 17\% associated with stop pair production.

For point~2.1, with the more massive gluino, there is a universal decay $\tilde{g} \to \tilde{t}_1 t$, with the subsequent decay of the stop to $\tilde{\chi}_1^{\pm} b$ 62\% of the time, and $\tilde{\chi}^0_{1,2} t$ for the remainder. As the three states $\tilde{\chi}_1^{\pm}$ and $\tilde{\chi}_{1,2}^0$ are highly degenerate, all three effectively represent missing transverse energy. For these gluino pair-production events, we therefore expect four {\em bona fide} b-jets and as many as four leptons in the final state from leptonic decays of $W$-bosons. Alternatively, we can expect up to four b-tagged jets and up to eight additional jets. If, instead, we consider associated production of a gluino and a light squark, the fact that the light-flavored squarks decay universally into a corresponding light quark and a gluino means the above analysis applies in this production channel as well, with perhaps one additional high-$p_T$ jet in the final state. Indeed, Table~\ref{table:sigprops} confirms this reasoning, with the percentage of events with at least one lepton for point~2.1 being 21.1\%, the percentage with at least one b-tagged jet being 85.3\%, and the peak in the jet multiplicity distribution being $N_{\rm jet} = 8$. It is therefore not surprising that the most effective discovery channel for this point is the six-jet (6jt+) channel, with an eventual confirmation in the single-lepton channel occuring much later in the lifetime of the~LHC. Specifically, we find that the first leptonic signal will arise in the six-jets plus muon channel of~\cite{Aad:2015mia} after 1100~fb$^{-1}$ of integrated luminosity.

Point~2.1 would appear to be a prime candidate for searches involving b-tagged jets. Given the high overall jet multiplicity for this point, it is not surprising that the strongest signal (or lowest $L_{\rm min}$) occurs for the seven-jet channels of Table~\ref{table:bjets2}, as demonstrated in Table~\ref{table:Lmin}. What is, perhaps, surprising is that there is no signal expected, even after 3000~fb$^{-1}$, for signature SRA2 involving at least six jets, two of which carry b-tags. Here is a case of signatures that are very effectively tailored to certain exclusive event categories -- that is to say, signatures that do their respective jobs very well. The b-jet based signatures of Table~\ref{table:bjets2} are based on gluino pair production, which is the dominant component of the production cross-section for point~2.1. The SRA signatures of Table~\ref{table:bjets} target stop pair production, followed by the decay $\tilde{t}_1 \to t \tilde{\chi}_1^0$. Stop pair production is only 17\% of the total, and the lightest stop decays to $t \tilde{\chi}_1^0$ only 38\% of the time, decaying to $b\tilde{\chi}_1^{\pm}$ the remaining 62\% of the time. Thus signature SRA2 captures only about 2.5\% of the total events. We note that the SRC channels on Table~\ref{table:bjets}, which are designed to target the $b\tilde{\chi}_1^{\pm}$ decays of the stop, do give a signal significance roughly twice that for the SRA channels, with SRC3 providing a five sigma excess in $L_{\rm min} = 1650\,{\rm fb}^{-1}$ of integrated luminosity.

%%%%%%%%%%%%%%%%% Final pair: point 7.1 Meff and DeltaPhi  %%%%%%%%%%%%%%%%%%
%
\begin{figure}[t]
\begin{center}
\includegraphics[width=0.51\textwidth]{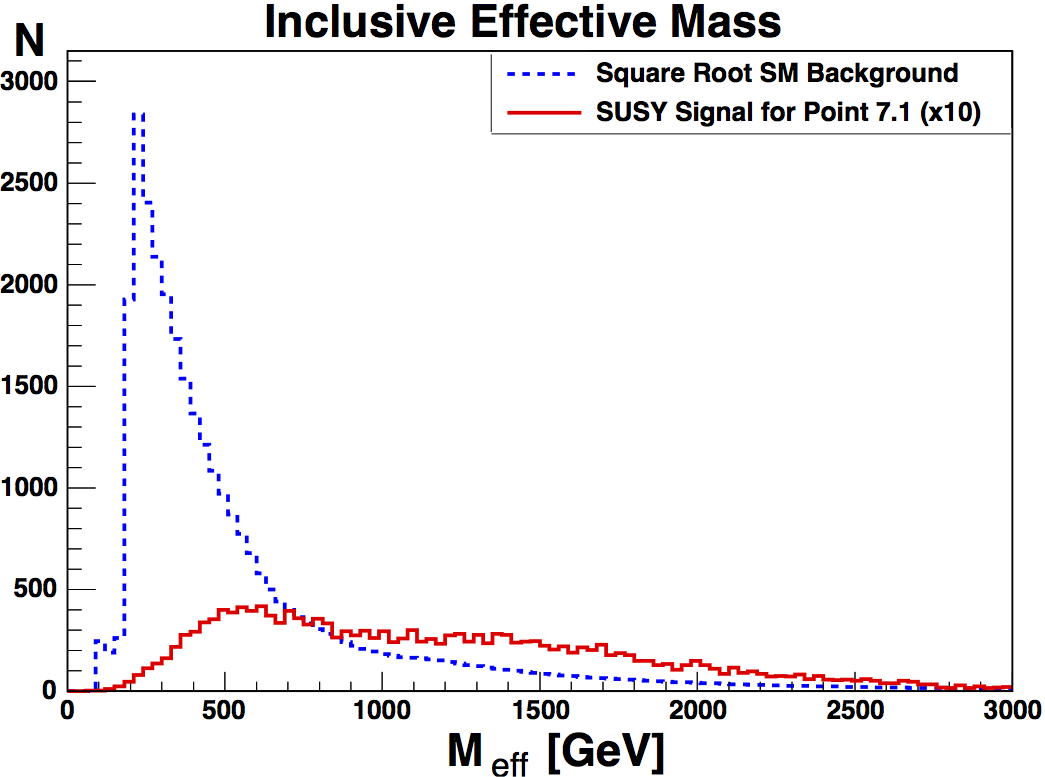}
\includegraphics[width=0.47\textwidth]{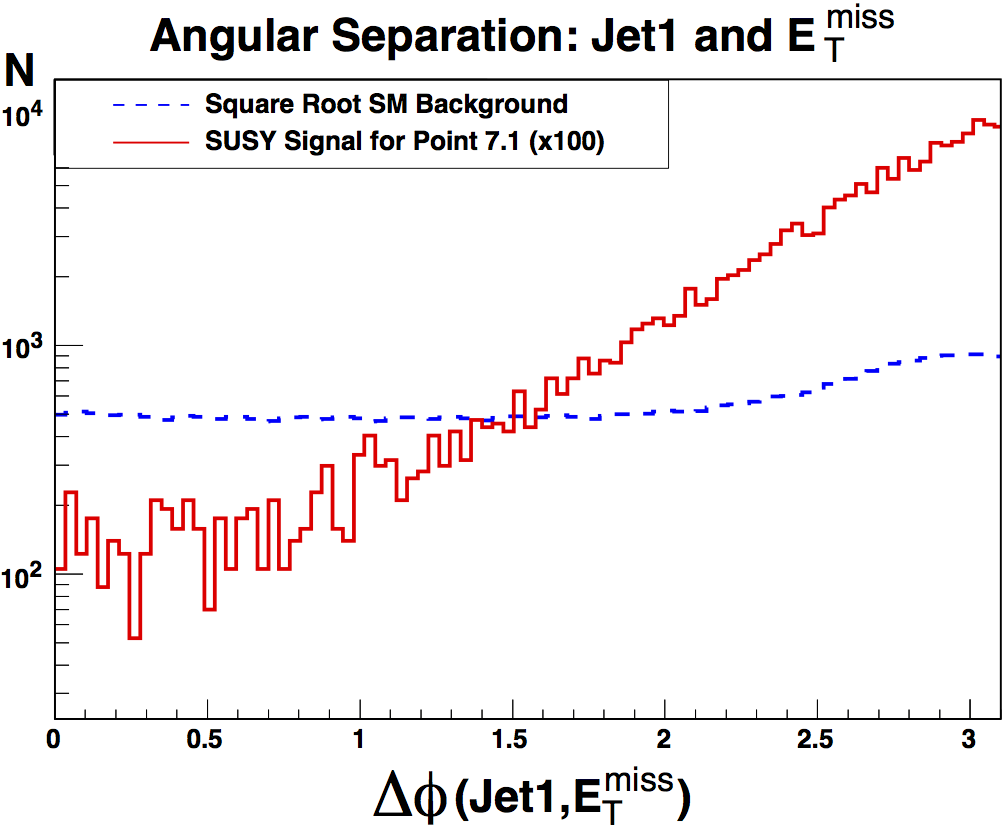}
\caption{Inclusive effective mass (left) and $\Delta \phi(Jet1,\slashed{E}_T)$ distribution (right) for point 7.1 versus combined Standard Model background. Distributions are for the signal (red, solid) prior to any event selection cuts, while the background distributions (blue, dashed) have a pre-cut of $\slashed{E}_T > 150\,{\rm GeV}$ applied. Background distributions represents the square root of the total counts in each bin, with the total data set normalized to 20~fb$^{-1}$. The signal has been augmented by a factor of~10 for the left panel, and a factor of~100 for the right panel, to allow for an easier comparison of the relative shapes of the distributions.}
\label{plot:final}
\end{center}
\end{figure}
%%%%%%%%%%%%%%%%%%%%%%%%%%%%%%%%%%%%%%%%%%%%%%%%%%%%%%

The situation for point~7.1 is nearly identical with regard to light squark decays, only now gluinos decay overwhelmingly into a gluon and one of the two lightest neutralinos, which are highly degenerate in mass. As $\tilde{g}\tilde{g}$ and $\tilde{g}\tilde{q}$ processes represent 94\% of the total SUSY production cross-section, we now expect zero leptons or b-tagged jets in the events, and at most three high~$p_T$ jets, with others arising from soft decay products of $\tilde{\chi}_2^0$ and/or initial and final state radiation. This is borne out by Table~\ref{table:sigprops} where we find zero events with a high-$p_T$, isolated lepton and very few with a b-tagged jet (consistent with the mis-tagging rate built into the Delphes detector simulator). We find that the strongest signal for this point is in the four-jet (4jl) channel, with an $L_{\rm min}$ of 65~fb$^{-1}$, though the dijet channel (2jt) is competitive with an $L_{\rm min}$ of 88~fb$^{-1}$.
In this case, the generic two-jet search performs much better than our monojet-motivated `unbalanced' two-jet signature (2jl-opt), which requires almost twice as much integrated luminosity to reach a five-sigma excess. While these two signatures are very similar, the relative efficacy can be understood in terms of competing cuts on $m_{\rm eff}$ versus the jet separation variable $\Delta \phi(Jet1,\slashed{E}_T)$, and their comparative effects on the signal versus the background sample. To improve signal-to-background, the 2jt~signal makes a cut on the inclusive effective mass of $m_{\rm eff} > 1600\,{\rm GeV}$, versus 800~GeV for the `unbalanced' two-jet signature (2jl-opt). Alternatively, the latter makes a cut on $\Delta \phi(Jet1,\slashed{E}_T) > \pi/2$, versus 0.4 for the signatures in Table~\ref{table:monojets}. These two quantities are plotted in the left and right panels, respectively, of Figure~\ref{plot:final}. The signal distribution (red, solid line) is shown prior to any event selection cuts, while the background distribution (blue, dashed line) has a pre-cut of $\slashed{E}_T > 150\,{\rm GeV}$ applied. Background distributions represent the square root of the total counts in each bin, with the total data set normalized to 20~fb$^{-1}$. The signal has been augmented by a factor of~10 for the left panel, and a factor of~100 for the right panel, to allow for an easier comparison of the relative shapes of the distributions. Clearly, both the $m_{\rm eff}$ cut and the $\Delta \phi$ cut prefer the signal distribution, but the stringent effective mass cut does so much more powerfully than the angular separation cut -- at least when considered in isolation. Thus, when the superpartner spectrum provides enough phase space to use large $m_{\rm eff}$ and or $\slashed{E}_T$ cuts to reduce the background, the classic multijet channels will be preferred. The angular separation cut can be a useful tool for those cases in which aggregate quantities such as $m_{\rm eff}$ are low, as in cases with a compressed superpartner spectrum.

Many pairs such as points~2.1 and~7.1 were generated in the course of our analysis, in which the relative masses of the gluino and light squarks are inverted, but with the overall superpartner scale roughly the same. The aggregation of such pairs allows us to make a very crude estimate of the reach (in the sense of a five sigma excess of signal over background in at least one search channel) in terms of the gluino mass and general squark mass for a broad array of DMM parameter sets. For the case $m_{\tilde{g}}>m_{\tilde{t}_1}$ we estimate a reach to be approximately $m_{\tilde{g}} \lappeq 1800\,{\rm GeV}$ in 100~fb$^{-1}$ of data, while for the case $m_{\tilde{t}_1}>m_{\tilde{g}}$ we estimate the reach to be $m_{\tilde{t}_1} \lappeq 1270\,{\rm GeV}$ in 100~fb$^{-1}$ of data.

\section{Dark Matter Detection in Deflected Mirage Mediation}
\label{sec:DM}

Even with the discovery of the Higgs, and increasingly stringent measurements of the dark matter relic density, model points with bino-like, wino-like and/or Higgsino-like LSPs remain from every combination of modular weights. One may now ask if any of these points, not yet excluded by searches for superpartners at the LHC, could be detected in the near future in dark matter direct detection experiments. We focus on direct detection here, as indirect detection signals (gamma rays, positrons, anti-protons, neutrino fluxes, etc.) tend to be well below estimated astrophysical backgrounds once the signal is scaled by the predicted thermal relic density. That is, wino-like and Higgsino-like LSPs in the DMM scenario tend to have thermal relic densities below that preferred by measurements of the CMB. Once any non-thermal mechanism for populating these LSPs is posited, the constraints from indirect detection become highly constraining~\cite{Fan:2013faa,Cohen:2013ama,Blinov:2014nla,Kane:2015qea}.

To date, discovery prospects for neutralino dark matter ($100\,\text{GeV} \leq m_{\chi} \leq 1000\,\text{GeV}$) have been dominated by the liquid xenon direct detection experiments: the Xenon100 Dark Matter Project in Gran Sasso, Italy~\cite{Aprile:2011dd}, and the South Dakota-based LUX experiment~\cite{Akerib:2012ys}. The former released data in~2012 for 224.6~live days of exposure on a 34~kg target~\cite{Aprile:2012nq}. In~2013, the LUX experiment released a preliminary result from 85.3~live days of exposure on a 118~kg target~\cite{Akerib:2013tjd}. In the near future, LUX will release data from approximately 300~days of exposure, while the extension of Xenon100 to the one ton level, Xenon1T, will follow soon thereafter \cite{Aprile:2012zx}.
We can therefore discuss the discovery prospects for dark matter in two stages. First we determine what, if any, parameter space is already in conflict with existing results from Xenon100 and LUX. We then project what part of the parameter space might yield a signal in future results from LUX, Xenon1T, or LZ, the next generation of the LUX experiment~\cite{Malling:2011va}.\footnote{We focus exclusively on those experiments which are sensitive to the spin-independent part of the scattering cross-section, as our investigations indicate that experiments sensitive to spin-dependent cross-sections, such as the PICO-2L experiment~\cite{Amole:2015lsj}, and its future upgrades, are always less sensitive to any given parameter point than the large-scale spin-independent measurements.} In what follows we will consider a subset of 258,225~DMM points, all of which satisfy $m_{\tilde{g}} \geq 1\,{\rm TeV}$, which can reasonably be expected to have passed the LHC supersymmetry searches at $\sqrt{s} = 8\,{\rm TeV}$.

\subsection{Bino-like LSPs}

%=(12)=============== Combined Bino-like LSP Xenon Comparison ==================
\begin{figure}[t]
\begin{center}
\includegraphics[width=0.49\textwidth]{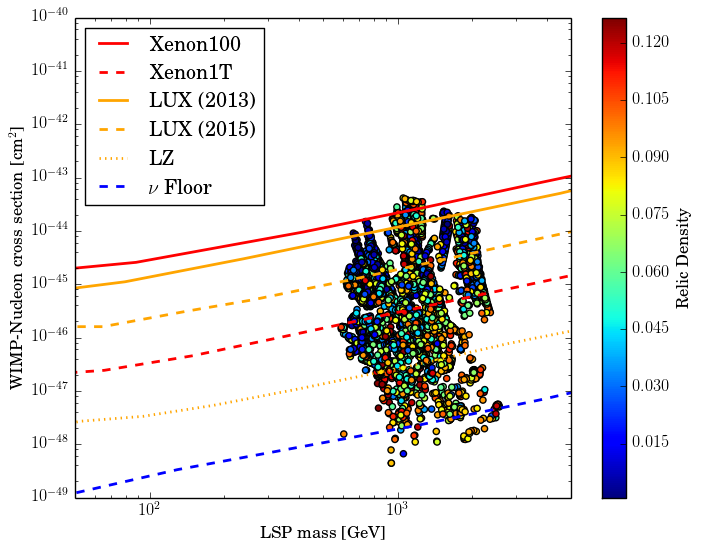}
\includegraphics[width=0.49\textwidth]{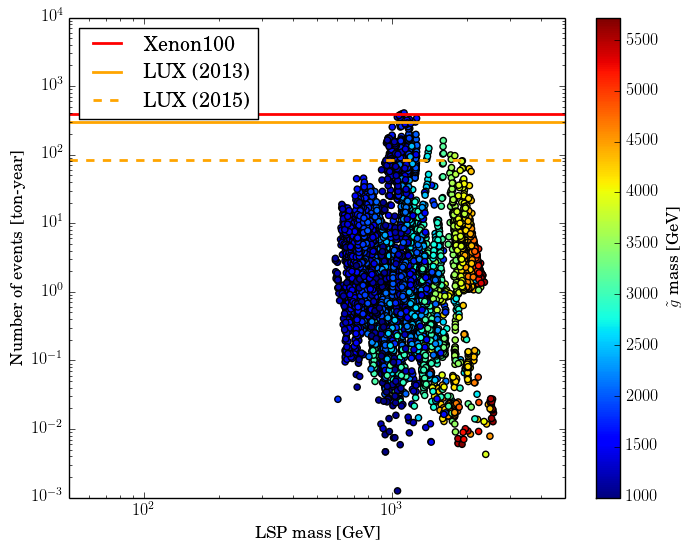}
\caption{The left plot shows the distribution in neutralino-nucleon scattering cross-sections versus neutralino mass for the bino-like segment of the DMM parameter space. The lines represents the current and future limits set by the recent results from Xenon100 and LUX, and future limits from LUX, Xenon1T, and LZ under the assumption that the relic density is saturated. The predicted thermal relic density is indicated by the color code. The right plot gives the rate of nuclear recoils, rescaled by the relic density, and  integrated over the recoil energy range of 5-25~keV, after one ton-year of exposure. Current limits from Xenon100 and current/future limits from LUX are represented as straight lines where 10~events would be observed. The color in the right figure indicates the gluino mass in GeV. Both plots aggregate all the cases with a bino-like LSP for all modular weight combinations.}
\label{fig:BinoXenonCompare}
\end{center}
\end{figure}
%==============================================================================

A nearly bino-like LSP can be found for nearly all modular weight combinations. For the purposes of discussing dark matter phenomenology, it is convenient to aggregate these modular weight combinations and consider the bulk properties of all bino-like neutralino cases as one phenomenologically similar region. For this combined region, the LSP is heavy, ranging from 590-2570~GeV. The left plot in Figure~\ref{fig:BinoXenonCompare} shows the familiar neutralino-nucleon cross-section versus LSP mass for all of the targeted scan regions with bino-like LSPs. The lines represent the results from various dark matter direct detection experiments under the assumption that the relic density constraints are saturated. The color scheme in the left panel gives the predicted thermal relic density for each point. Clearly, many of these points would need to rely on some non-thermal production mechanism for this figure to be valid.

More realistic, perhaps, is the right panel in Figure~\ref{fig:BinoXenonCompare}, which gives the number of expected events for an exposure of 300~days for 1000~kg of liquid Xenon ({\em i.e.} one ton-year), within the recoil energy range of 5-25~keV. In this case, we have renormalized the count rate to the expectation for the predicted relic density. That is, we have scaled the prediction by the ratio $(\Omega h^2)_{\rm pred}/0.12$. In this panel, current limits from Xenon100 and current/future limits from LUX are represented as straight lines where 10~events would be observed.

The 2013 LUX data for LSPs in the appropriate mass range, corresponding to a fiducial volume of 118~kg and an exposure of 85~days~\cite{Akerib:2013tjd} has already begun to cut into the bino-like parameter space, but only marginally so. 
While there are a handful of points with very large cross sections, the bulk of the bino-like parameter space in the DMM scenario is currently outside the reach of these experiments. The 2015 run of LUX, Xenon1T and LZ expect to improve the limiting cross-section on WIMP-nucleon scattering by orders of magnitude. For many of these points, more than $O(1)$ events per ton-year are expected, which could lead to significant signals in future direct detection experiments. Note that upcoming LUX~data should begin to eliminate models with characteristic gluino masses up to $\order(3\TeV)$ (see color key in right panel of Figure~\ref{fig:BinoXenonCompare}).

However, for bino-like LSPs in the DMM paradigm, that improvement is likely to leave a significant portion of the parameter space unexplored, including many of the points with a Planck-preferred relic density of $\Omega_{\chi}h^2 \simeq 0.12$. This data set also includes points below the cross section for coherent neutrino scattering~\cite{Cushman:2013zza}, the blue line on Figure~\ref{fig:BinoXenonCompare}.
For all of the points below the neutrino floor, the LSP is nearly degenerate with either the stop or the gluino giving us significant coanihillation effect in the early universe. Many of these points should still be accessible at the LHC or a future 100~TeV collider.

\subsection{Wino-like and Higgsino-like LSPs}

\begin{figure}[t]
\begin{center}
\includegraphics[width=0.49\textwidth]{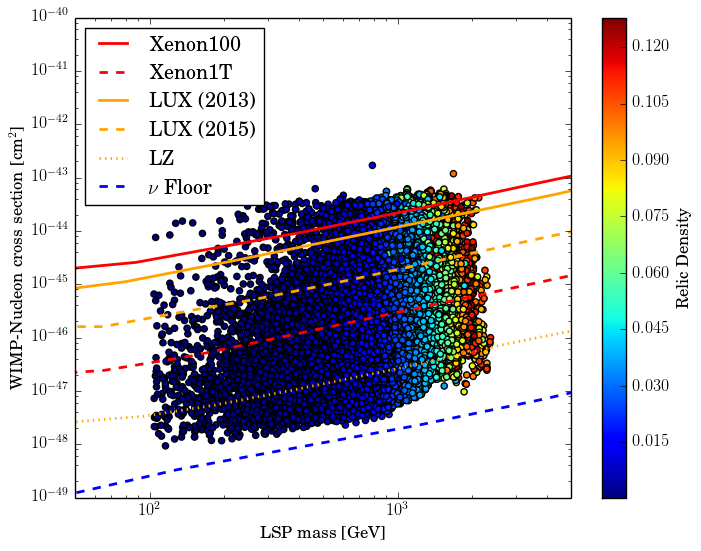}
\includegraphics[width=0.49\textwidth]{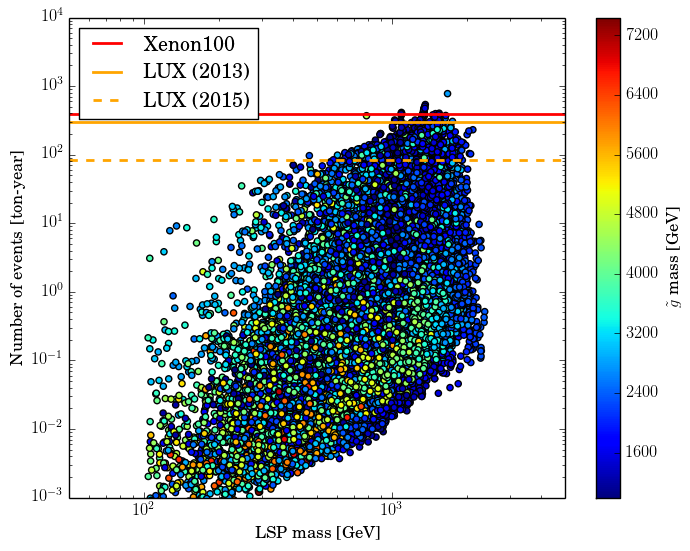}
\caption{The left plot shows the distribution in neutralino-nucleon scattering cross-sections versus neutralino mass for the wino-like segment of the DMM parameter space. The lines represents the current and future limits set by the recent results from Xenon100 and LUX, and future limits from LUX, Xenon1T, and LZ under the assumption that the relic density is saturated. The predicted thermal relic density is indicated by the color code. The right plot gives the rate of nuclear recoils, rescaled by the relic density, and  integrated over the recoil energy range of 5-25~keV, after one ton-year of exposure. The color in the right figure indicates the gluino mass in GeV. Both plots aggregate all the cases with a wino-like LSP for all modular weight combinations.}
\label{fig:WinoXenonCompare}
\end{center}
\end{figure}

\begin{figure}[th]
\begin{center}
\includegraphics[width=0.49\textwidth]{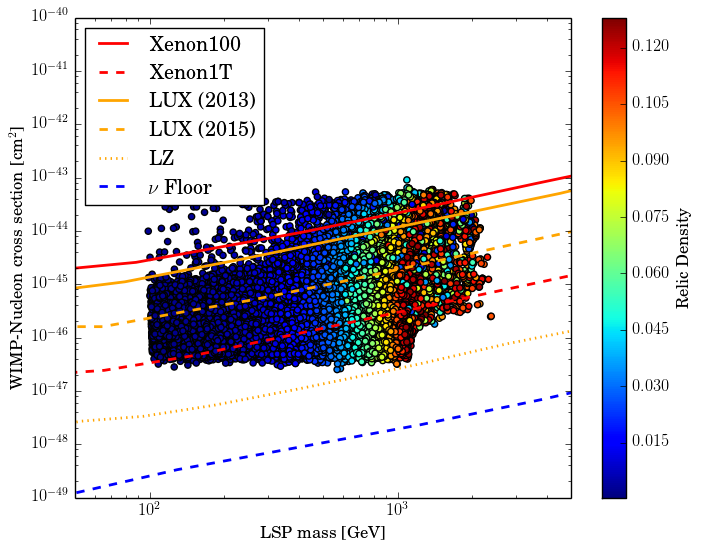}
\includegraphics[width=0.49\textwidth]{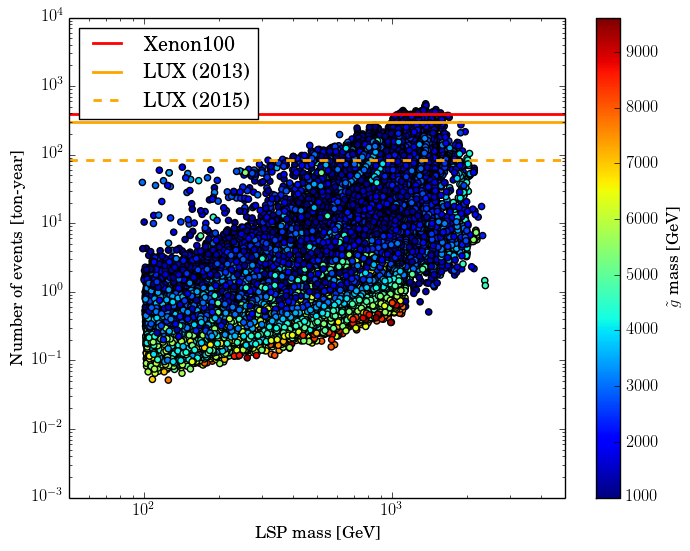}
\caption{The left plot shows the distribution in neutralino-nucleon scattering cross-sections versus neutralino mass for the Higgsino-like segment of the DMM parameter space. The lines represents the current and future limits set by the recent results from Xenon100 and LUX, and future limits from LUX, Xenon1T, and LZ under the assumption that the relic density is saturated. The predicted thermal relic density is indicated by the color code. The right plot gives the rate of nuclear recoils, rescaled by the relic density, and  integrated over the recoil energy range of 5-25~keV, after one ton-year of exposure. The color in the right figure indicates the gluino mass in GeV. Both plots aggregate all the cases with a Higgsino-like LSP for all modular weight combinations.}
\label{fig:HinoXenonCompare}
\end{center}
\end{figure}

In contrast to pure mirage mediation, as in KKLT, a wino-like LSP can be found for many modular weight combinations in DMM. As was done in the previous subsection, we combine our analysis for all modular weights.  The left plot in Figure~\ref{fig:WinoXenonCompare} shows the neutralino-nucleon cross-section versus LSP mass for all of the targeted scan regions with wino-like LSPs, analogously to Figure~\ref{fig:BinoXenonCompare}. The lines again represent the results from various dark matter direct detection experiments under the assumption that the relic density constraints are saturated. The right panel gives the number of expected events for an exposure of 300~days for 1000~kg of liquid Xenon ({\em i.e.} one ton-year), within the recoil energy range of 5-25~keV, scaled by the ratio $(\Omega h^2)_{\rm pred}/0.12$.
The wino-like region extends from roughly the LEP chargino limit of 103.5~GeV, to 2.5 TeV, spanning much of the region from $10^{-44}\,\rm{cm}^2$ to $10^{-48}\,\rm{cm}^2$. Note that the predicted thermal relic density is a strong function of the LSP mass, saturating the Planck limit at approximately 2~TeV.

In the alternative case for which the neutralino is almost purely Higgsino-like, we again aggregate all eight combinations of modular weights that admit a Higgsino-like LSP into a single region. For this combination, the LSP can be as heavy as 2.4~TeV, which is slightly lighter than the maximum value in the bino- or wino-like case. However, LSPs as light as 100~GeV are also present. The neutralino-nucleon scattering cross section is similarly spread over a wide range of values. The entirety of the parameter space has a cross section of between $10^{-46}\,\rm{cm}^2$ and $10^{-43}\,\rm{cm}^2$.

As with the previous two figures, the left plot in Figure~\ref{fig:HinoXenonCompare} shows the neutralino-nucleon cross-section versus LSP mass for all of the targeted scan regions with Higgsino-like LSPs, analogously to Figure~\ref{fig:BinoXenonCompare}. The lines again represent the results from various dark matter direct detection experiments under the assumption that the relic density constraints are saturated. The right panel gives the number of expected events for an exposure of 300~days for 1000~kg of liquid Xenon ({\em i.e.} one ton-year), within the recoil energy range of 5-25~keV, scaled by the ratio $(\Omega h^2)_{\rm pred}/0.12$.

While more of the wino-like and Higgsino-like parameter space is being contrained by current experiments than in the bino-like case, the bulk of the viable points lie outside the reach of Xenon100 and LUX, (see the right panels of Figures~\ref{fig:WinoXenonCompare} and~\ref{fig:HinoXenonCompare}). The anticipated release of new~LUX data will begin to probe models with relatively light gluinos in the Higgsino-like LSP regime, thus largely overlapping with anticipated early LHC~results. In the more limited wino-like regime, however, some model points within reach of LUX correspond to gluino masses of 3~-~4~TeV, likely outside of the LHC reach even after 3000~fb$^{-1}$. LUX~claims a future background expectation of approximately 1~event per ton-year at these recoil energies~\cite{Akerib:2012ys}. It is therefore plausible to expect an even larger fraction of spectra with a Higgsino-like LSP may be within reach in the near future. Note that all of the DMM parameter space with a wino-like or Higgsino-like LSP should give signals well-above the background signal from coherent neutrino scattering. In fact, it is conceivable that the entire parameter space with Higgsino-like LSPs in the DMM model will be definitely probed by the LZ~experiment in future years.

\section{Conclusions}

If there is any single paradigm for supersymmetry breaking that could claim to be considered a consensus within the string phenomenology community, as of this writing, it would undoubtedly be the mirage mediation scenario popularized in the period following the celebrated paper detailing moduli-stabilization in certain Type~IIB flux compactifications by Kachru et al. (KKLT) in 2003~\cite{Kachru:2003aw}. The pattern of soft terms, which was soon referred to as mixed modulus-anomaly mediation~\cite{Choi:2005uz}, or more simply ``mirage mediation''~\cite{LoaizaBrito:2005fa}, is not unique to Type~IIB constructions. In fact, the so-called mirage pattern of gaugino masses were first identified in heterotic string constructions that achieved acceptable moduli phenomenology using the technique of K\"ahler stabilization~\cite{Gaillard:2000fk,Binetruy:2000md,Nelson:2002fk,Nelson:2003tk}. In the years following the KKLT publication, ever more manifestations of the mirage pattern of guagino masses were motivated, culminating in the original papers describing deflected mirage mediation~\cite{Everett:2008qy,Everett:2008ey}.

Given the ubiquity of the mirage mediation paradigm, and its general acceptance as a realistic outcome of moduli stabilization and supersymmetry breaking in a variety of string theoretic contexts, it is absolutely natural to begin an in-depth study of the implications of LHC data on string-motivated models on this subset of theories. The first two papers on this series were conducted in the heterotic~\cite{Kaufman:2013pya} and original Type~IIB contexts~\cite{Kaufman:2013oaa}, so a natural completion to this study is the present work.

Deflected mirage mediation models offer the broadest possible paradigm for investigating supersymmetry breaking of any ultraviolet-complete theory, in that they allow for similar-sized contributions form gravity mediation, anomaly mediation and gauge mediation. The present work expands upon earlier treatments of the parameter space of DMM models~\cite{Altunkaynak:2010tn,Altunkaynak:2010xe,Holmes:2009mx} by updating the constraints implied by collider-based superpartner searches, dark matter search constraints and, critically, the measurement of the Higgs boson mass. The latter has profound implications for all supersymmetric theories, and this is particularly acute for the models of the KKLT/fluxed Type~IIB paradigm. 

In the previous work~\cite{Kaufman:2013oaa}, which studied the simplest mirage mediation models motivated by flux compactifications of Type~IIB string theory, it was found that the relatively large CP-even Higgs mass of $m_h \simeq 125\,{\rm GeV}$ puts very strong constraints on the allowed parameter space. Some combinations of modular weights for the matter and Higgs sectors are very hard to reconcile with all current experimental constraints, while other can persist only for very special ranges of parameters like $\alpha_m$, $M_0$ and $\tan\beta$. Overall, points in the parameter space that would give a sufficient Higgs boson mass would tend to imply superpartners that are so massive as to avoid detection at the LHC, even after 3000~fb$^{-1}$ of integrated luminosity. This is also in evidence in the current work, in the form of the KKLT `base points' of Table~\ref{table:KKLTBenchmarks}, none of which are estimated to generate a discovery at the LHC.

Yet the inclusion of a sector which mediates supersymmetry breaking via gauge interactions radically alters this prediction, allowing for much lower superpartner masses -- particularly for the gluinos and quarks -- while still satisfying the requirement that $m_h \simeq 125\,{\rm GeV}$. In fact, it is possible to achieve gluino and squark masses so low that a discovery would have been made in the previous LHC runs at $\sqrt{s} = 8\, {\rm TeV}$.  Roughly speaking, we find a reach of $m_{\tilde{g}}\lappeq 600\GeV$ for $\Delta m(\tilde{g},\tilde{\chi}_1^0)\lappeq 50\,{\rm GeV}$, and $m_{\tilde{g}}\lappeq 900\GeV$ for more sizeable mass gaps. Moreover, DMM~corrections tend to alter the masses of $SU(3)$-charged objects in a correlated way, tending to compress both the gluino and the stop toward the lightest neutralino mass. The modular weight combinations most likely to produce spectra detectable at $\sqrt{s}=8\TeV$ are the $(n_M,n_H)= (0,\,0)$, $(0,\,0.5)$ and $(0.5,\,0)$ cases, with low messenger scale and $\alpha_g >0$ -- precisely the most interesting cases from the point of view of string model building. This gives hope that the next round of LHC data-taking will probe deeply into this rich parameter space.

We have estimated the reach of the LHC at $\sqrt{s} = 14\,{\rm TeV}$ center-of-mass energies within the DMM parameter space by sampling parameter combinations that give the lowest possible values of key superpartner masses. For the case $m_{\tilde{g}}>m_{\tilde{t}_1}$ we estimate a reach to be approximately $m_{\tilde{g}} \lappeq 1800\,{\rm GeV}$ in 100~fb$^{-1}$ of data, while for the case $m_{\tilde{t}_1}>m_{\tilde{g}}$ we estimate the reach to be $m_{\tilde{t}_1} \lappeq 1270\,{\rm GeV}$ in 100~fb$^{-1}$ of data. Much of the $\alpha_g>0$ parameter space will be probed, including those regions around the theoretically motivated area of $\alpha_m \simeq 1$. The most likely discovery channel will be in the low-multiplicity jets plus missing transverse energy channel, with lepton vetoes, but corroborating signals should be expected in various channels utilizing b-tagged jets, or those channels which emphasize a `lopsided' two or three-jet event, which will resemble the classic `monojet' signature. In fact, the presence of these corroborating signals will be precisely the indication that a compressed spectrum is present.

The search strategies we employ were defined for applicability at $\sqrt{s} = 8\TeV$. Surely, the kinematic cuts can be adjusted to more fully optimize the signal-to-background. Some suggestions were identified in the course of discussing the strengths and weakness of various b-jet based signatures, and various monojet-like signatures, through various case studies involving the DMM points of Table~\ref{table:Partners}. One can undoubtedly do even better, and we encourage our experimental and theoretical colleagues to consider such top-down motivated models for honing signal definitions in the forthcoming LHC run. There has been much interest of late in two opposite extremes: the study of so-called `simplified models', which posit a very simple superpartner spectrum with large mass gaps generating energetic decay products, and compressed-spectrum models which are motivated from the bottom-up in terms of some abstract sense of `naturalness'. The former are popular with the experimental community, while the latter seem to be enjoying popularity with model-builders. The DMM paradigm allows a unified, ultraviolet-complete and string-motivated framework that spans both extremes. We therefore hope that studies such as this one will serve as motivation to continue to refine search strategies to maximize the impact of the coming LHC data. 

\section*{Acknowledgements}

The authors would like to thank Ian-Woo Kim for assistance, early in the project, with the modified, two-stage SOFTSUSY software package that implements the DMM model. BK would like to thank Andrew Spisak for technical assistance with ROOT scripts. The work of BK and BDN was supported by a grant from the National Science Foundation, PHY-1314774. LE and TG are supported by the U.S. Department of Energy under the contract DE-FG-02-95ER40896. LE acknowledges the support and hospitality of the Enrico Fermi Institute at the University of Chicago.

%\pagebreak
\appendix

\section{Anomalous dimensions}
\label{Anomalous}
At one loop, the anomalous dimensions are given by 
\begin{eqnarray}
\gamma_i = 2 \sum_a g_a^2 c_a(\Phi_i) - \frac{1}{2}\sum_{lm} |y_{ilm}|^2,
\label{gammaexp}
\end{eqnarray}
in which $c_a$ is the quadratic Casimir, and $y_{ilm}$ are the normalized Yukawa couplings.  Here we will consider only the Yukawa couplings of the third generation $y_t$, $y_b$, and $y_\tau$.  For the MSSM fields $Q$, $U^c$, $D^c$, $L$, $E^c$, $H_u$ and $H_d$,
the anomalous dimensions are
\begin{eqnarray}
\gamma_{Q,i} &=& \frac{8}{3} g_3^2 + \frac{3}{2} g_2^2 + \frac{1}{30} g_1^2
- (y_t^2 + y_b^2) \delta_{i3}\nonumber \\
\gamma_{U,i} &=& \frac{8}{3} g_3^2 + \frac{8}{15} g_1^2
- 2 y_t^2 \delta_{i3},\;\; %\nonumber \\
\gamma_{D,i} =
%&=& 
\frac{8}{3} g_3^2 + \frac{2}{15} g_1^2
- 2 y_b^2 \delta_{i3},\nonumber \\
\gamma_{L,i} &=& \frac{3}{2} g_2^2 + \frac{3}{10} g_1^2
-y_\tau^2 \delta_{i3}, \;\; % \nonumber \\
\gamma_{E,i} =
%&=& 
\frac{6}{5} g_1^2
-2 y_\tau^2 \delta_{i3}, \nonumber \\
\gamma_{H_u} &=& \frac{3}{2} g_2^2 + \frac{3}{10} g_1^2
-3 y_t^2, \;\; 
%\nonumber \\
\gamma_{H_d} =
%&=& 
\frac{3}{2} g_2^2 + \frac{3}{10} g_1^2
- 3 y_b^2 - y_{\tau}^2,
\end{eqnarray}
respectively.
Above $M_{\rm mess}$, the beta function of the gauge couplings
changes because of the messenger fields.  However, $\gamma_i$ does not change according to
Eq.~(\ref{gammaexp}), and hence $\gamma'_i = \gamma_i$.  The $\dot{\gamma}_i$'s are given by the expression
\begin{eqnarray}
\dot{\gamma}_i=2\sum_a g_a^4b_a c_a(\Phi_i) - \sum_{lm} |y_{ilm}|^2b_{y_{ilm}},
\end{eqnarray}
in which $b_{y_{ilm}}$ is the beta function for the Yukawa coupling $y_{ilm}$.  The 
$\dot{\gamma}_i$'s are given by
\begin{eqnarray}
\dot\gamma_{Q,i}
&=& \frac{8}{3} b_3 g_3^4 + \frac{3}{2} b_2 g_2^4 + \frac{1}{30} b_1 g_1^4
- (y_t^2 b_t + y_b^2 b_b ) \delta_{i3} \nonumber \\
\dot\gamma_{U,i}
&=& \frac{8}{3} b_3 g_3^4 + \frac{8}{15} b_1 g_1^4
- 2 y^2_t b_t \delta_{i3}, \;\; % \nonumber \\
\dot\gamma_{D,i}=
%&=& 
\frac{8}{3} b_3 g_3^4 + \frac{2}{15} b_1 g_1^4
- 2 y^2_b b_b \delta_{i3} \nonumber \\
\dot\gamma_{L,i}
&=& \frac{3}{2} b_2 g_2^4 + \frac{3}{10} b_1 g_1^4
 - y_\tau^2 b_\tau \delta_{i3},\;\; 
 %\nonumber \\
\dot\gamma_{E,i}=
%&=& 
\frac{6}{5} b_1 g_1^4
 - 2 y_\tau^2 b_\tau \delta_{i3} \nonumber \\
\dot\gamma_{H_u}
&=& \frac{3}{2} b_2 g_2^4 + \frac{3}{10} b_1 g_1^4
 - 3 y^2_t b_t,\;\; % \nonumber \\
\dot\gamma_{H_d}
%&=&
=\frac{3}{2} b_2 g_2^4 + \frac{3}{10} b_1 g_1^4
 - 3 y_b^2 b_b - y^2_\tau b_\tau,  \label{dotgammaexp}
\end{eqnarray}
where
%\begin{eqnarray}
$b_t = 6 y_t^2 + y_b^2 -\frac{16}{3} g_3^2 - 3 g_2^2 - \frac{13}{15} g_1^2$, %\;\;
%\nonumber \\
$b_b =
%&=& 
y_t^2 + 6 y_b^2 + y_\tau^2 -\frac{16}{3} g_3^2 - 3 g_2^2
- \frac{7}{15} g_1^2$ %, \nonumber \\
and $b_\tau = 3 y_b^2 + 4 y_\tau^2 - 3 g_2^2 -\frac{9}{5} g_1^2$.
%\end{eqnarray}
$\dot\gamma^\prime_i$ is obtained by replacing $b_a$ with $b'_a = b_a + N$
in Eq.~(\ref{dotgammaexp}). 

Finally, ${\theta_i}$, which appears in the mixed modulus-anomaly term in the soft scalar mass-squared parameters, is given by
\begin{eqnarray}
\theta_i = 4 \sum_a g_a^2 c_a(Q_i) - \sum_{i,j,k} |y_{ijk}|^2
( 3- n_i -n_j- n_k).
\end{eqnarray}
For the MSSM fields, they take the form
\begin{eqnarray}
\theta_{Q,i} &=& \frac{16}{3} g_3^2 + 3 g_2^2 + \frac{1}{15} g_1^2
-2 ( y_t^2 (3-n_{H_u}-n_Q -n_U) + y_b^2 (3-n_{H_d} -n_Q - n_D )) \delta_{i3},
\nonumber \\
\theta_{U,i} &=& \frac{16}{3} g_3^2 + \frac{16}{15} g_1^2
- 4 y_t^2 (3-n_{H_u} - n_Q - n_U ) \delta_{i3}\nonumber \\
\theta_{D,i}&=& 
\frac{16}{3} g_3^2 + \frac{4}{15} g_1^2
- 4 y_b^2 (3- n_{H_d} - n_Q - n_D ) \delta_{i3}, \nonumber \\
\theta_{L,i} &=& 3 g_2^2 + \frac{3}{5} g_1^2
-2 y_\tau^2 ( 3- n_{H_d} - n_L - n_E ) \delta_{i3}\nonumber \\
\theta_{E,i} &=& 
\frac{12}{5} g_1^2
- 4 y_\tau^2 (3-n_{H_d} -n_L -n_E ) \delta_{i3}, \nonumber \\
\theta_{H_u} &=& 3 g_2^2 + \frac{3}{5} g_1^2
- 6 y_t^2 ( 3- n_{H_u} - n_Q - n_U )\nonumber \\
\theta_{H_d} &=& 
3 g_2^2 + \frac{3}{5} g_1^2
-6 y_b^2 ( 3- n_{H_d} - n_Q - n_D )
-2 y_\tau^2 ( 3- n_{H_d} - n_L - n_E ).
\end{eqnarray}
As in the case of $\gamma_i$, $\theta^\prime_i$ is the same as $\theta_i$.

%\pagebreak
%\section{Bibliography}

\end{document}